\documentclass[20pt,a4paper,notitlepage]{article}

\usepackage[T1]{fontenc}
\usepackage{amsmath}
\usepackage{amsfonts}
\usepackage{amssymb}
\usepackage{graphicx}
\usepackage[margin=1in]{geometry}
\usepackage[varg]{txfonts}
\usepackage{hyperref,natbib}
\usepackage{bm,caption,subcaption}
\usepackage{authblk}
\usepackage{xcolor}
\allowdisplaybreaks

\title{A theoretical study on the dynamics of a compound vesicle in linear shear flow}
\author{Kumari Priti Sinha and Rochish M Thaokar \\Department of Chemical Engineering, Indian Institute of Technology Bombay, Powai, Mumbai-400076, India. \\ E-mail: pritisinha.026@gmail.com, rochish@che.iitb.ac.in}

\begin{document}

\maketitle

\begin{center} \textbf{\Large Abstract} \end{center} 
The dynamics of a nucleate cell in shear flow is of great relevance in cancer cells and circulatory tumor cells where they dominate the dynamics of blood. Buoyed by the success of Giant Unilamellar vesicles in explaining the dynamics of anucleate cells such as Red Blood cells, compound vesicles have been suggested as a simple model for nucleate cells. In this work, a theoretical model is presented to study the  deformation and dynamics of a compound vesicle in linear shear flow using
small deformation theory and spherical harmonics with higher order approximation to the membrane forces. A coupling of viscous and membrane stresses at the membrane interface of the two vesicles results in highly nonlinear
shape evolution equations for the inner and the outer vesicle which are solved numerically. The results indicate that size of the inner vesicle ($\chi$) does not affect the tank-treading dynamics of the outer vesicle. The inner vesicle admits a greater inclination angle than the outer vesicle. However, the transition to trembling/swinging and tumbling is significantly affected. The inner and outer vesicle exhibit identical dynamics in most of the modified viscosity ($\Lambda_{an}$)-shear rate ($S$) parameter space. At moderate size of the inner vesicle ($\chi$), a swinging mode is observed for the inner vesicle while the outer vesicle exhibits tumbling. The inner vesicle also exhibits modification of the TU mode to IUS (Intermediate Tumbling Swinging) mode. Moreover, synchronization of the two vesicles at higher $\chi$ and a capillary number sensitive motion at lower $\chi$ is observed in the tumbling regime. These results are in accordance with the few experimental observations reported by \cite{steinberg2014PRL}. A reduction in the inclination angle is observed with an increase in size of the inner vesicle ($\chi$) when the inner vesicle is a solid inclusion. Additionally a very elaborate phase diagram is presented in the $\Lambda_{an}-S$ parameter space, which could be tested in future experiments or numerical simulations.

\section{Introduction}
The importance of the understanding of the dynamics of biological cells such as Red Blood Cells (erythrocytes) and White Blood Cells (leukocytes) needs no elaboration. For example, the flow of RBCs (\cite{fischer1978SCIENCE, skalak1982JFM, tay1984BJ, pozrikidis2003ABE, goldsmith2004PRSL, vlah2009CRP, vla2013CRP, dupire2012PNAS, dodson2011PRE, sui2008POF}) and their resulting morphologies in blood stream has been well investigated.
Towards this, the studies on Giant unilamellar vesicles, GUVs, in shear flow have provided a very simple model for understanding RBC mechanics. The GUVs (\cite{lipowsky1995text}) are droplets separated from aqueous exterior by a two-dimensional, incompressible, bilayer lipid membranes, and exhibit great resemblance to anucleate cells such as red blood cells (RBCs). The inner and outer fluids are typically Newtonian.

 A single unilamellar vesicle shows different steady and unsteady dynamic responses to linear shear flow which have been categorized as specific regimes of flow. These include  the tank-treading regime (TT), where a vesicle is deformed into an ellipsoidal shape and orients itself at a fixed inclination angle with the direction of flow and the membrane rotates in a tank-treading manner; the trembling(TR)/Vacillating Breathing(VB)/swinging(SW) regimes, where a vesicle oscillates about a positive inclination angle and undergoes deformation as well; and lastly the tumbling (TU) regime, where a vesicle undergoes periodic flipping motion about the flow direction. These dynamical modes depend upon different system parameters such as a viscosity contrast between the internal and the external fluids, the shear rate, or more specifically the ratio of elongational and rotational component of the linear flows, and the membrane excess area. In their equilibrium as well as nonequilibrium dynamical states, the vesicles preserve their area and volume through the constraint of membrane incompressibility and membrane impermeability (fixed internal volume). Starting from \cite{skalak1982JFM} who presented a first theoretical study on red blood cell dynamics in shear flow by fixing the initial shape as an ellipsoid, thereby admitting only TT and TU modes, a single vesicle under shear flow has been extensively studied in the literature using experiments (\cite{haas1997PRE, steinberg2005PRL, podgorski2006EPJE, steinberg2006PRL,  steinberg2007PRL,  misbah2009CRP,steinberg2009PNAS,  vsteinberg2009PRL, steinberg2011POF, seifert2014ACIS}), theory (\cite{fujitani1995FDR, haas1997PRE, seifert1999EPJB, misbah2004PRE2, misbah2006PRL, vlahovska2007PRE, podgorski2007EPJE, misbah2007PRE, lebedev2007PRL, lebedev2008JOP, seifeert2008EPJE, misbah2009CRP, misbah2009PRE, misbah2012PRE, Kaoui2012SM, seifert2013PRL, vla2013CRP, guedda2014PRE, seifert2014ACIS}), and simulations (\cite{lipowsky1996PRL,  misbah2004PRE1, noguchi2004PRL, noguchi2005PRE, cmisbah2012PRE}) to understand its highly non-linear dynamical response. The dynamics predicted by a vesicle in shear flow has been validated to be quite similar to a RBC in shear flow (\cite{Annie2007PRL, vlahov2009CRP, dupire2012PNAS}).

Motivated by the success of the GUV as a model system to understand RBC dynamics in shear flow, a natural choice to model nucleate cells such as white blood cells (leukocytes) is a compound vesicle (\cite{schmid1980BLOOD, vlah2011PRL,kaoui2013SM, steinberg2014PRL}). A compound vesicle is a bilamellar closed vesicular system in which two vesicles are concentric and separated by an aqueous suspension. One can then assume the inner vesicle to mimic the nucleus with the outer vesicle can represent the cell membrane.

Although the volume concentration of leukocytes in very small compared to RBCs in normal blood, their role is important in the microcirculation during blood flow and in blood rheology, especially in cancer tissues when their concentration increases significantly. Apart from the presence of a nucleus in leukocytes,  their larger size ($15-20 \mu m$)  compared to $\sim8 \mu m$ of a RBCs results in a greater resistance to flow than that due to RBCs. The leukocytes show abnormal flow characteristics in patients with ischaemia (\cite{nash1988, nash1991}) and therefore their study in fluid flow  has attracted interest in the last few decades (\cite{Moazzam1997, Dong1999, Chang2000, Takeishi2014, Mulki2014}).

Another instance of relevance is Circulation Tumor Cells (CTCs) (\cite{catherine2014, katarzyna2012, ramdane2013}) which have structural similarity with a compound vesicle due to the presence of a cell nucleus that is surrounded by cell actin cortex (comparatively larger size greater than 8 micrometer) as compared to haematopoietic cells which have a small deformability. Their very low concentration makes early cancer detection difficult. After their detachment from the primary tumor site, they enter into blood stream to reach distant secondary sites where they again grow and form colonies. At the early stage of cancer most of CTCs travel as isolated cells after their release into flow stream and undergo shape deformation due to hydrodynamic force and shear exerted by flow stream in blood vessels during there translocation. Thus analysis of the dynamical behavior of CTCs when they take part in the circulation system as a single entity is important in early detection of cancer.

Towards this, a compound vesicle in shear flow can be considered to be a simple model to understand the individual behavior of leukocytes or CTCs  under external shear flow.

\cite{vlah2011PRL} presented 2D boundary integral simulation and 3D analytical results on a compound vesicle system in which the inclusion is a highly viscous solid particle. They showed that the hydrodynamic interactions become important on increasing the size of the inclusion which increases the effective internal viscosity, subsequently leading to a TT-TU transition at much lower critical viscosity ratios as compared to single vesicles. A strong VB or SW mode was observed when the inclusion inside is ellipsoidal.  

\cite{kaoui2013SM} investigated the dynamics of 2-D bilamellar vesicle at low and high capillary number using the Lattice Boltzmann method and Immersed Boundary method, and investigated the effect of increasing the size of the inner vesicle on the apparent internal viscosity of the outer vesicle. Their results are in agreement with those of \cite{vlah2011PRL}. A transition from TU to TR was reported at higher $Ca$, especially at higher $\chi$. Similar simulation studies have been conducted on compound elastic capsules (\cite{Luo2016POF}) wherein, a transition from swinging to tumbling was seen to occur at vanishing viscosity mismatch through increasing the inner capsule size alone to a critical value regardless of the initial shape of the nonspherical compound capsule (i.e., prolate or oblate). 

Recently, \cite{steinberg2014PRL} conducted the only experimental study to date on a compound vesicle under shear flow. In this work, they present the dynamics of a compound vesicle for a given set of flow parameters ($S$ and $\Lambda_{an}$) when the inner vesicle size is small or large. They observed that in the TT regime, the inclination angle of the outer vesicle was lower than the inner vesicle, but the transition $\Lambda^{crit}_{an}$ was found to be irrespective of the volume fraction. This is in sharp contrast to the studies of \cite{vlah2011PRL} and \cite{kaoui2013SM} who find sensitive dependence of the TT-TU transition on the size of the inner vesicle ($\chi$). Although in the experiments of the Steinberg group, the TT dynamics was found to be nearly independent of the filling fraction, the TU and TR regimes were significantly affected by the size of the inner vesicle ($\chi$). A SW mode, instead of the TR mode was experimentally observed, in agreement with simulations (\cite{vlah2011PRL, kaoui2013SM}). De-synchronization and synchronization of the inner and outer vesicles was observed at lower and higher $\chi$, respectively.

The experiments of Levant and Steinberg are few, only 6 data points are available for comparison. While the study of \cite{vlah2011PRL} is restricted to a solid inclusion, that of \cite{kaoui2013SM} uses 2-D simulations and thereby, very few simulations have been conducted in their work. It is therefore necessary to understand this very important system of great biological relevance in greater details. Several questions need answers: the discrepancy of the Steinberg's group experiments versus simulations in the TT regimes, the appearance of the SW mode, the synchronization and de-synchronization observed in the experiments of the Steinberg group, but more importantly a much more elaborate phase diagram with a wider range of $\Lambda_{an}$ and $S$ needs to be explored. 

In this work, we present analytical calculations to higher order perturbations in the membrane forces using spherical and vector spherical harmonics, akin to previous works (\cite{misbah2007PRE, lebedev2007PRL, misbah2009PRE, Sinha2018PRE}). 
 
In section two of this work we solve the Stokes equations for the fluid flow part to get, the velocity and the pressure fields in each regime of the three regions that is outer, annular and inner core and calculate the hydrodynamic (shear) stresses. A second order analysis in the small deformation regime is conducted to calculate the membrane forces due to initial asphericity of the vesicle shape. The evolution equations for the membrane surface due to advection by surrounding fluid is obtained by solving the Stokes equation with the velocity and stress boundary conditions at the two surfaces. The dynamical evolution equations are derived which predict modes such as: TT, SW, and TU. In the last section we present results which show the role of different system parameters in predicting different types of modes and their transition, and finally make some comparisons of our theory with the experimental and simulation data in the literature. As a limiting case we also discuss the condition when the viscosity of the inner vesicle is very large such that it behaves as a solid particle and compare the outcomes of this analysis with \cite{vlah2011PRL} simulation and analytic results.      \\  
 
 \section{Mathematical model}

\begin{figure}[tbp]
 \centering
 \includegraphics[width=.6\textwidth]{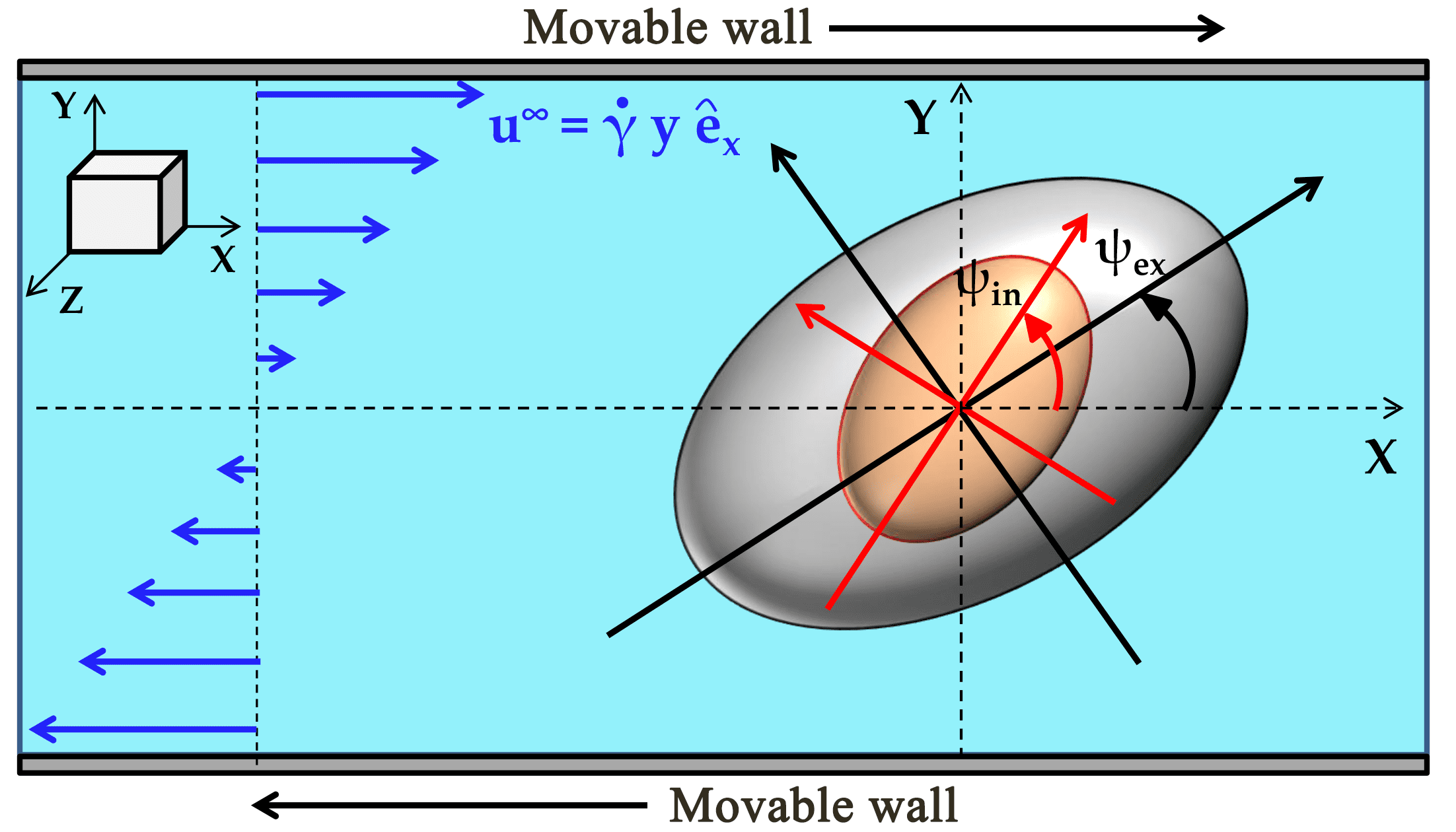}
 \caption{Schematic view of a compound vesicle subjected to linear shear flow}
 \label{compvesicle-schematic}
 \end{figure}

The system (figure \ref{compvesicle-schematic}) consists of a compound vesicle suspended in an aqueous medium and is subjected to a linear shear flow. The imposed shear flow is defined by the velocity field 
\begin{equation}
\bm{u^\infty}= s (x \bm{\hat{e}_y}+y \bm{\hat{e}_x})+ \omega (y \bm{\hat{e}_x}-x \bm{\hat{e}_y})
\end{equation}
where $s$ and $\omega$ are the strengths of the elongational and rotational component (vorticity) of shear flow in the plane of shear which is X-Y plane (for simple shear flow,  $s=\omega=\dot{\gamma}/2$ where $\dot{\gamma}$ is the shear rate). The vorticity of the imposed flow is in the Z-direction, $\bm{\hat{e}_x}$ and $\bm{\hat{e}_y}$ are unit vectors along X and Y directions. $\psi_{in}$ and $\psi_{ex}$ are the inclination angles of the inner and the outer vesicle measured from direction of flow through the central axis, i.e. the X axis. The initial shapes of the two vesicles, the inner as well as the outer vesicles, are quasi-spherical with excess area $\Delta_{in}$ and $\Delta_{ex}$ associated with each vesicle respectively. $\chi= R_{in}/R_{ex}$ represents the radius ratio of the inner and the outer vesicle in undeformed state. The fluids in each of the three regions are incompressible and Newtonian and have different viscosity: $\mu_{in}$, $\mu_{an}$ and $\mu_{ex}$ for inner, annular and exterior fluids respectively. Their ratios are represented by two new variables: $\lambda_{in}=\mu_{in}/\mu_{ex}$ for viscosity contrast of inner and exterior fluid, $\lambda_{an}=\mu_{an}/\mu_{ex}$ for viscosity contrast of annular and exterior fluid, and $\lambda_{ex}=\mu_{ex}/\mu_{ex}=1$. The viscosity of the lipid bilayer membrane which bounds the vesicle is assumed to be very small ($\sim10^{-9}$ Ns/m) and the viscous dissipation in the membrane is neglected. Our model is based on the assumption that the energy associated with the different deformation modes is much higher than the thermal energy (around $25\kappa_b T$) and hence the effect of thermal noise is also neglected. \\

\subsection{Dimensionless parameters}

The characteristic length scale used for non-dimensionalization is the radius of the external vesicle $R_{ex}$, the time scale is the inverse shear rate $\dot{\gamma}^{-1}$, all the stress are scaled by $\dot{\gamma}\mu_{ex}$, the velocity scale is $\dot{\gamma} R_{ex}$, and the membrane tension is scale by $\kappa_b/R_{ex}^2$.

\subsection{Description of the velocity field}

The velocity fields ($u_{in}, u_{an}, u_{ex}$) as well as the pressure fields ($p_{in}, p_{an}, p_{ex}$) in each of the regions are described by the continuity equations and the nondimensional Stokes equations in low Reynolds number limit as,
\begin{align}
&\nabla p_{j}-\lambda_{j}\nabla^2 \boldsymbol{u_{j}}=0 &\\
&\nabla.\boldsymbol{u_{j}}=0&
\end{align}
where, $j=in, an, ex$ to represent the inner, annular and the outer (exterior) region.\\

The velocity fields induced (\cite{vlahovska2007PRE}) in the three regions are given by the solution of Stokes equation (\cite{lamb1932text})
\begin{align}
&\ \bm{u_{\bm {in}}}= C_{lm0}^g \bm{u_{\bm{lm0}}^g}+C_{lm1}^g \bm{u_{\bm{lm1}}^g}+C_{lm2}^g \bm{u_{\bm{lm2}}^g} &\\
&\ \bm{u_{\bm{an}}}=(C_{lm0}^{da} \bm{u_{\bm{lm0}}^d}+C_{lm1}^{da} \bm{u_{\bm{lm1}}^d}+C_{lm2}^{da} \bm{u_{\bm{lm2}}^d})+(C_{lm0}^{ga} \bm{u_{\bm{lm0}}^g}+C_{lm1}^{ga} \bm{u_{\bm{lm1}}^g}+C_{lm2}^{ga} \bm{u_{\bm{lm2}}^g})&\\
&\ \bm{u_{\bm{ex}}}=(C_{lm0}^{d} \bm{u_{\bm{lm0}}^d}+C_{lm1}^{d} \bm{u_{\bm{lm1}}^d}+C_{lm2}^d \bm{u_{\bm{lm2}}^d})+(C^\infty_{lm0} \bm{u_{\bm{lm0}}^g}+C^\infty_{lm1} \bm{u_{\bm{lm1}}^g}+C^\infty_{lm2} \bm{u_{\bm{lm2}}^g}) &
\end{align}  
where coefficients $C_{lm0}^g, C_{lm1}^g, C_{lm2}^g$ are the coefficients of the growing harmonics for velocity eigen functions in the inner region, $C_{lm0}^{ga}, C_{lm1}^{ga}, C_{lm2}^{ga}, C_{lm0}^{da}, C_{lm1}^{da}, C_{lm2}^{da}$
are the growing and decaying harmonics in the annular region and $C_{lm0}^d, C_{lm1}^d, C_{lm2}^d$ are the decaying harmonics in the outer region. All these unknown coefficients are determined by using the velocity continuity, the membrane incompressibility and the stress balance conditions at the two interfaces of the two vesicles. Note that viscosity in the Navier Stokes equation is accounted for the stress balance.

$C^\infty_{lm0}, C^\infty_{lm1}, C^\infty_{lm2}$ are the coefficients associated with the applied unperturbed external flow (when $u_{ex}=u^{\infty}$ ) and they depend on the elongational ($s$) as well as the rotational ($\omega$) components of the applied flow described as (for $l=2$, $m=\pm2$ mode) 
\begin{align}
&\ C^\infty_{2\pm22}=\mp2 i \left(\sqrt{\frac{2 \pi}{15}}\right) s & \\
&\ C^\infty_{2\pm20}=\mp2 i \left(\sqrt{\frac{\pi}{5}}\right) s & \\
&\ C^\infty_{1\pm10}=2 i \left(\sqrt{\frac{2\pi}{3}}\right) \omega &
\end{align}
$u_{lm0}^g, u_{lm1}^g, u_{lm2}^g$ and $u_{lm0}^d, u_{lm1}^d, u_{lm2}^d$ are the growing and decaying velocity eigen functions (\cite{vlahovska2007PRE}).
\begin{align}
&\ u_{lm0}^g=\frac{1}{2} r^{l-1}\left(-(l+1)+(l+3)r^{2}\right)y_{lm0}
-\frac{1}{2}r^{l-1}[l(l+1)]^{1/2}\left(1-r^{2}\right)y_{lm2} & \\
&\ u_{lm1}^g=r^{l}y_{lm1} &\\
&\ u_{lm2}^g=\frac{1}{2}r^{l-1}(3+l)\left(\frac{l+1}{l}\right)^{1/2}\left(1-r^{2}\right)y_{lm0} +\frac{1}{2}r^{l-1}\left(l+3-(l+1)r^{2}\right)y_{lm2}& \\
&\ u_{lm0}^d=\frac{1}{2} r^{-l}\left(2-l+lr^{-2}\right)y_{lm0}+\frac{1}{2}r^{-l}[l(l+1)]^{1/2}\left(1-r^{-2}\right)y_{lm2} &\\
&\ u_{lm1}^d=r^{-l-1}y_{lm1} &\\
&\ u_{lm2}^d=\frac{1}{2}r^{-l}(2-l)\left(\frac{l}{l+1}\right)^{1/2}\left(1-r^{-2}\right)y_{lm0}
+\frac{1}{2}r^{-l}\left(l+(2-l)r^{-2}\right)y_{lm2}&
\end{align} 
Here $y_{lm0}, y_{lm1}, y_{lm2}$ are vector spherical harmonics which are defined as
\begin{align}
&\ y_{lm0}=\frac{1}{\sqrt{l(l+1)}}\frac{\partial Y_{lm}}{\partial \theta} \hat{e}_\theta+\frac{im}{\sqrt{l(l+1)}}\frac{Y_{lm}}{\sin \theta}\hat{e}_\phi &\\
&\ y_{lm1}=-\frac{m}{\sqrt{l(l+1)}}\frac{Y_{lm}}{\sin\theta} \hat{e}_\theta-\frac{i}{\sqrt{l(l+1)}}\frac{\partial Y_{lm}}{\partial \theta}\hat{e}_\phi &\\
& y_{lm2}=\hat{r}Y_{lm} &
\end{align}
with scalar spherical harmonics
\begin{equation}
\ Y_{lm}(\theta, \phi)=\sqrt{\frac{2l+1}{4\pi}\frac{(l-m)!}{(l+m)!}}(-1)^m P_{lm}(\cos\theta)e^{im\phi}
\end{equation} 

\subsection{Solution for hydrodynamic stress}

The dimensional hydrodynamic stresses in the core, annular, and external ($j=in, an, ex$) regions of a compound vesicle are given by
\begin{equation}
\ \bm{\tau}^{h, j}=-p_{j}\bm{I}+\mu_{j} \left[\nabla u_{j}+(\nabla u_{j})^T\right]
\end{equation}
where $\textbf{I}$ is the identity matrix and superscript $T$ represents the transpose of the matrix, $p$ and $u$ are the pressure and the velocity fields, respectively, and $\mu$ is the fluid viscosity. The tractions due to the hydrodynamic stresses exerted on the inner and outer surface of inner vesicle ($r=\chi$), similarly on the inner and outer surface of outer vesicle ($r=1$) are given by
\begin{align}
&\ \tau^{h,in}.\bm{\hat{e}_r}=-p_{in} \bm{\hat{e}_r}+\mu_{in} Z_{in} \qquad \mathrm{at\ r=\chi} \label{hstress-1}&\\
&\ \tau^{h,an}.\bm{\hat{e}_r}=-p_{an} \bm{\hat{e}_r}+\mu_{an} Z_{an} \qquad \mathrm{at\ r=\chi, 1} \label{hstress-2}&\\
&\ \tau^{h,ex}.\bm{\hat{e}_r}=-p_{ex} \bm{\hat{e}_r}+\mu_{ex} Z_{ex} \qquad \mathrm{at\ r=1} \label{hstress-3}&
\end{align}
with
\begin{equation}
\  Z_{j}=\bm{\hat{e}_r}.\left(\nabla u_{j}+(\nabla u_{j})^T\right)=r \frac{d}{dr}\left(\frac{u_{j}}{r}\right)+\frac{1}{r}\nabla((u_{j}.\bm{\hat{e}_r}) r)
\end{equation}
where $j=in, an, ex$. The expressions for $Z_{in}|_{r=\chi}, Z_{an}|_{r=\chi,1}, Z_{ex}|_{r=1}$ are provided in the Appendix \ref{App1}.\\
The pressure field in each region can also be expressed in terms of the growing/decaying harmonics since it is a solution to the Laplace equation ($\nabla^2 p=0$)
\begin{align}
&\ p_{in}=A^g_{in} r^l Y_{lm}&\\
&\ p_{an}=A^g_{an} r^l Y_{lm}+A^d_{an} r^{-l-1} Y_{lm} &\\
&\ p_{ex}=A^g_{ex} r^l Y_{lm}+A^d_{ex} r^{-l-1} Y_{lm} &
\end{align}
 where $A^g_{in}, A^g_{an}, A^d_{an}, A^g_{ex}, A^d_{ex}$ are the pressure coefficients obtained by solving the momentum equations for core, annular and external fluids (i.e., $\nabla p_j=\mu_{j} \nabla^2 u_{j}$). Their full expressions are provided in the Appendix \ref{App2}.\\
The substitution of Z values ($Z_{in}, Z_{an}, Z_{ex}$) and pressure coefficients ($A^g_{in}, A^d_{an}, A^g_{an}, A^d_{ex}, A^g_{ex}$) from Appendix \ref{App1} and \ref{App2} into the equations (\ref{hstress-1}) to (\ref{hstress-3}) gives the net hydrodynamic traction on the surface of the inner vesicle (at $r=\chi$) due to the core fluid and the annular fluid in a matrix form as (from eq.(\ref{hstress-1}) and eq.(\ref{hstress-2}) for $r=\chi$)
\begin{align}
\tau^{h,in}.\bm{\hat{e}_r}|_{r=\chi}=
\begin{pmatrix} C^g_{lm0} & C^g_{lm1} & C^g_{lm2} \end{pmatrix}
  \begin{pmatrix}
    \chi^{l-2}(1-l^2+l(l+2)\chi^2) & 0 & \frac{\chi^{l-2}(-3-l(l+2)(\chi^2-1)}{\sqrt{\frac{l}{l+1}}}  \\
    0 & (l-1)\chi^{l-1} & 0  \\
    \frac{\chi^{l-2}(-3\chi^2+l(l-1)(\chi^2-1)}{\sqrt{\frac{l}{l+1}}}  & 0 & \frac{\chi^{l-2}(l(l-1)(l+3)+(3+4l-l^3)\chi^2}{l}
  \end{pmatrix}
\begin{pmatrix} y_{lm0} \\ y_{lm1} \\ y_{lm2} \end{pmatrix}  
\end{align}
\begin{align}
\tau^{h,an}.\bm{\hat{e}_r}|_{r=\chi}=
\begin{pmatrix} C^{ga}_{lm0} & C^{ga}_{lm1} & C^{ga}_{lm2} \end{pmatrix}
  \begin{bmatrix}
    \chi^{l-2}(1-l^2+l(l+2)\chi^2) & 0 & \frac{\chi^{l-2}(-3-l(2+l)(-1+\chi^2))}{\sqrt{\frac{l}{l+1}}}  \\
    0 & (l-1)\chi^{l-1} & 0  \\
    \frac{\chi^{l-2}(-3\chi^2+l(l-1)(-1+\chi^2))}{\sqrt{\frac{l}{l+1}}} & 0 & \frac{\chi^{l-2}(l(l-1)(l+3)+(3+4l-l^3)\chi^2)}{l} 
  \end{bmatrix}
  \begin{pmatrix} y_{lm0} \\ y_{lm1} \\ y_{lm2} \end{pmatrix}+\nonumber\\
  \begin{pmatrix} C^{da}_{lm0} & C^{da}_{lm1} & C^{da}_{lm2} \end{pmatrix}\nonumber\\
  \begin{bmatrix}
    \chi^{-l-3}(-l(l+2)+(l^2-1)\chi^2) & 0 & \sqrt{\frac{l}{l+1}}\chi^{-l-3}(4-\chi^2+l^2(\chi^2-1))\\ 
    0 & -(2+l)\chi^{-l-2} & 0 \\
    \sqrt{\frac{l}{l+1}}\chi^{-l-3}(2+\chi^2-l(l+3)(\chi^2-1)) & 0 & \frac{\chi^{-l-3}(-4+l(-4+\chi^2+l(1+l-(3+l)\chi^2)))}{l+1} 
  \end{bmatrix}
  \begin{pmatrix} y_{lm0} \\ y_{lm1} \\ y_{lm2} \end{pmatrix}
\end{align}
and the tractions on the surface of the outer vesicle (at $r=1$) due to the annular fluid and the outer fluid as (from eq.(\ref{hstress-2}) and eq.(\ref{hstress-3}) for $r=1$)
\begin{align}
\tau^{h,an}.\bm{\hat{e}_r}|_{r=1}=&
  \begin{pmatrix} C^{ga}_{lm0} & C^{ga}_{lm1} & C^{ga}_{lm2} \end{pmatrix}
  \begin{bmatrix}
    1+2l & 0 & -3 \sqrt{\frac{l+1}{l}}   \\
    0 & -l+1 & 0  \\
    -3 \sqrt{\frac{l+1}{l}}  & 0 & 1+\frac{3}{l}+2l 
  \end{bmatrix}
  \begin{pmatrix} y_{lm0} \\ y_{lm1} \\ y_{lm2} \end{pmatrix} + \nonumber\\&  
    \begin{pmatrix} C^{da}_{lm0} & C^{da}_{lm1} & C^{da}_{lm2} \end{pmatrix}
\begin{bmatrix}
    -1-2l & 0 & 3\sqrt{\frac{l}{l+1}}  \\
    0 & -2-l & 0 \\
    3\sqrt{\frac{l}{l+1}} & 0 & -1-2l-\frac{3}{l+1}  
  \end{bmatrix}
  \begin{pmatrix} y_{lm0} \\ y_{lm1} \\ y_{lm2} \end{pmatrix}  
\end{align}
\begin{align}
\tau^{h,ex}.\bm{\hat{e}_r}|_{r=1}=&
  \begin{pmatrix} C^\infty_{lm0} & C^\infty_{lm1} & C^\infty_{lm2} 
  \end{pmatrix}
  \begin{bmatrix}
    1+2l & 0 & -3 \sqrt{\frac{l+1}{l}}  \\
    0 & -l+1 & 0  \\
    -3 \sqrt{\frac{l+1}{l}} & 0 & 1+\frac{3}{l}+2l 
  \end{bmatrix}
  \begin{pmatrix} y_{lm0} \\ y_{lm1} \\ y_{lm2} \end{pmatrix} + \nonumber\\&  
    \begin{pmatrix} C^d_{lm0} & C^d_{lm1} & C^d_{lm2} \end{pmatrix}
\begin{bmatrix}
    -1-2l & 0 & 3\sqrt{\frac{l}{l+1}}  \\
    0 & -2-l & 0 \\
    3\sqrt{\frac{l}{l+1}} & 0 & -1-2l-\frac{3}{l+1}  
  \end{bmatrix}
  \begin{pmatrix} y_{lm0} \\ y_{lm1} \\ y_{lm2} \end{pmatrix}  
\end{align}
For both the inner as well as the outer vesicle the components of $y_{lm0}$ and $y_{lm2}$ represents the tangential and the normal stresses respectively. The jump in the hydrodynamic stresses across the membrane of the two vesicles is balanced by the respective bending and the tension forces at the membrane interfaces.

\subsection{Small deformation analysis and membrane stress}
To describe the membrane forces in a vesicle deformed under external shear flow, the nondimensional shape of a vesicle is expanded using spherical harmonics ($Y_{lm}(\theta, \phi)$). Any point on the surface of a vesicle is expressed as (\cite{vlahovska2007PRE})
\begin{equation}
\ R_s^{in,ex}(\theta, \phi, t)= \alpha_{in,ex}+\sum_{l=2}^{\infty}\sum_{m=-l}^{l} f_{lm}(t)^{in,ex} Y_{lm}(\theta, \phi) \label{surface}
\end{equation}
where $f_{lm}(t)^{in,ex}$ is the time-dependent nondimensional amplitude of the deformation modes associated with the inner and the outer vesicle, respectively, and $R_s^{in,ex}$ are the radial positions of points on the surfaces of the two vesicles. Note all length scales are nondimensionalized by $R_{ex}$. The volume conservation of the fluids in the vesicles can be used to determine $\alpha_{in,ex}$ in eq.(\ref{surface}) and are provided in Appendix \ref{App3} to first order approximation.\\
Under the asymptotic limit of small deviation from a spherical shape, the relation between the amplitude of the shape perturbation and the excess area ($\Delta$) is given by the area conservation constraint (provided in Appendix \ref{App3}) which results in (\cite{vlahovska2007PRE})
\begin{equation}
\ \Delta_{j}=\sum_{lm}a(l)f_{lm}^j f_{lm}^{j*}
\end{equation}
where $*$ represents complex conjugate and
\begin{equation}
\ a(l)=\frac{(l+2)(l-1)}{2}
\end{equation}
From the above equations we obtain  $\Delta^j=2\left(2 f_{22}^j f_{2-2}^j+(f_{20}^j)^2\right)$ where $j=ex,in$. It suffices to consider only the $m=-2,0,2$ modes since the symmetry breaking $m=1$ mode does not contribute to vesicle deformation.\\
Bending resistance offered by a vesicle under shear flow is given by $\tau^{mem}=\tau^{mem}_{2m2}+\tau^{mem}_{2m0}$. Here normal ($\tau^{mem}_{2m2}$) and tangential ($\tau^{mem}_{2m0}$) membrane stress are
\begin{align}
&\ \tau^{mem}_{2m2}=-Ca (4H(H^2-K)+2 \nabla_s^2 H)+2\sigma H  \label{mnstr}&\\
&\ \tau^{mem}_{2m0}=-Ca\nabla_s \sigma \label{mtstr}&
\end{align}
where $Ca=\dot{\gamma}\mu_{ex}R_{ex}^3/\kappa_b$ is the flow capillary number, membrane tension ($\sigma=\tilde{\sigma}/(\kappa_b/R_{ex}^2)$) is a sum of uniform ($\sigma_u$) and nonuniform tension ($\sigma_{nu}$) i.e., $\sigma=\sigma_u+\sigma_{nu}$ with $\sigma_{nu}=\sum_{lm}\sigma_{lm} Y_{lm}(\theta, \phi)$ which varies along the surface with $\theta$ and $\phi$. The full expression for the mean curvature $H$ and the Gaussian curvature $K$ are provided in Appendix \ref{App4} and \ref{App5} in terms of spherical harmonics up to second order correction in the shape amplitudes. The uniform tension is obtained by using the area constraint ($\dot{\Delta}_{in,ex}=0$, i.e., $\left(\sum_{lm} a(l)f_{lm}^{j *}\dot{f}_{lm}^j=0\right)$) on each vesicle and the nonuniform tension is obtained by using the tangential stress balance. In the first order approximation in shape perturbation, the $H(H^2-K)$ term is identically zero, but the term is non-zero in the second order approximation. The tension ($\sigma$) in normal membrane stress balance contains both the uniform and the nonuniform parts while the tangential membrane stress balance contains only the non-uniform part of the tension (Appendix \ref{App4} and \ref{App5}).

\subsection{Membrane incompressibility condition}
The membrane incompressibility condition indicates that the surface divergence of the velocity field vanishes on the vesicle interface. That is
\begin{equation}
\ \nabla_s.u^j=0
\end{equation}
In more general form after expansion in a spherical coordinate system
\begin{equation}
\ \frac{du_\theta^j}{d\theta}+u_{\theta}^j \left(\frac{\cos\theta}{\sin\theta}\right)+\left(\frac{1}{\sin\theta}\right)\frac{du_\phi^j}{d\phi}=-2u_r^j
\end{equation}
where $j=in, an, ex$\\
The relation between the velocity coefficients are obtained by solving the above equation for the outer and inner vesicle at $r=1, \chi$, respectively.

\subsection{Overall stress balance}
At the surface of the inner and outer vesicles, the membrane stresses up to second order approximation (provided in Appendix \ref{App4} and \ref{App5}) are balanced by the jump in the hydrodynamic stress. Thus the overall tangential stress balances across the inner and outer vesicles is given by
\begin{align}
&\ \lambda_{an}\tau_{2m0}^{h,an}-\lambda_{in} \tau_{2m0}^{h,in}=\tau_{2m0}^{mem,in} &\\
&\ \tau_{2m0}^{h,ex}-\lambda_{an} \tau_{2m0}^{h,an}=\tau_{2m0}^{mem,ex} &
\end{align}
Similarly the overall normal stress balance are
\begin{align}
&\ \lambda_{an}\tau_{2m2}^{h,an}-\lambda_{in} \tau_{2m2}^{h,in}=\tau_{2m2}^{mem,in} &\\
&\ \tau_{2m2}^{h,ex}-\lambda_{an} \tau_{2m2}^{h,an}=\tau_{2m2}^{mem,ex} &
\end{align}
in the above equations, for the inner and outer vesicles, the tangential stresses ($\tau_{lm0}$) are obtained from $y_{lm0}$ components of all the traction vectors and the normal stresses ($\tau_{lm2}$) are obtained from the $y_{lm2}$ components. The solution of the overall tangential stress balance provides the non-uniform tension distribution ($\sigma_{lm}^{in}, \sigma_{lm}^{ex}$) over inner and outer vesicles for all the three deformation modes $m=-2,0,2$ (full expressions are provided in Appendix \ref{App6}). The normal stress balances give the coefficients of the tangential velocity components ($C_{lm0}^g, C_{lm0}^{ga}$) while the incompressibility conditions (which relate the normal and tangential velocity coefficients) yield the normal velocity components ($C_{lm2}^g, C_{lm2}^{ga}$) (Appendix \ref{App6}).

\subsection{Kinematic condition for vesicle shape evolution} \label{kinematiceq}
A coupling of fluid stresses in the two vesicles, mediated by the hydrodynamic flow and the elastic membrane stresses due to deformation on surface of the inner and the outer vesicle results in the evolution of the coupled dynamic modes ($f_{2\pm2}, f_{20}$) of the two vesicles. The velocities of the adjacent fluids which are continuous across and normal to the vesicle surface give the evolution equation for the vesicle shape. Thus the non-linear evolution equations for the two vesicles are of the form (detailed expression is provided in Appendix \ref{App7}) 
\begin{align}
&\ \frac{\partial f_{2m}^{in}}{\partial t}=\frac{1}{2}\left(C_{2m2}^{g} (5-3 \chi^2)+\sqrt{6}C_{2m0}^{g}(\chi^2-1)\right)+\omega_{in} \frac{im}{2}f_{2m}^{in} \label{evoleq-1} &\\
&\ \frac{\partial f_{2m}^{ex}}{\partial t}=C_{2m2}^{ga}+\omega_{ex} \frac{im}{2}f_{2m}^{ex} \label{evoleq-2} &
\end{align}
here $C_{lm2}^g$ and $C_{lm2}^{ga}$ (full expression provided in Appendix \ref{App6} for $m=-2,0,2$) are functions of the viscosity contrast ($\lambda_{in}, \lambda_{an}$), capillary number ($Ca$), isotropic tensions ($\sigma_0^{in}, \sigma_0^{ex}$ to be determined using area conservation constraint), and the different modes of deformation ($f_{lm}^{in}, f_{lm}^{ex}$). Here $\omega_{ex}=\omega_{in}=\dot{\gamma}/2=1/2$ due to the non-dimensionalization used. The above two equations are highly nonlinear equations and solved numerically using Mathematica 10.\\
From the area constraint $\dot{\Delta}_{in,ex}=0$
\begin{equation}
\sum_{lm} \dot{f}_{lm}^j f_{lm}^{j*}=0
\end{equation}
Solving the above equations using the two evolution equations (\ref{evoleq-1}) and (\ref{evoleq-2}) yield the isotropic tensions ($\sigma_0^{in}, \sigma_0^{ex}$) in the inner and outer vesicle.
The evolution of shape of the vesicles is more conveniently described by the inclination ($\psi$) of the vesicle in the flow direction and by it's elliptical deformation ($R$, length of long axis) under flow. This is easier described by decomposing $f_{lm}$ into deformation amplitudes ($R_{in}, R_{ex}$) of the vesicles and inclination angles ($\psi_{in}, \psi_{ex}$) of the long axis of the deformed vesicle by using
\begin{equation}
\ f_{2\pm 2}^{j}=R_{j}(t)e^{\mp i\omega \psi_{j}(t)} 
\end{equation} 
Substitution of the above equations in the evolution equations eq.(\ref{evoleq-1}) and eq.(\ref{evoleq-2}) and separating the real and imaginary parts of the two equations provides equations for the deformation ($dR_{in}/dt, dR_{ex}/dt$) and the inclination angles ($d\psi_{in}/dt, d\psi_{ex}/dt$), respectively. The equations are lengthy and therefore not provided in this manuscript. However, in the $\chi \rightarrow 0$ limit one obtains the single vesicle results which are in agreement with the theoretical works   (\cite{misbah2006PRL,lebedev2007PRL}).\\
Further substitution of $R_j=\frac{\epsilon_j}{2}\cos\theta_j$ in the evolution equation (where $\epsilon_j=\sqrt{A_j}$) for $dR_{in}/dt, dR_{ex}/dt$ and $d\psi_{in}/dt, d\psi_{ex}/dt$ yields the final evolution equations in terms of the two dynamic variables $\psi$ and $\theta$, where $\theta$ evolution equation describes the vesicle shape and $\psi$ evolution equation yields the vesicle orientation.  \\
In the steady tank-treading regime, the deformation as well as the orientation of the inner and the outer vesicle is determined by equating 
$d\psi_{in}/dt=0, d\psi_{ex}/dt=0$.\\  
 	
\subsection{Results and discussion}
We present results for a compound vesicle in simple shear flow, wherein two vesicles are concentrically positioned at the geometric center of the coordinate system. The imposed shear flow induces deformation and flow in the outer vesicle, which, due to the hydrodynamic coupling through the annular fluid, induces deformation and motion in the inner vesicle. The size of the inner vesicle is varied from $\chi$= 0.1 - 0.9. Along with the size of inner vesicle ($\chi$), the dynamic modes of the two vesicles depend upon three other non-dimensional parameters: the viscosity contrasts across the two vesicles $\lambda_{an}$, $\lambda_{in}$, (we vary $\lambda_{an}$=1-22 while it is assumed that there is no viscosity contrast across the inner vesicle, $\lambda_{in}=1$) and the excess area $\Delta_{in}=\Delta_{ex}$ (which we keep fixed at 0.2 for both the vesicles, within the small deformation approximation). The capillary number is varied from $Ca$ = 0.01-2 and the flow is simple shear flow, that is the strengths of the rotational component, $\omega$ and the extensional part $s$ are equal. The results are generated by varying the viscosity ratios and the capillary number. However, the results are presented in the parameter space used by \cite{lebedev2007PRL} and the Steinberg group (\cite{vsteinberg2009PRL, steinberg2011POF, steinberg2014PRL}) ($\Lambda$ and $S$) using  the relation  $S = 14 \pi Ca/(6
\sqrt{3}\Delta)$ and $\Lambda_{an}$ ($\Lambda_{an} = (32 + 23 \lambda_{an})\sqrt{\Delta_{ex}}/(8\sqrt{30\pi})$, to make easy comparisons with the experimental results of the Steinberg group. Thus although all calculations are conducted for simple shear flow they are applicable for general linear flows through this transformation. All the results are presented in the non-dimensional parameter space of $\lambda_{an}-Ca$ or $\Lambda_{an}-S$. The value of $\Lambda_{in}$ is not varied or discussed. Such an approach is necessary since unlike \cite{lebedev2007PRL}, where the evolution equations have dependency on only two independent parameters, $S$ and $\Lambda$, the complexity of the lengthy equations in the case of a compound vesicle does not admit a natural appearance of these two independent parameters.
\begin{figure} [h] %[tbp]
    \hspace{0.0cm}
    \begin{subfigure}[b]{0.28\linewidth}       \includegraphics[width=\linewidth]{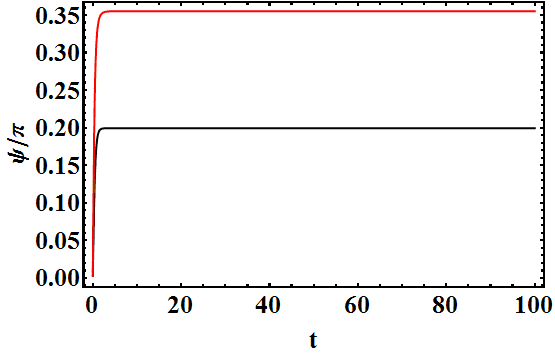}
     \caption{$\chi=0.3, \lambda_{an}=1, \Lambda_{an}=0.517$}
     \end{subfigure}
     \hspace{1.2cm}
      \begin{subfigure}[b]{0.28\linewidth}
      \includegraphics[width=\linewidth]{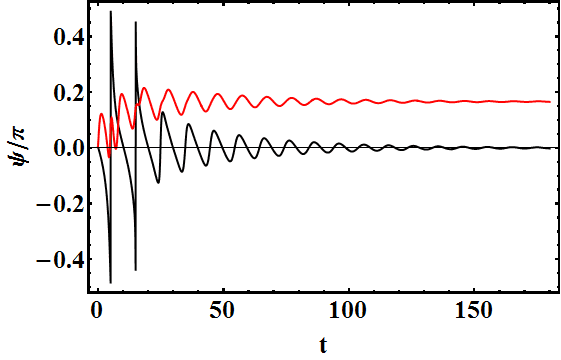}
      \caption{$\chi=0.3, \lambda_{an}=8, \Lambda_{an}=1.244$}
     \end{subfigure} 
 \hspace{1.2cm}
      \begin{subfigure}[b]{0.28\linewidth}
      \includegraphics[width=\linewidth]{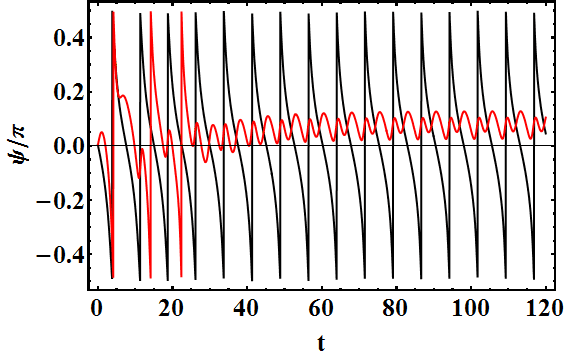}
      \caption{$\chi=0.3, \lambda_{an}=12, \Lambda_{an}=1.774$}
     \end{subfigure} 
    \hspace{0.12cm}
    \begin{subfigure}[b]{0.28\linewidth}       \includegraphics[width=\linewidth]{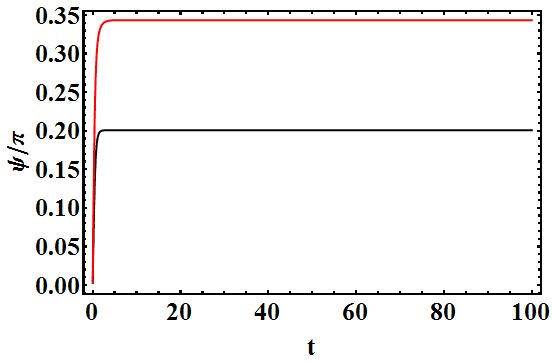}
     \caption{$\chi=0.5, \lambda_{an}=1, \Lambda_{an}=0.517$}
     \end{subfigure}
     \hspace{0.01cm}
      \begin{subfigure}[b]{0.28\linewidth}
      \includegraphics[width=\linewidth]{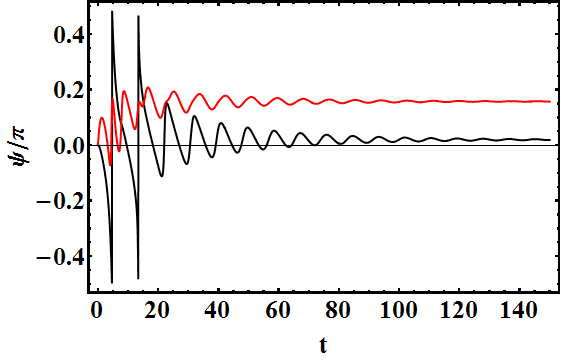}
      \caption{$\chi=0.5, \lambda_{an}=8, \Lambda_{an}=1.244$}
     \end{subfigure} 
 \hspace{0.12cm}
      \begin{subfigure}[b]{0.28\linewidth}
      \includegraphics[width=\linewidth]{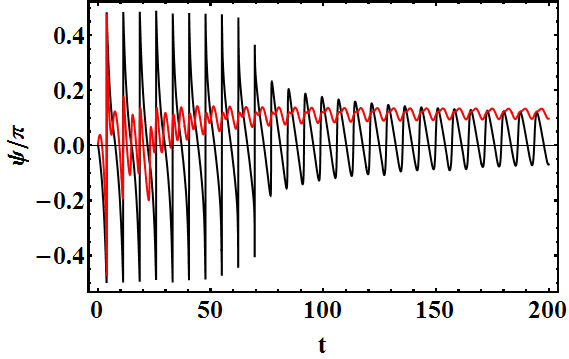}
      \caption{$\chi=0.5, \lambda_{an}=12, \Lambda_{an}=1.774$}
     \end{subfigure} 
     \hspace{0.12cm}
     \begin{subfigure}[b]{0.28\linewidth}       \includegraphics[width=\linewidth]{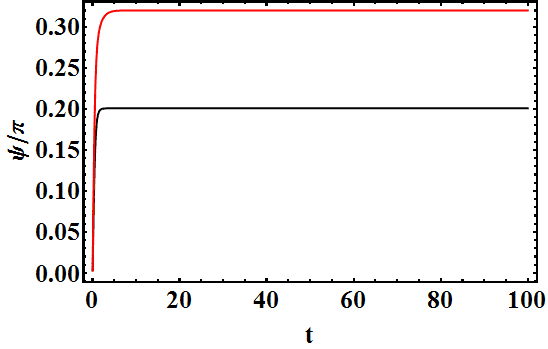}
      \caption{$\chi=0.7, \lambda_{an}=1, \Lambda_{an}=0.517$}
      \end{subfigure}
      \hspace{0.02cm}
       \begin{subfigure}[b]{0.28\linewidth}
       \includegraphics[width=\linewidth]{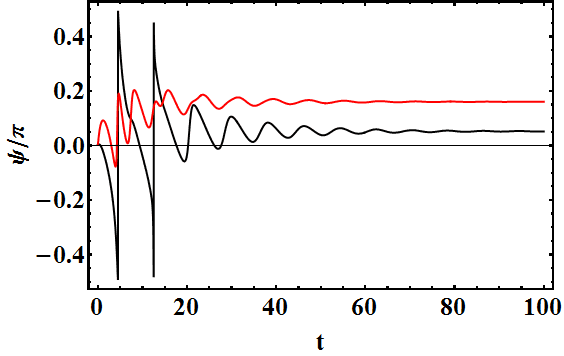}
       \caption{$\chi=0.7, \lambda_{an}=8, \Lambda_{an}=1.244$}
      \end{subfigure} 
  \hspace{0.12cm}
       \begin{subfigure}[b]{0.28\linewidth}
       \includegraphics[width=\linewidth]{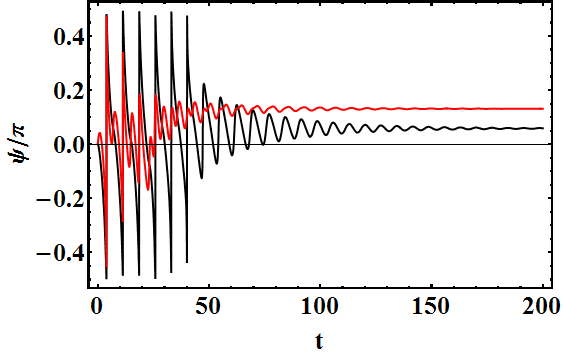}
       \caption{$\chi=0.7, \lambda_{an}=12, \Lambda_{an}=1.774$}
      \end{subfigure} 
    \hspace{0.12cm}
      \begin{subfigure}[b]{0.28\linewidth}       \includegraphics[width=\linewidth]{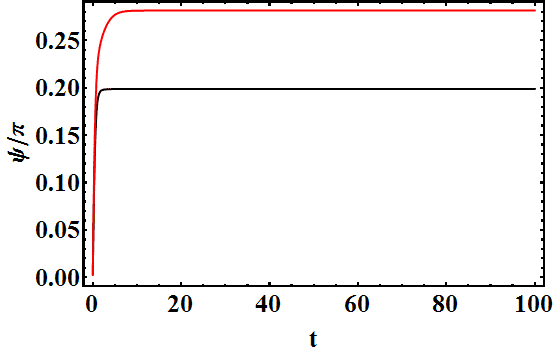}
       \caption{$\chi=0.9, \lambda_{an}=1, \Lambda_{an}=0.517$}
       \end{subfigure}
       \hspace{1.2cm}
        \begin{subfigure}[b]{0.28\linewidth}
        \includegraphics[width=\linewidth]{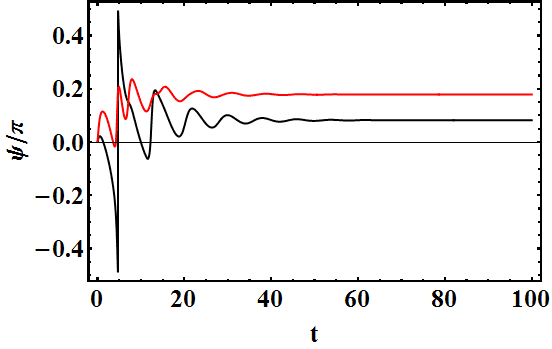}
        \caption{$\chi=0.9, \lambda_{an}=8, \Lambda_{an}=1.244$}
       \end{subfigure} 
   \hspace{1.2cm}
        \begin{subfigure}[b]{0.28\linewidth}
        \includegraphics[width=\linewidth]{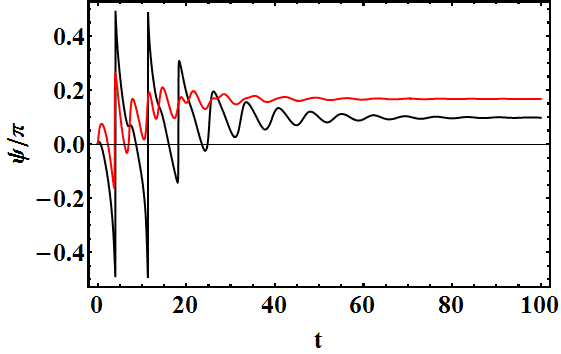}
        \caption{$\chi=0.9, \lambda_{an}=12, \Lambda_{an}=1.774$}
       \end{subfigure}                      
 \caption{Variation of inclination angle of the outer (black) and the inner (red) vesicles with time for $\lambda_{in}=1, \Delta_{in}=\Delta_{ex}=0.2, Ca=1$ using the higher order theory.}
    \label{psivst_higher}
\end{figure}

\begin{figure} [h!] %[tbp]
\centering
        \begin{subfigure}[b]{0.6\linewidth}       \includegraphics[width=\linewidth]{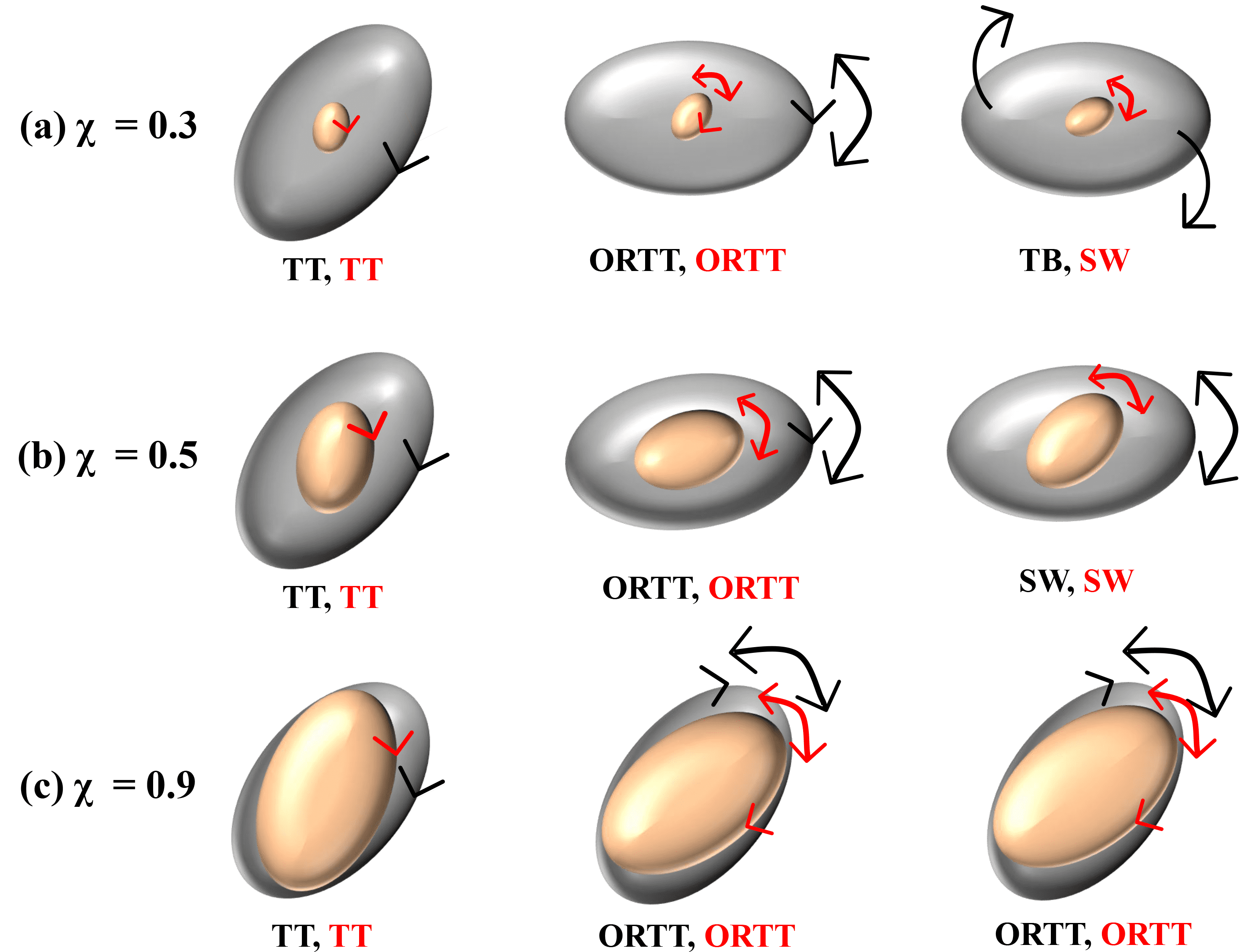}
     \end{subfigure}                  
 \caption{Schematic representation of the dynamic modes observed in figure-2 for $\chi=0.3, 0.5, 0.9$ ($\Delta_{in}=\Delta_{ex}=0.2, Ca=1$).}
    \label{vesdynamics}
\end{figure}
We also highlight the difference between the results obtained when the higher order spherical harmonics are either introduced, or not, in the membrane forces, similar to \cite{lebedev2007PRL}, and \cite{noguchi2007PRL} the analysis of a single vesicle. To avoid further complexity of the resulting equations, we do not consider higher order corrections in the hydrodynamic stress ((\cite{misbah2007PRE,  misbah2009PRE}). The shape evolution equations obtained for the inner and the outer vesicle are lengthy and highly non-linear, because of the coupling of stresses (membrane and hydrodynamic), and continuity of velocity fields in the three regions (core, annular, and exterior) at the surface of the inner and the outer vesicle. Therefore, for simplicity, we present the full expressions for the evolution equations of the two vesicles in the manuscript in terms of variables $f_{lm}^{in,ex}, \sigma_0^{in,ex}$ only (provided in Appendix \ref{App7}). The results from both linear as well as higher order theories are presented for comparison to bring out the role of the higher order membrane forces.

\subsubsection{Dynamic modes of a compound vesicle}

Figure \ref{psivst_higher} (and Appendix \ref{App8}-fig. \ref{psivst_leading}) present the time evolution of the inclination angle for the inner and the outer vesicle using the higher and linear order theories, respectively, for four different sizes of the inner vesicle ($\chi$ = 0.3, 0.5, 0.7, 0.9), and by varying the viscosity contrast across the outer vesicle ($\lambda_{an}$ = 1, 8, 12). No viscosity contrast is set across the inner vesicle i.e. $\lambda_{in}$ = 1 in most of the studies reported here, except, a separate study wherein the inner vesicle is considered as a limiting case of a solid sphere with $\lambda_{in}$= 1000.

When there is no viscosity contrast across the outer vesicle ($\lambda_{an}=1, \Lambda_{an}=0.517$), (fig \ref{psivst_higher}-a,d,g,j) the two vesicles show TT motion for both the inner and the outer vesicle for all the four size of the inner vesicle ($\chi$). On increasing the viscosity contrast ($\lambda_{an}=8, \Lambda_{an}=1.244$) the two vesicles in the TT regime, show transition to a new regime, wherein the vesicles exhibit initial oscillations before attaining a fixed orientation with respect to the flow direction (fig \ref{psivst_higher}-b,e,h,k). We call this the "ORTT (oscillatory relaxation to tank-treading)" mode, in line with \cite{lebedev2007PRL}. This is uniformly observed across all the $\chi$ considered in this study. For still higher values of $\lambda_{an}$ ($\Lambda_{an}$) (fig \ref{psivst_higher}-c,f,i,l) and for small $\chi$, the outer vesicle undergoes a transition to TU regime since at a low $\chi$, it is akin to a single vesicle with viscosity contrast beyond its critical value. As the size of the inner vesicle, $\chi$ is increased, the outer vesicle shows transition from TU to SW regime, followed by the ORTT regime for high $\chi$ value. The inner vesicle on the other hand exhibits SW  at low $\chi$ and undergoes a transition to ORTT as $\chi$ is increased. Thus, at higher values of $\lambda_{an}$ ($\Lambda_{an}$), the dynamics of the inner and the outer vesicle is dependent upon the size of the inner vesicle ($\chi$) and the two vesicles can exhibit different dynamical regimes. It is pertinent to re-emphasize here that the variation of $\lambda_{an}$, allows us to realise different values of $\Lambda_{an}$.

These results are in agreement with the experimental results of \cite{steinberg2014PRL} which show very poor dependence of the dynamics on the size of the inner vesicle ($\chi$), at low values of $\Lambda_{an}$. On the other hand a sensitive dependence on the size of the inner vesicle ($\chi$) is observed at higher values of $\Lambda_{an}$ in their experiments. 

The leading order theory (Appendix \ref{App8}-fig \ref{psivst_leading}) predicts similar results as the higher order theory for small $\lambda_{an}$ ($\Lambda_{an}$) (fig \ref{psivst_leading}-a,d,g,j) but shows a highly suppressed ORTT regime, restricted to only very high $\chi$. The swinging regime is also suppressed. This is in accordance with the linear order theory for single vesicles, which also shows an absence of the TR regime for instance. Apart from these regimes a new mode, called "IUS", appears at intermediate $\chi$ values (this mode is discussed in subsection \ref{phaseD}).

Figure \ref{vesdynamics} shows the summary of these regimes as a schematic presentation for various dynamic modes of the inner and the outer vesicle from higher order theory.

\subsubsection{Tank-treading regime}
\begin{figure} [tbp] %[tbp]
    \centering
    \begin{subfigure}[b]{0.49\linewidth}       \includegraphics[width=\linewidth]{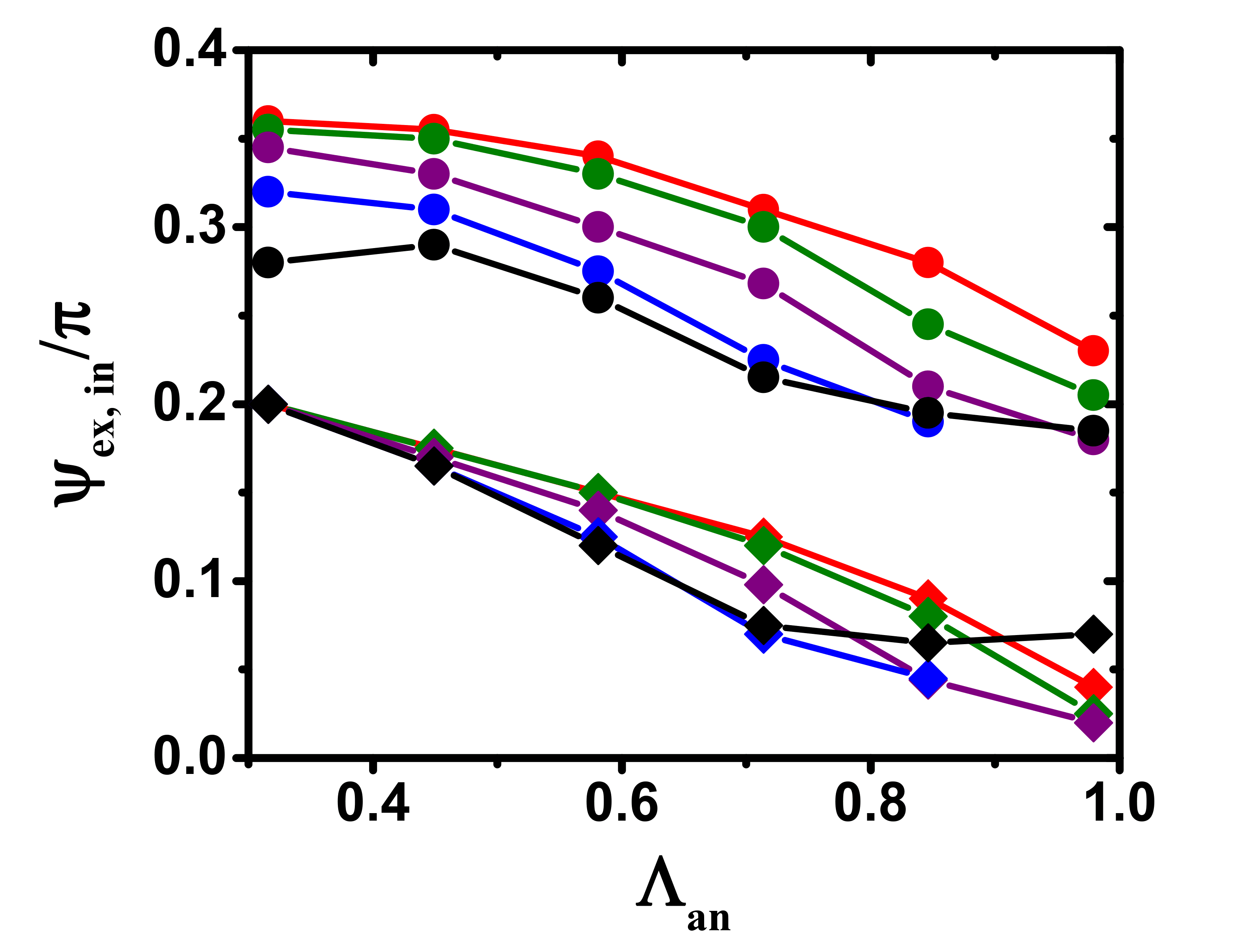}
     \caption{$\Delta_{in}=\Delta_{ex}=0.2$}
     \end{subfigure} 
     \hspace{0.12cm}
     \begin{subfigure}[b]{0.49\linewidth}       \includegraphics[width=\linewidth]{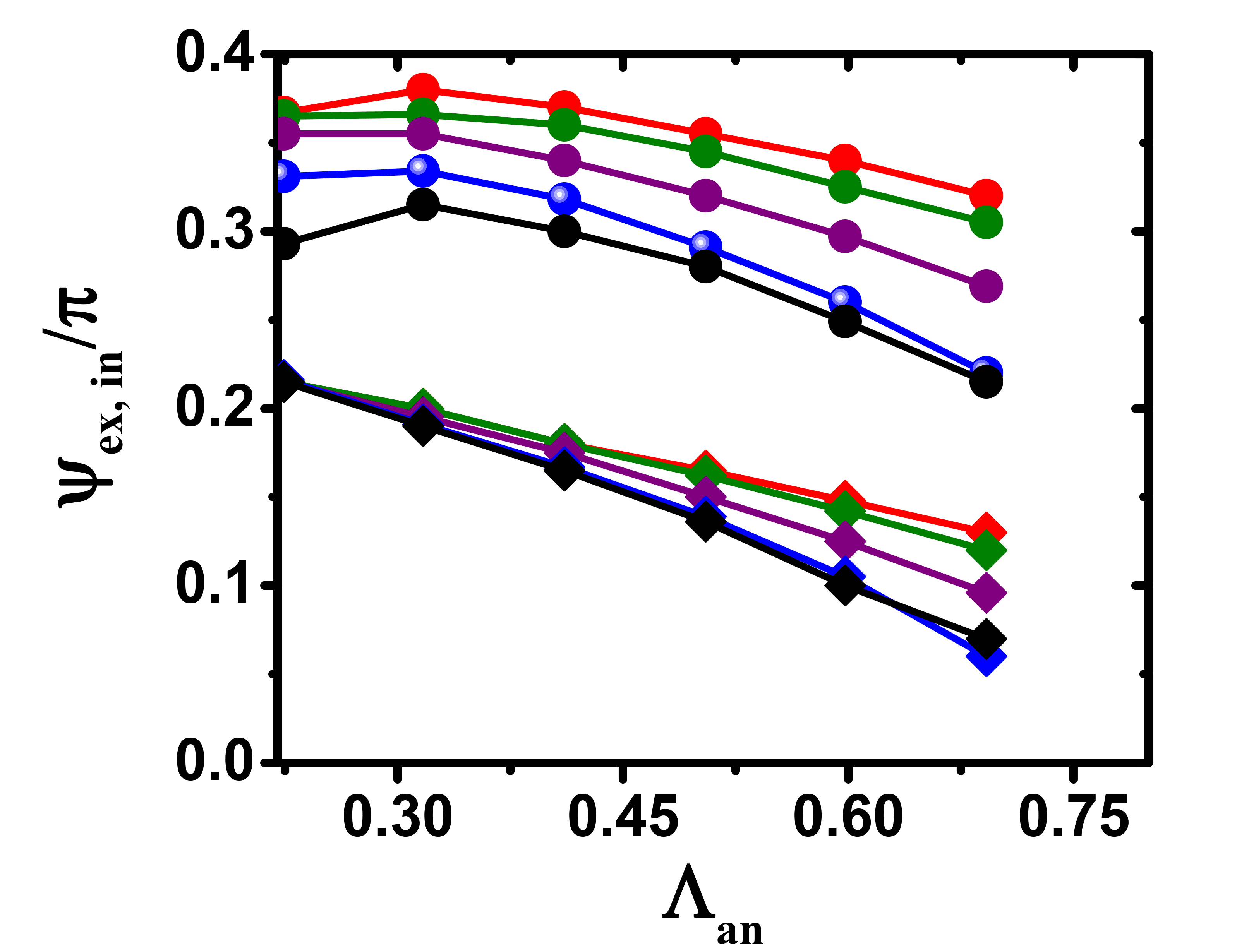}
      \caption{$\Delta_{in}=\Delta_{ex}=0.1$}
      \end{subfigure}
      \hspace{0.12cm}
       \begin{subfigure}[b]{0.49\linewidth}
       \includegraphics[width=\linewidth]{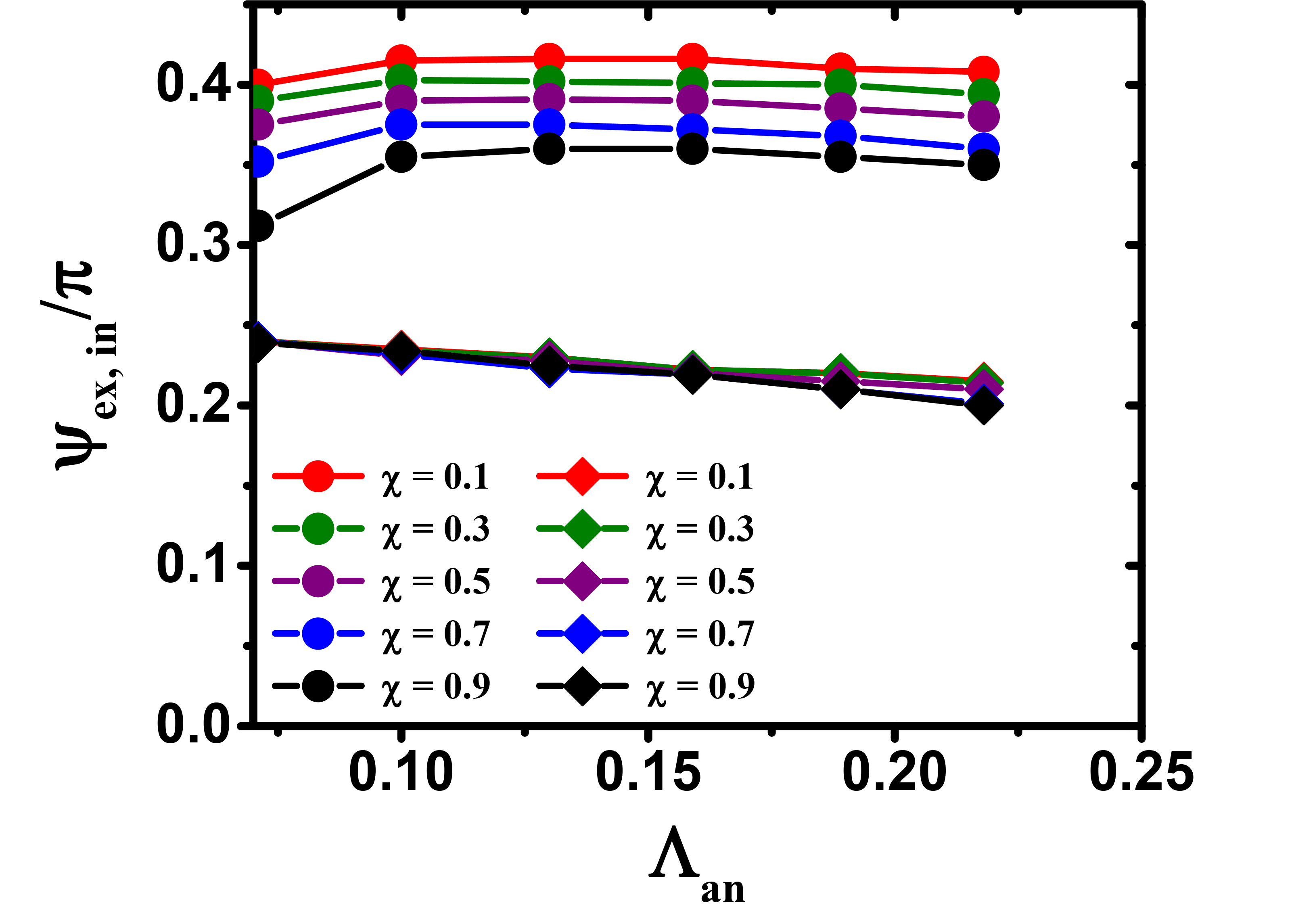}
       \caption{$\Delta_{in}=\Delta_{ex}=0.01$}
      \end{subfigure}
       \hspace{0.12cm}
        \begin{subfigure}[b]{0.49\linewidth}
        \includegraphics[width=\linewidth]{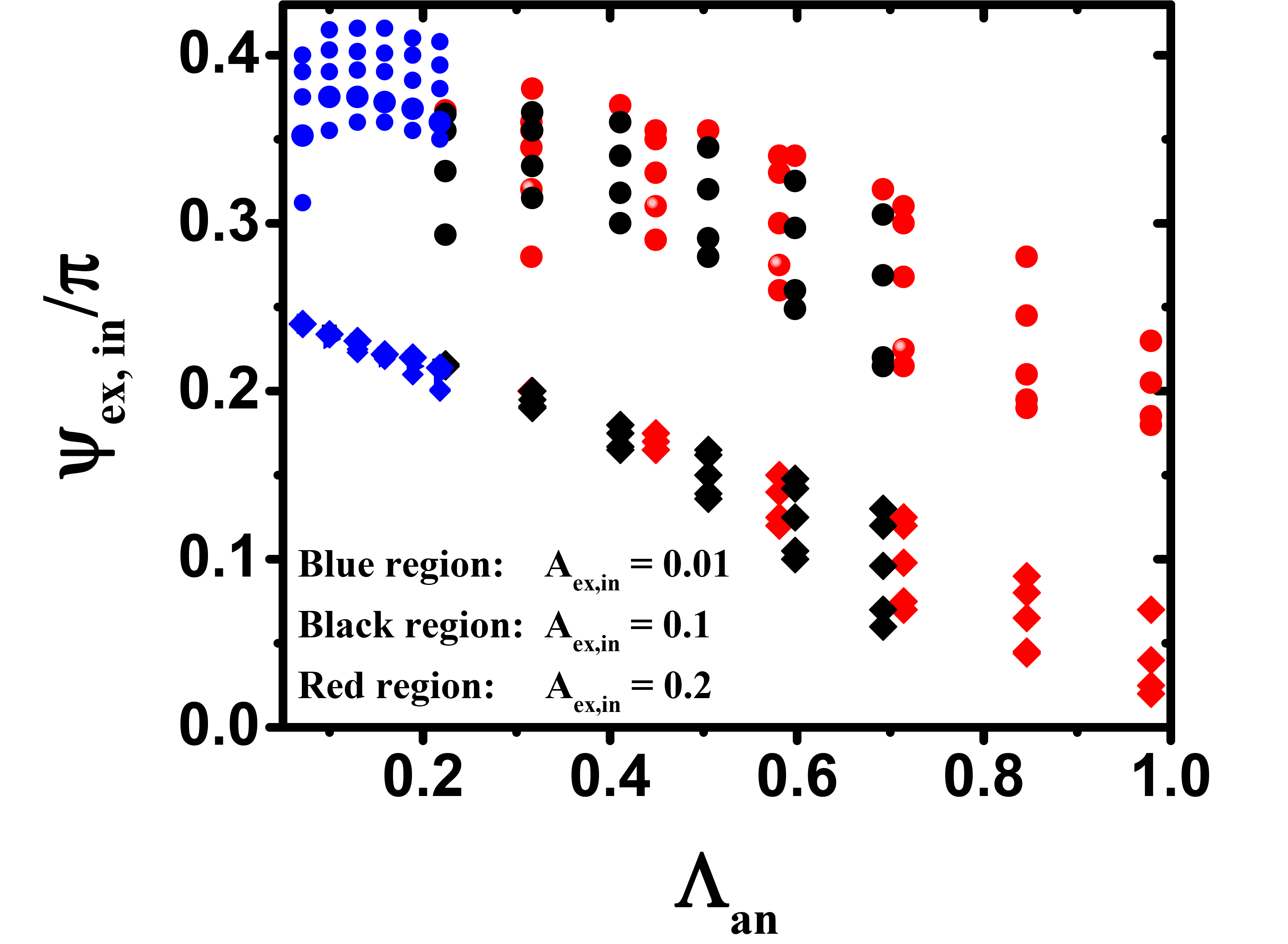}
        \caption{$\Delta_{in}=\Delta_{ex}=0.01, 0.1, 0.2$}
       \end{subfigure}                       
 \caption{Variation of the inclination angle of the inner (solid circles) and the outer (solid diamonds) vesicle with $\Lambda_{an}$ across the outer vesicle for different $\chi$, and the effect of excess area on variation of inclination angle from higher order theory. The $\Lambda_{an}$ is realized by varying $\lambda_{an}$ ($Ca=1$).} 
    \label{psivsoutvis1}
\end{figure}
\begin{figure} [tb] %[tbp]
    \centering
    \begin{subfigure}[b]{0.6\linewidth}
          \includegraphics[width=\linewidth]{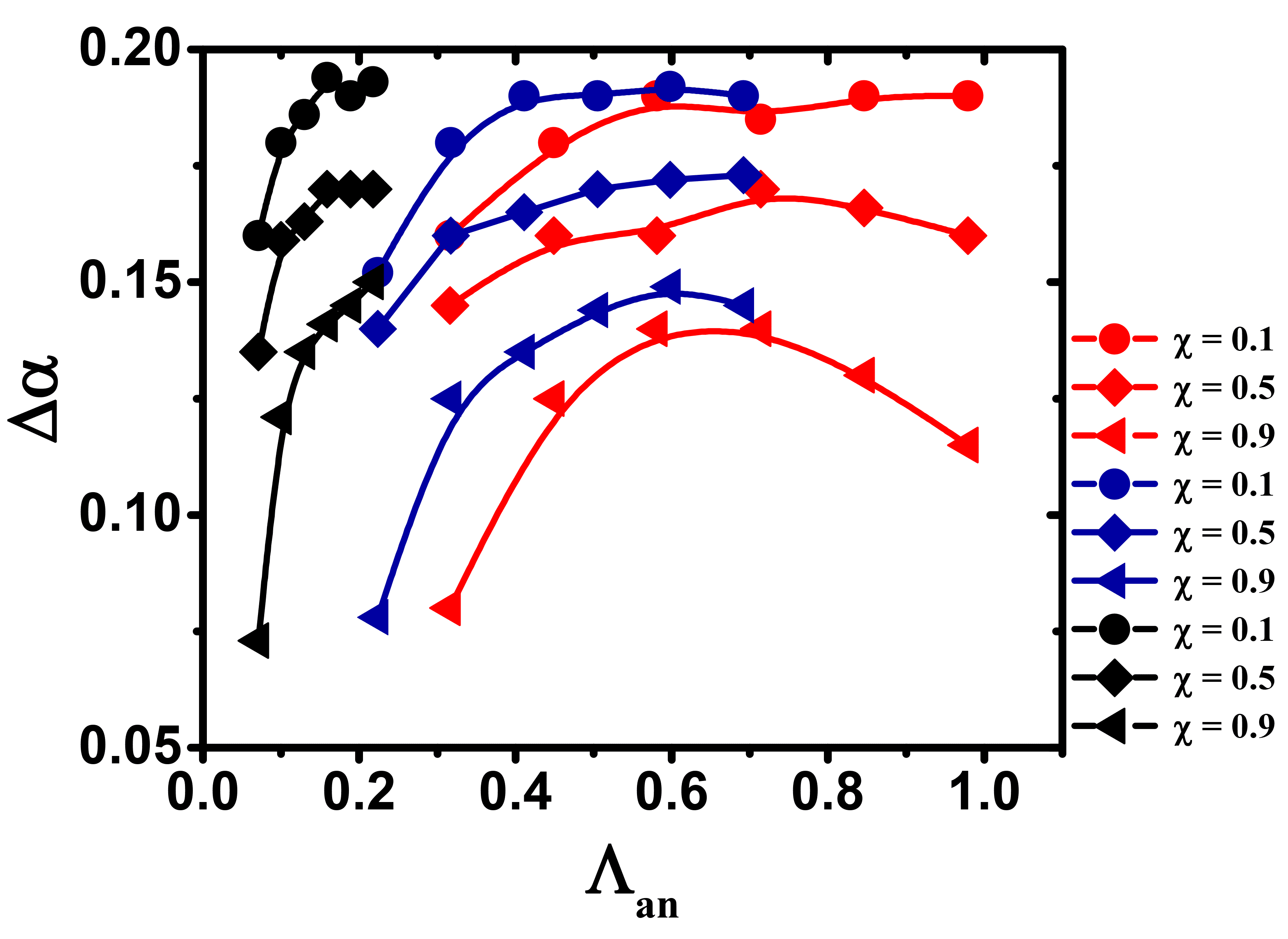}
         \end{subfigure}                        
 \caption{Relative angle ($\Delta\alpha$) between the main axis of the inner and the outer vesicle in TT regime for three different excess areas: $\Delta_{in, ex}$=0.2 (red), 0.1 (blue), 0.01 (black), using higher order theory with $Ca=1$.}
    \label{angledifference}
\end{figure}
We now discuss in detail the TT regime, wherein figure \ref{psivsoutvis1} shows the variation of the inclination angles of the inner and the outer vesicle as predicted by the higher and the leading order theories (Appendix \ref{App8}-fig\ref{inclinationangle_leadord}). The higher order theory (fig \ref{psivsoutvis1}a) shows that :
(1) The inclination angles of both the vesicles decrease with an increase in $\Lambda_{an}$ for a given $\chi$ which is in accordance with the  single vesicle theory.
(2) The inclination angle of the inner vesicle is always higher than that of  the outer vesicle for any $\Lambda_{an}$ and $\chi$.
(3) The outer vesicle shows a weak dependence of the inclination angle on the size of the inner vesicle ($\chi$).
(4) The inner vesicle shows a greater dependence of the inclination angle on the size of the inner vesicle ($\chi$), and the inclination angle increases with a reduction in $\chi$. (5) The $\Lambda_{an}$ at which the $\psi$ becomes zero is termed the critical $\Lambda_{an}^{crit}$ for transition of the system from TT to another appropriate regime. Figure \ref{psivsoutvis1}(a) shows that the critical $\Lambda_{an}^{crit}$ is higher for the inner vesicle as compared to the outer vesicle.

The above observations are in accordance with the experiments of the Steinberg group (\cite{steinberg2014PRL}). Our results indicate that there is very weak dependence of the critical $\Lambda_{an}^{crit}$ on $\chi$ for the outer vesicle, while the inner vesicle shows a stronger dependence (fig \ref{psivsoutvis1}a,b,c,d). Our results are in contradiction to \cite{vlah2011PRL} and \cite{kaoui2013SM} who show much higher sensitivity on the inner radius in transition from TT to TU. It should be noted that the results of \cite{vlah2011PRL} are valid for a solid inclusion, whereas the results of \cite{kaoui2013SM} are from a 2 D analysis. Moreover, the results of the Steinberg group (\cite{steinberg2014PRL}) do indicate a much strong dependence of the inclination angle of the inner vesicle on the size of the inner vesicle ($\chi$), similar to our predictions.

Similar to the higher order theory, the leading order theory (Appendix \ref{App8}-fig\ref{inclinationangle_leadord}) also shows that the inclination angle for the inner vesicle is higher than the outer vesicle for any $\chi$, while the critical $\Lambda_{an}^{crit}$ decreases with $\chi$.

It is appropriate here to verify the similarity of results presented on $\lambda_{an}-Ca$ plots with $\Delta_{ex}$ as a variable, with the results on $\Lambda_{an}-S$, where $\Delta_{ex}$ is absorbed in $\Lambda_{an}$ and $S$ (fig \ref{psivsoutvis1}a,b,c). Towards this, results are generated with different excess area of the vesicles. It should be noted that the excess area in our study is the same for the inner and the outer vesicle. When the excess area is reduced systematically, $\Delta_{ex} = \Delta_{in}$ = 0.2, 0.1, 0.01 (fig \ref{psivsoutvis1}a,b,c), the results on the $\Lambda_{an}$ scale seem to be insensitive to $\Delta_{in, ex}$ if they are represented with $\Lambda_{an}$ and $S$ as parameters (which absorbs $\Delta_{ex}$). Fig \ref{psivsoutvis1}d shows a combined plot where data using all the three excess areas are plotted together and substantiates the insensitivity of the $\psi$ vs $\Lambda_{an}$ behavior to $\Delta_{ex}$. Effect of excess area on inclination angle of two vesicle is also plotted in Appendix \ref{App8}-fig\ref{PSIvsA} for low and high $\lambda_{an}$ in the TT regime. The figure shows that the inclination angle varies weakly with excess area at low $\lambda_{an}$ while a much stronger decrease in inclination angle with excess area is seen at higher $\lambda_{an}$. The difference angle $\Delta\alpha=\psi_{in}-\psi_{ex}$ in in the TT regime is shown in figure \ref{angledifference} and shows that the difference increases with an increase in $\Lambda_{an}$ and then plateaus or decreases at still higher $\Lambda_{an}$, On the other hand the difference angle $\Delta\alpha$ increase with a decrease in $\Delta_{in,ex}$ and a decrease in $\chi$.
This indicates that stronger hydrodynamic interaction (at higher $\chi$, higher $\Lambda_{an}$) leads to a strong coupling of the motion of the two vesicles and thereby lower $\Delta\alpha$. Similarly, a stronger rotational component of shear ($\Lambda_{an}$) tends to increase the difference angle between the vesicles. 

\subsubsection{Discussion on compound vesicle phase diagram} \label{phaseD}
\begin{figure} [tbp] %[tbp]
\centering
    \hspace{0.cm}
    \begin{subfigure}[b]{0.49\linewidth}
          \includegraphics[width=\linewidth]{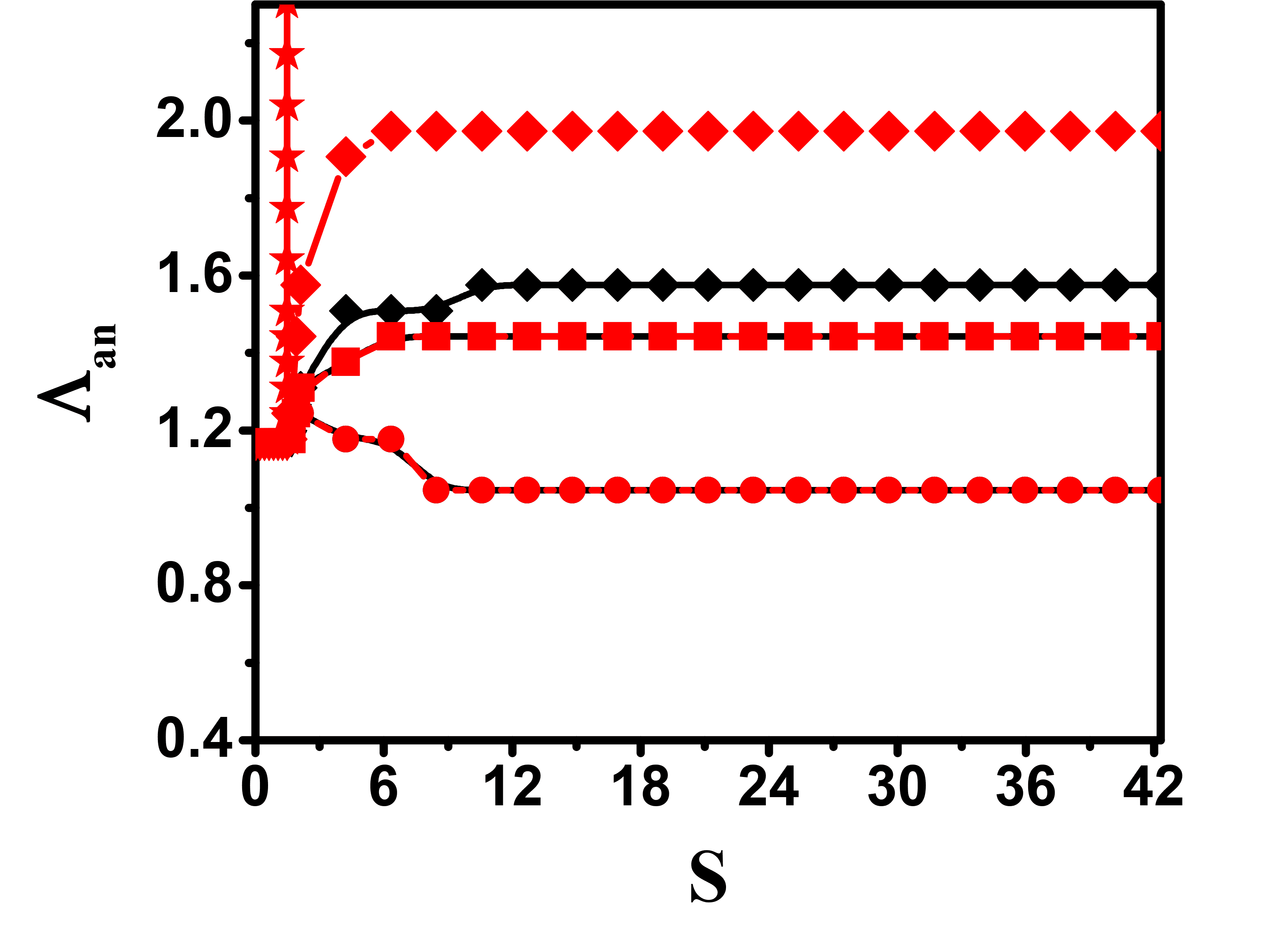}
          \caption{$\chi = 0.1$}
         \end{subfigure} 
         \hspace{0.cm}
    \begin{subfigure}[b]{0.49\linewidth}
          \includegraphics[width=\linewidth]{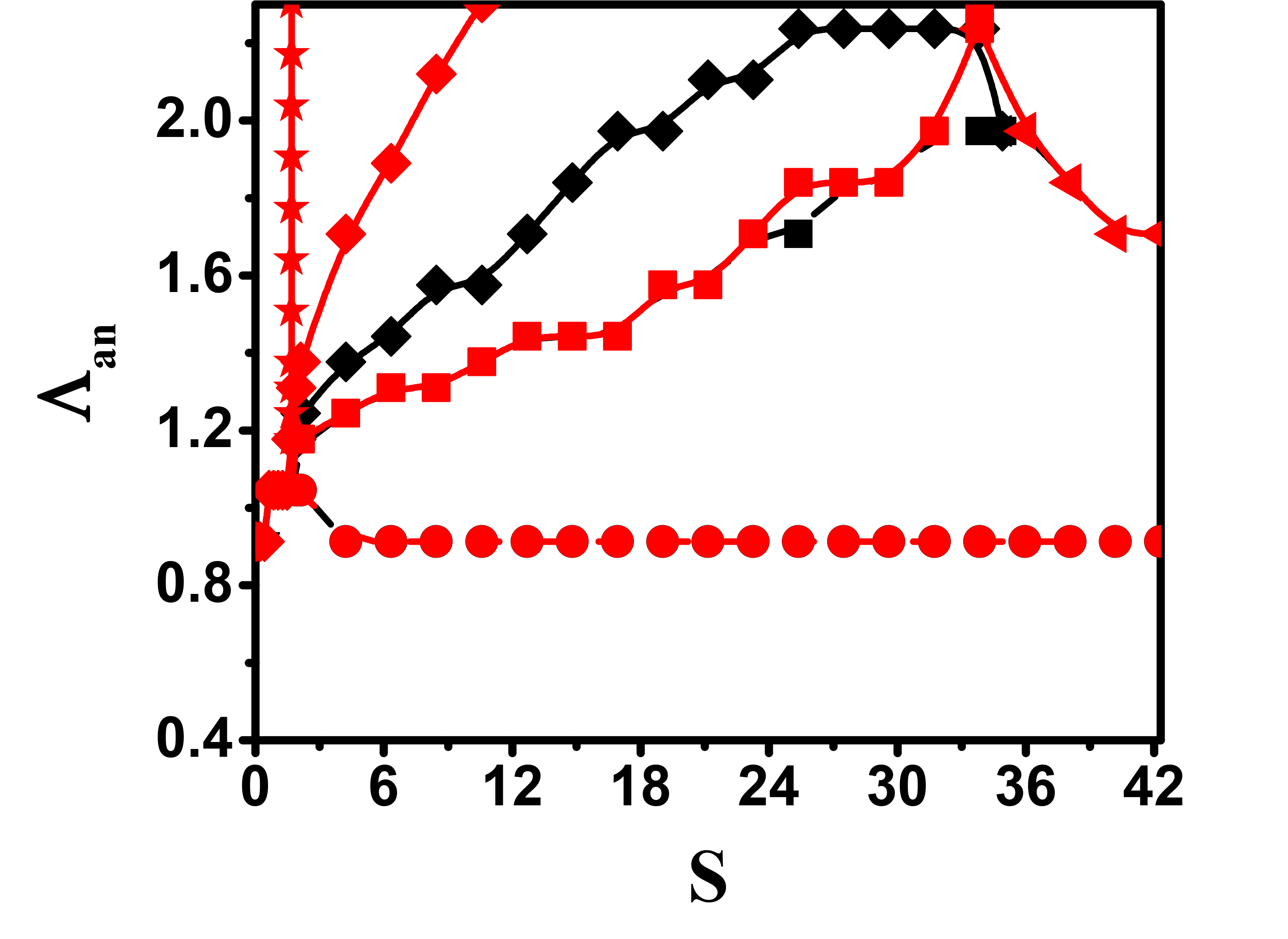}
          \caption{$\chi = 0.5$}
         \end{subfigure}  
            \hspace{0.cm}
    \begin{subfigure}[b]{0.45\linewidth}
   \includegraphics[width=\linewidth]{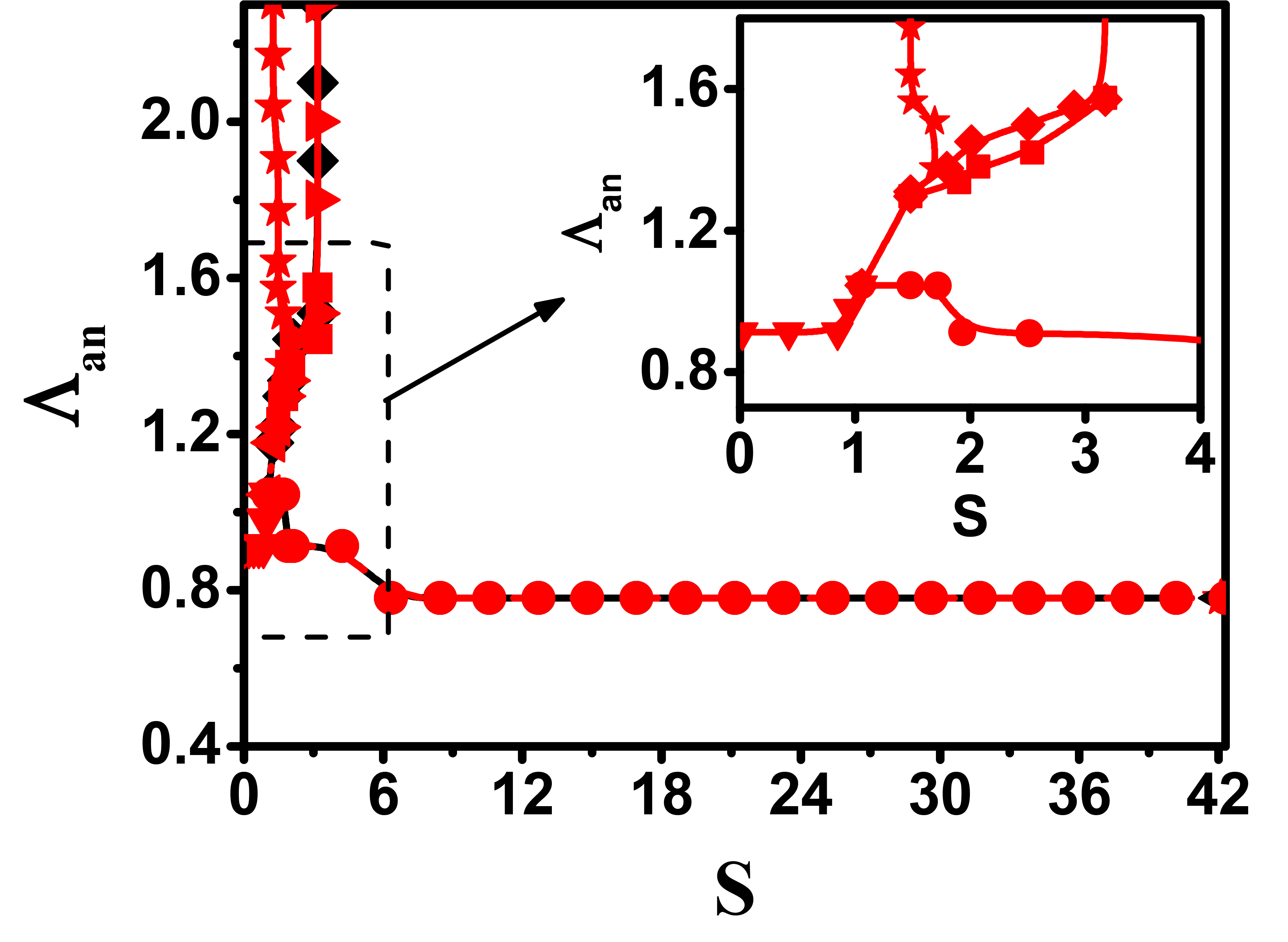}
    \caption{$\chi = 0.9$}
     \end{subfigure}  
     \hspace{2.cm}
         \begin{subfigure}[b]{0.38\linewidth}
        \includegraphics[width=\linewidth]{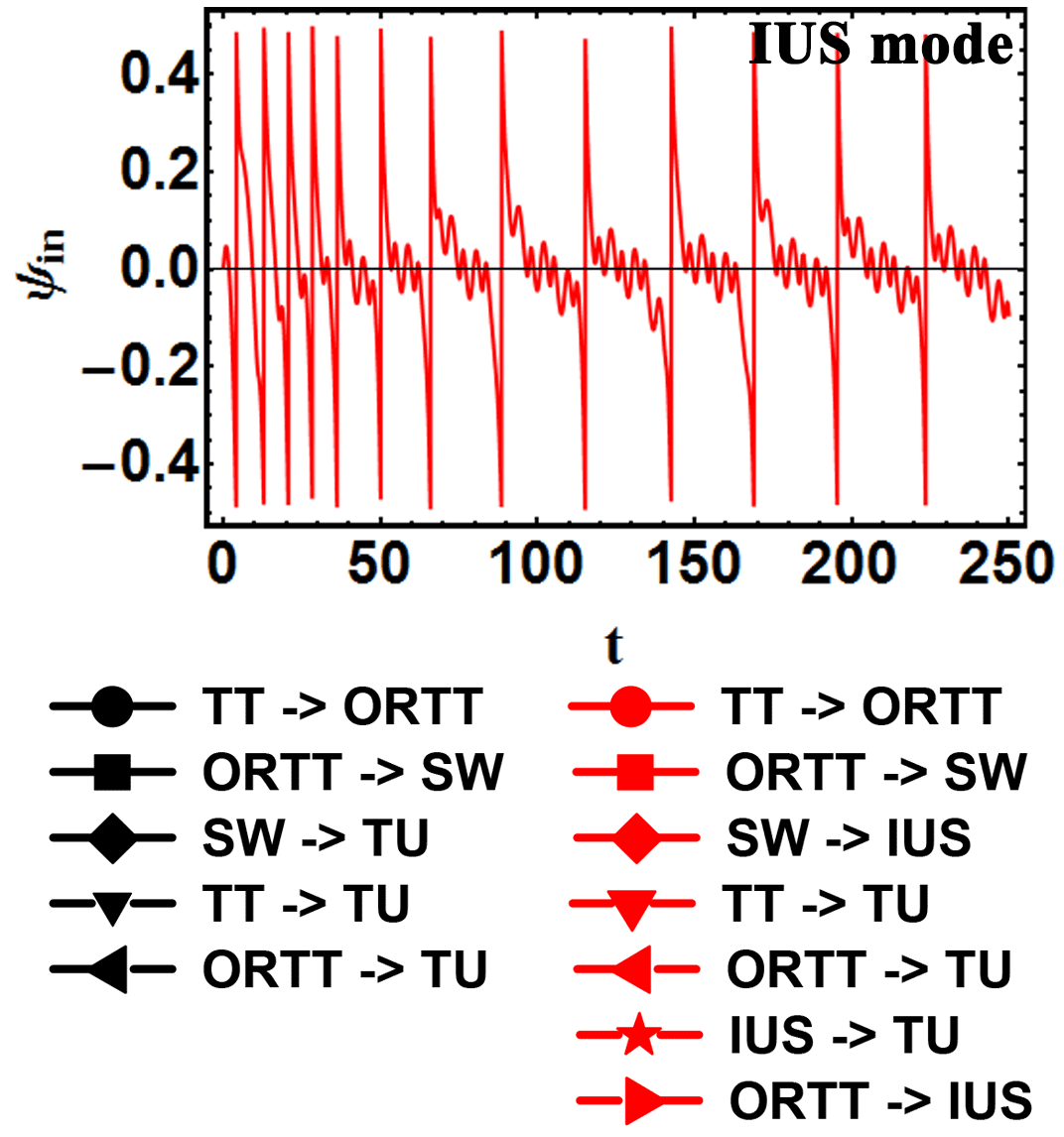}
                   \end{subfigure}                 
 \caption{Combined phase diagram for the inner (red) and the outer (black) vesicle on $\Lambda_{an}-S$ co-ordinates ($\Delta_{in}=\Delta_{ex}=0.2, \lambda_{in}=1$) using higher order theory. All legends presents dynamic mode (TT, ORTT, TU, SW, IUS) transition from bottom to top, and from right to left.}
    \label{combinedPD}
\end{figure}
\begin{figure} [tbp] %[tbp]
\centering
    \begin{subfigure}[b]{0.6\linewidth}       \includegraphics[width=\linewidth]{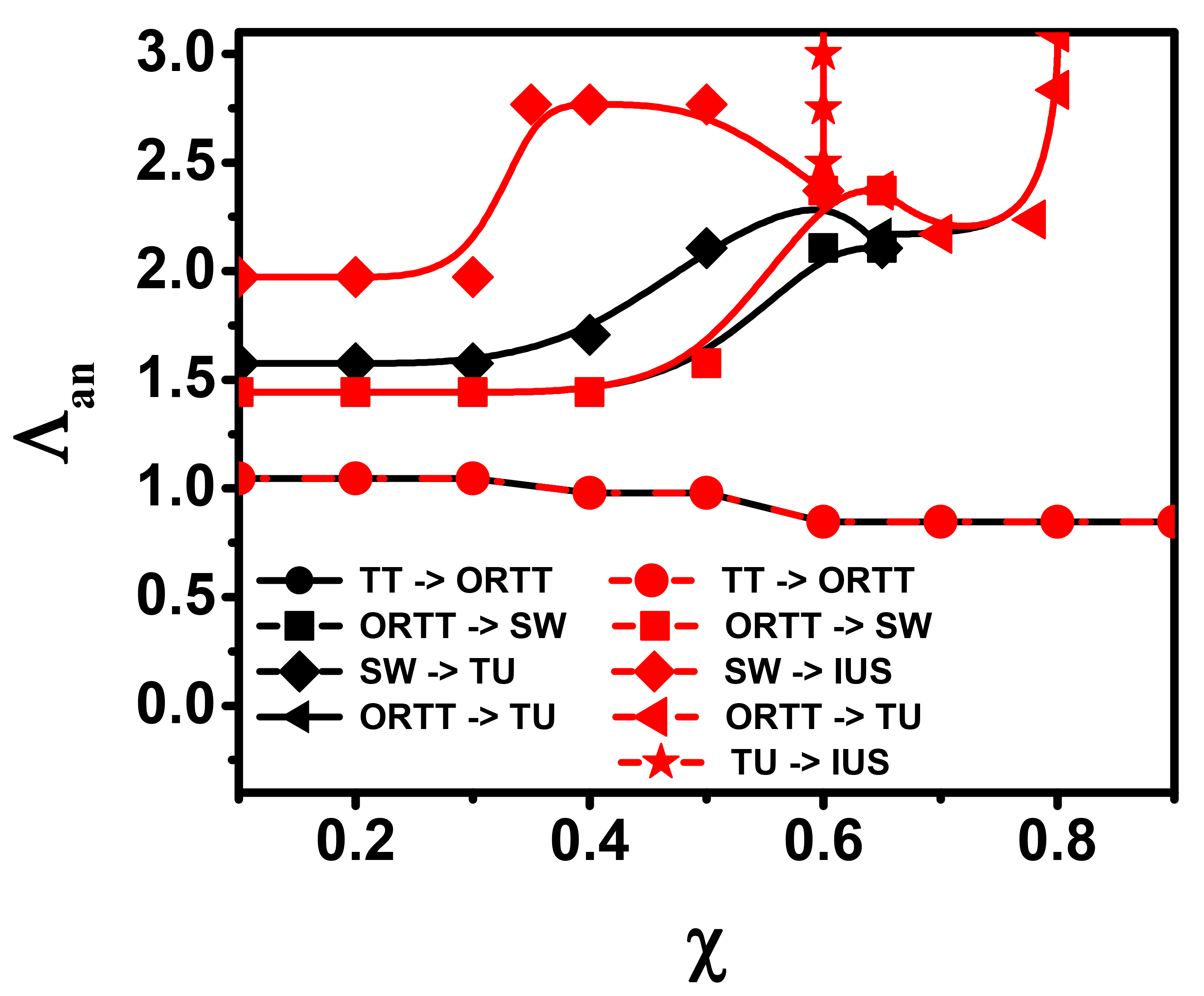}
     \end{subfigure}              
 \caption{$\Lambda_{an}$ vs $\chi$ phase diagram for the inner (red line) and the outer vesicle (black line) in linear shear flow ($\Delta_{in}=\Delta_{ex}=0.2, \lambda_{in}=1, Ca=1$) using higher order theory. All legends presents transition from bottom to top, and from right to left.}
    \label{viscoutvschi}
\end{figure}
A single vesicle undergoes transition to TU and TR regimes on change of $\Lambda$ and $S$. To understand the dependency of the different regimes of vesicle dynamics on the $Ca$ (plotted here as $S$), a higher order theory needs to be conducted. Fig \ref{combinedPD} shows $\Lambda_{an}$ vs $S$ phase diagram for a compound vesicle as obtained from the higher order theory. The outer and inner vesicle phase diagrams (fig \ref{combinedPD}-a,b,c) are constructed for three different $\chi$ values.

For small $\chi$ (fig \ref{combinedPD}a), the phase  boundaries of the inner and outer vesicle almost completely overlap, The behavior of both the vesicles is similar to the  single vesicle theory, and exhibits a ORTT-SW as well as SW-TU (IUS) transition boundaries, which saturate at high $S$ values (or high $Ca$) (\cite{misbah2007PRE, lebedev2007PRL, misbah2009PRE}). Compared to the outer vesicle, the inner vesicle shows SW over a wider range of $\Lambda_{an}$. An important point to note here is the appearance of the SW mode, characterized by oscillations of the inclination axis along a positive angle, (unlike in VB or TR modes in which the oscillations are along the shear direction). This is unlike a single vesicle which shows the presence of a TR mode, while the SW mode is absent. 

It should be noted here that we follow the definition for VB modes (\cite{misbah2006PRL, misbah2007PRE, misbah2009PRE}) as asymmetric oscillations around flow direction and breathing dynamics of the shapes, TR (\cite{steinberg2006PRL, lebedev2007PRL, bagchi2012PRE}) as highly deformed shapes with concavities and lobes and deformation in the shear flow while the SW mode (\cite{noguchi2007PRL, kaoui2013SM}) as the oscillations around non-zero positive angle, where shape does not deform in small deformation limit.

At an intermediate $\chi$ (fig \ref{combinedPD}b), the two vesicles show similar phase boundaries except the SW-TU (IUS) transition. With an increase in $\chi$, the phase boundaries seem to shift counterclockwise, thereby promoting the ORTT regime, and restricting the TU and the SW regimes to lower regions of the phase diagram. For the outer vesicle, ORTT-SW and SW-TU phase boundaries are inclined, therefore a transition from SW to TU can be affected by an increase in $S(Ca)$, e.g. at $\Lambda_{an}$=1.6, a transition to TU can be seen at around $S$=12. This bears some resemblance to the results of \cite{misbah2009PRE} which show a SW-TU transition with $S$ for a fixed $\Lambda_{an}$.

The behavior of the phase diagram in the low $S$ (or $Ca$) region for fig \ref{combinedPD}a,b is presented in Appendix \ref{App8} fig\ref{inset-phasediag}a, and fig\ref{inset-phasediag}b, respectively by zooming out that region. A new regime appears in this phase diagram (fig \ref{combinedPD}) which is an intermediate regime between TU and SW, we call it as "IUS". In a lot of ways, it can be considered to be a modification of the TU regime. The IUS mode was previously observed in capsule studies in shear flow by \cite{Vla2011JFMCapsules}. Interestingly, in this present study on a compound vesicle, the appearance of this mode is limited only to the inner vesicle, the outer vesicle does not show such dynamics. Thus in the inner vesicle phase space a significant region of phase diagram is now occupied by the IUS mode in all three phase diagrams. For high $\chi$ (fig \ref{combinedPD}c), the TU, IUS, and SW regimes are significantly suppressed to a very small region in the $\Lambda_{an}$-$S$ phase space which is zoomed out in the inset in the same figure. This inset also shows that the SW region appears in insignificantly small region. A major area of the phase space is now occupied by the TT and the ORTT regimes. The phase diagrams in the $\chi=0.5$ and $\chi=0.9$ are remarkably different, indicating a strong influence of the hydrodynamic mediated vesicle interaction.
\begin{figure} [tbp] %[tbp]
    \centering
 \begin{subfigure}[b]{0.49\linewidth}
   \includegraphics[width=\linewidth]{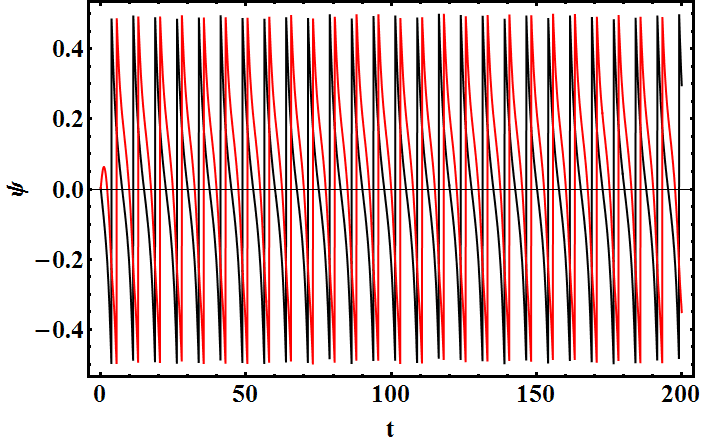}
  \caption{$\chi=0.1, Ca=0.05, S=1.05$}
   \end{subfigure}
\hspace{0.12cm}     
 \begin{subfigure}[b]{0.49\linewidth}
 \includegraphics[width=\linewidth]{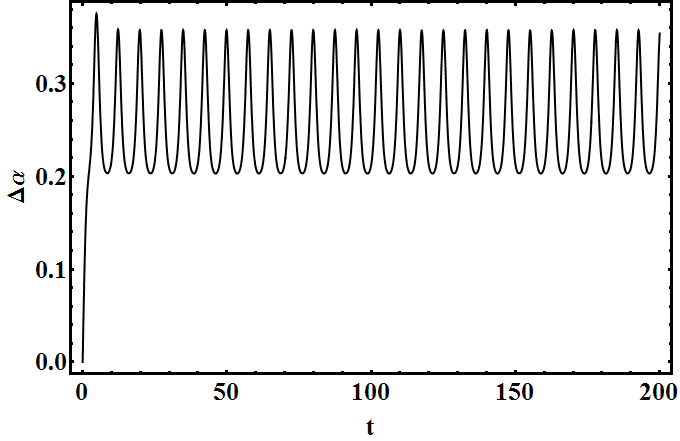}
  \caption{$\chi=0.1, Ca=0.05, S=1.05$}
 \end{subfigure}   
  \hspace{0.12cm}     
 \begin{subfigure}[b]{0.49\linewidth}
 \includegraphics[width=\linewidth]{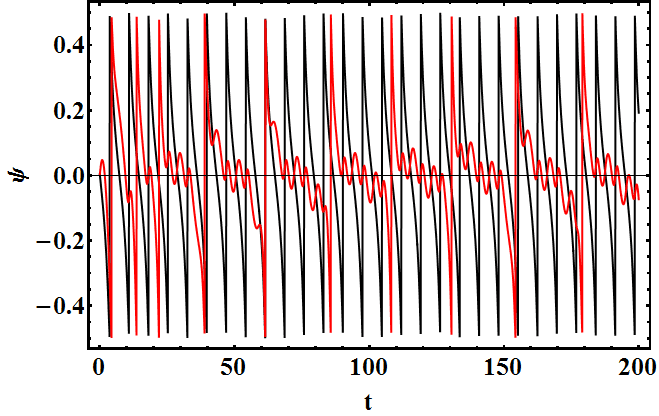}
  \caption{$\chi=0.1, Ca=0.5, S=10.57$}
 \end{subfigure}                    
   \hspace{0.12cm}     
 \begin{subfigure}[b]{0.49\linewidth}
\includegraphics[width=\linewidth]{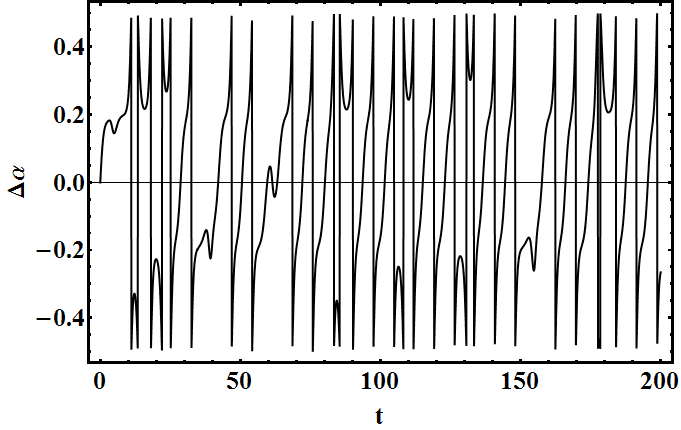}
  \caption{$\chi=0.1, Ca=0.5, S=10.57$}
  \end{subfigure} 
 \begin{subfigure}[b]{0.49\linewidth}
 \includegraphics[width=\linewidth]{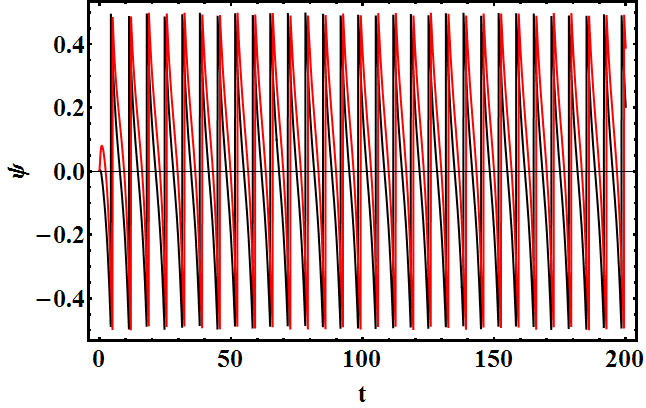}
  \caption{$\chi=0.9, Ca=0.05, S=1.05$}
   \end{subfigure}
    \hspace{0.12cm}  
    \begin{subfigure}[b]{0.49\linewidth}
   \includegraphics[width=\linewidth]{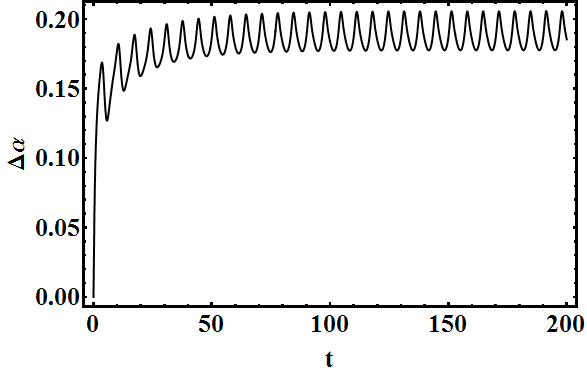}
      \caption{$\chi=0.9, Ca=0.05, S=1.05$}
     \end{subfigure}                             
 \caption{Tumbling motion of the inner (red) and the outer (black) vesicle at $\chi=0.1$ for (a) $Ca=0.05$, (c) $Ca=0.5$; and at $\chi=0.9$ for $Ca=0.05$; (b),(d),(f)- presents the relative angle ($\Delta\alpha$) between the main axis of the inner and outer vesicle in TU regime for (a), (c), and (e), respectively ($\Delta_{in}=\Delta_{ex}=0.2, \lambda_{an}=14, \lambda_{in}=1$).}
    \label{TUdelta}
\end{figure}
The results available from literature for comparison are the experiments of the Steinberg group (\cite{steinberg2014PRL}) in which although the values of $S$ used are exceptionally high, the values of $\Lambda_{an}$ are comparable to our work. The experimental results of the Steinberg group indicate that the TT regime is observed at high $S$ and small $\Lambda_{an}$, the TR regime is at lower $S$ and higher $\Lambda_{an}$, while the TU regime is at much higher $\Lambda_{an}$ and lower $S$. These seem to be the general features of the 6 experimental conditions reported by the Steinberg group (\cite{steinberg2014PRL}), three each for the two size of the inner vesicle of $\chi=0.35$ and $\chi=0.6$.

It should be noted that the excess area in the experiments of the Steinberg group is very large, to the tune of 1.0 or even higher. On the contrary the excess area in the present work is restricted to around 0.2 on account of the asymptotic nature of the analysis in the excess area as the small parameter. The salient features in experiments are the SW mode observed in the TR regime for inner vesicle for smaller $\chi=0.35$ whereas wrinkling is observed at $\chi=0.6$. 

The only other results are by \cite{kaoui2013SM} who predict a transition from TU to SW at a constant $\Lambda_{an}$ with respect to $S$.

Our results are in agreement with the experimental results of the Steinberg group (\cite{steinberg2014PRL}), wherein the inner and outer vesicles can be in different regimes at an intermediate size of the inner vesicle ($\chi$) (See fig \ref{combinedPD}b) where the outer vesicle can tumble when the inner vesicle can admit a SW motion. The two vesicles admit the same regime at higher $\chi$'s.

Another way to present the dynamics is on a phase diagram with $\Lambda_{an}$ and $\chi$ as coordinates using both higher ((fig \ref{viscoutvschi}) and linear order theories (Appendix \ref{App8}-fig \ref{viscoutvschi_leading}). For a given $S$ (or $Ca$) (fig \ref{viscoutvschi}) the phase diagrams of the two vesicles are identical in the TT regime, as well as for all the regimes at higher $\chi$. At intermediate $\chi$ though, the vesicles can be in different regimes at the same $\Lambda_{an}$ and $S$, for e.g. while the outer vesicle can exhibit TU, the inner can admit SW, as discussed earlier, and which is also in accordance with the experiments of the Steinberg group (\cite{steinberg2014PRL}). A significant difference is seen between the leading order and the higher order theories. Expectedly, the linear theory (Appendix \ref{App8}-fig \ref{viscoutvschi_leading}) does not admit the SW regime. Moreover, the leading order theory does not show dependence on the capillary number (or S) and can therefore be considered to be the limiting case of ($S$=0 or $Ca$=0) the higher order theory.
\begin{figure} [tp] %[tbp]
    \centering
 \begin{subfigure}[b]{0.32\linewidth}
   \includegraphics[width=\linewidth]{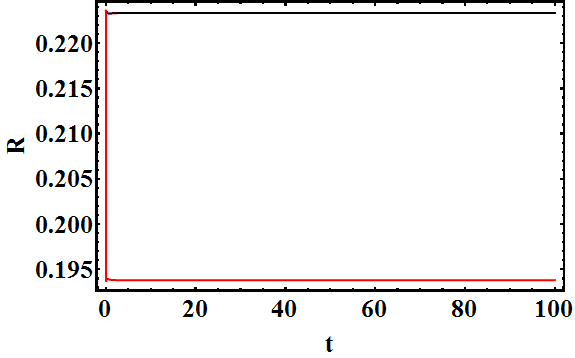}
          \caption{$\chi=0.1, \Lambda_{an}=0.32, S=21.15$}
          \end{subfigure} 
       \hspace{0.12cm}     
    \begin{subfigure}[b]{0.32\linewidth}
\includegraphics[width=\linewidth]{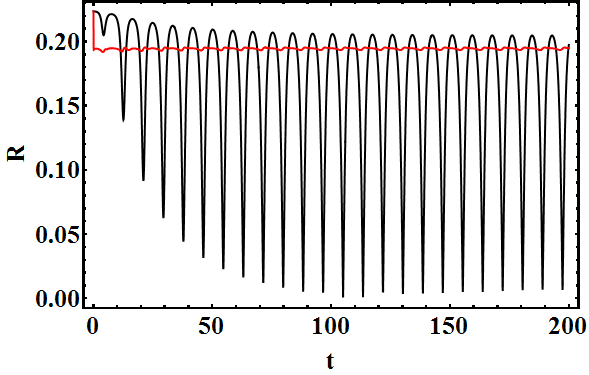}
    \caption{$\chi=0.1, \Lambda_{an}=1.51, S=21.15$}
     \end{subfigure}  
 \hspace{0.12cm}     
     \begin{subfigure}[b]{0.32\linewidth}
 \includegraphics[width=\linewidth]{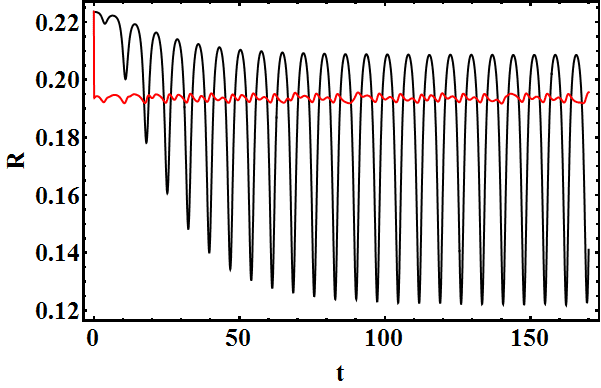}
     \caption{$\chi=0.1, \Lambda_{an}=2.04, S=21.15$}
      \end{subfigure} 
        \hspace{0.12cm}     
     \begin{subfigure}[b]{0.32\linewidth}
 \includegraphics[width=\linewidth]{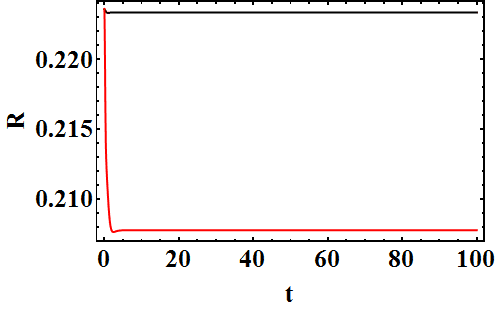}
     \caption{$\chi=0.5, \Lambda_{an}=0.32, S=21.15$}
      \end{subfigure}  
  \hspace{0.12cm}     
      \begin{subfigure}[b]{0.32\linewidth}
  \includegraphics[width=\linewidth]{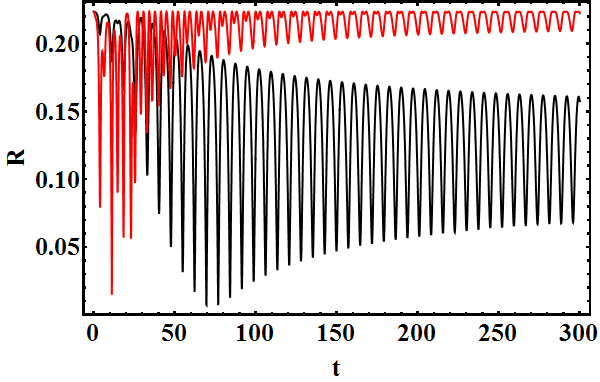}
      \caption{$\chi=0.5, \Lambda_{an}=1.77, S=21.15$}
       \end{subfigure} 
     \hspace{0.12cm}     
         \begin{subfigure}[b]{0.32\linewidth}
     \includegraphics[width=\linewidth]{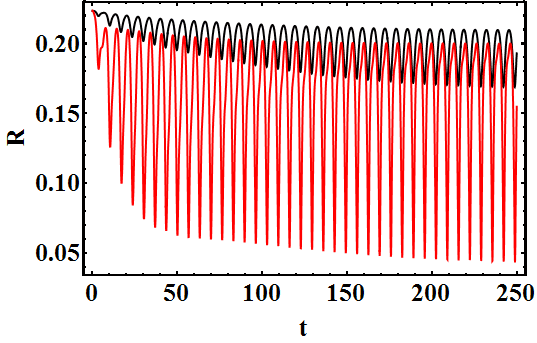}
         \caption{$\chi=0.5, \Lambda_{an}=3.09, S=21.15$}
          \end{subfigure}
          \hspace{0.12cm}     
       \begin{subfigure}[b]{0.32\linewidth}
   \includegraphics[width=\linewidth]{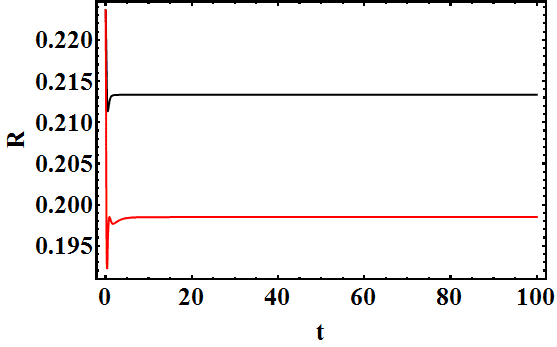}
       \caption{$\chi=0.9, \Lambda_{an}=0.32, S=2.11$}
        \end{subfigure}  
    \hspace{0.12cm}     
        \begin{subfigure}[b]{0.32\linewidth}
    \includegraphics[width=\linewidth]{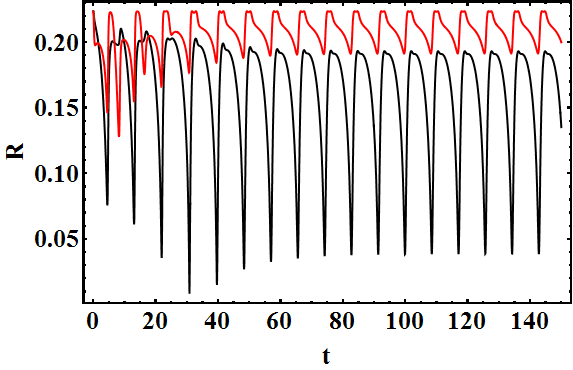}
        \caption{$\chi=0.9, \Lambda_{an}=1.51, S=2.11$}
         \end{subfigure} 
       \hspace{0.12cm}     
           \begin{subfigure}[b]{0.32\linewidth}
       \includegraphics[width=\linewidth]{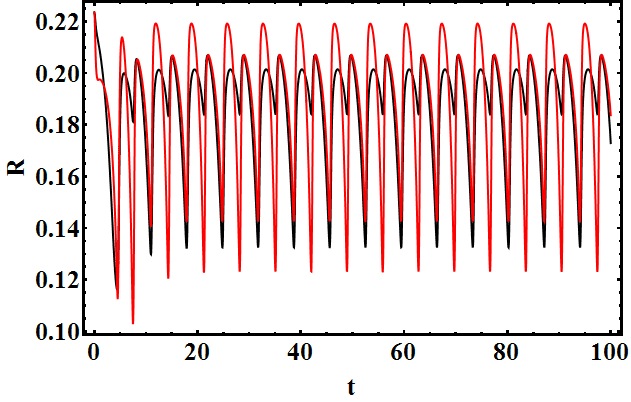}
           \caption{$\chi=0.9, \Lambda_{an}=1.77, S=2.11$}
            \end{subfigure}                          
\caption{Variation of amplitude of shape deformation (R) of the inner (red) and the outer (black) vesicle with time for three different sizes of the inner vesicle $\chi=0.1, 0.5, 0.9$ ($\Delta_{in}=\Delta_{ex}=0.2$).}
    \label{deformationR}
\end{figure}
\subsubsection{Desynchronization in tumbling regime}

An important observation was made by the Steinberg group (\cite{steinberg2014PRL}) in their experiments in the TU regime at various sizes of the inner vesicle ($\chi$). It was observed that the dynamics at low $\chi$ is de-synchronized. Specifically, the angle $\Delta\alpha=\psi_{in}-\psi_{ex}$ changes with time. However at a higher $\chi$, a perfect synchronous dynamics was observed between the inner and the outer vesicles in the TU regime. Our calculation are presented in fig \ref{TUdelta}b which shows the $\Delta\alpha$ for the TU dynamics (of both the vesicles) shown in fig \ref{TUdelta}a at low $\chi$. The figure indicates that the difference angle is nearly constant over time at low $S$ (or $Ca$). However in the small $\chi$ limit, at a slightly higher $S$, while the outer vesicle shows TU, the inner vesicle undergoes transition to IUS (fig \ref{TUdelta}c). Therefore, $\Delta\alpha$ (fig \ref{TUdelta}d) in this case significantly fluctuates. Thus the difference angle ($\Delta\alpha$) is highly sensitive to the $S$ in the low $\chi$ regime. On the other hand, our calculations  show that the angle $\Delta\alpha$ in the high $\chi$ regime (fig \ref{TUdelta}e,f), although not equal to zero, (unlike the experiments by the Steinberg group) fluctuates weakly, and appears synchronized.

\subsubsection{Compound vesicle shape deformation}

Figure \ref{deformationR} shows the time dependent shape deformation of the inner and the outer vesicle for the three different sizes of the inner vesicle, $\chi=0.1, 0.5, 0.9$. The deformation amplitude ($R$) in the TT regime (fig \ref{deformationR}a,d,g) is higher for the outer vesicle that possibly results in a lower inclination angle due to a higher torque. Figure \ref{deformationR}b,e,h show, that in the SW regime, the overall shape deformation of the inner vesicle is always higher than the outer vesicle, although the amplitudes of the shape oscillations are smaller compared to the outer vesicle. In TT and SW regime results are independent of the size of the inner vesicle ($\chi$). In the TU regime (fig \ref{deformationR}c,f,i) a greater sensitivity of the deformation to the size ($\chi$) of inner vesicle is observed. The oscillation amplitudes are higher for the outer vesicle, where as the average deformation is lower, as compared to the inner vesicle at low $\chi$. At intermediate $\chi$, an exact reversal is seen. The outer vesicle now has higher deformation but lower amplitudes. At very high values of $\chi$, the deformation and amplitudes of the two vesicles are comparable.  

\subsubsection{Limiting case: Inner vesicle as a solid inclusion}

\begin{figure} [tbp] %[tbp]
    \centering
    \begin{subfigure}[b]{0.48\linewidth}
          \includegraphics[width=\linewidth]{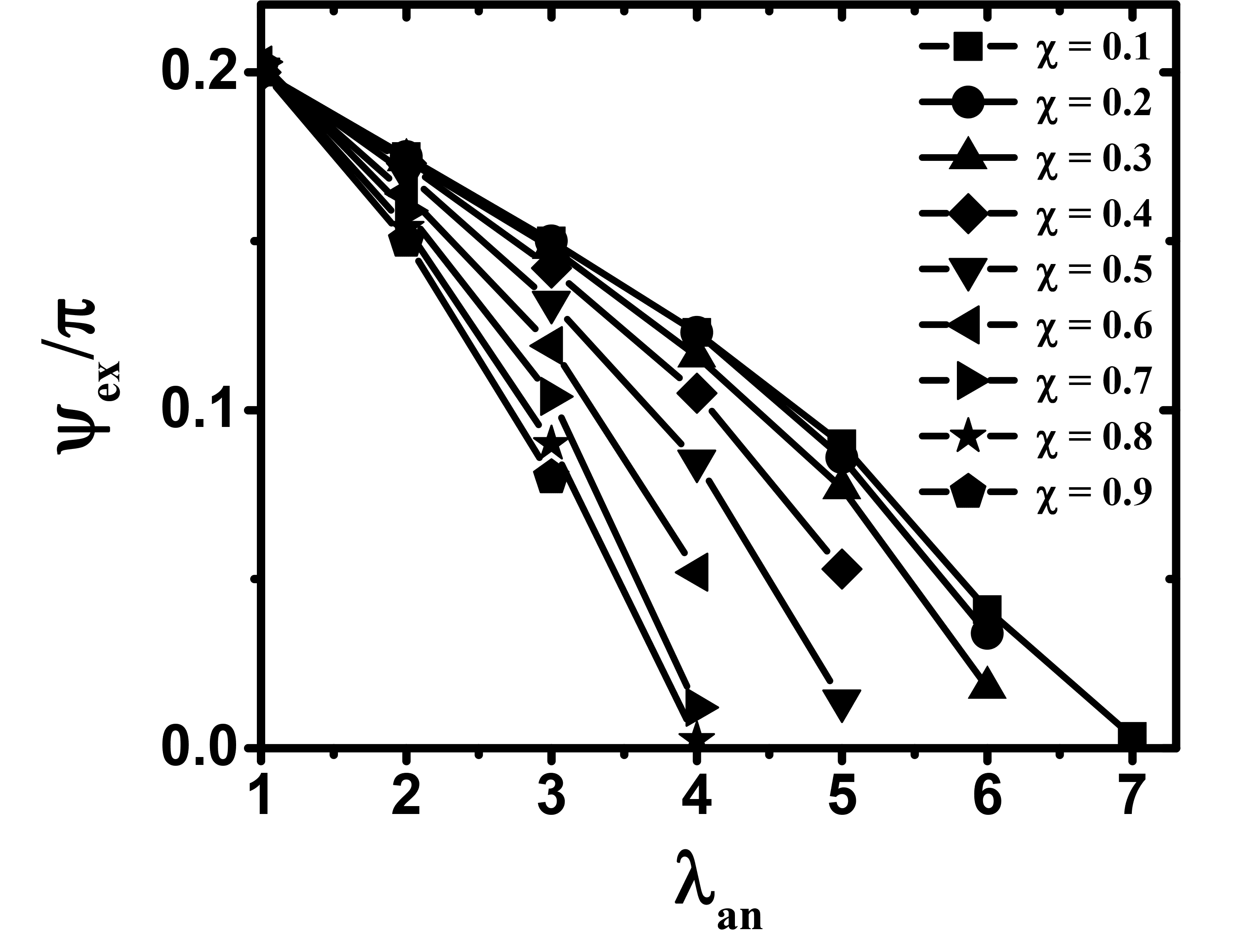}
          \caption{}
         \end{subfigure}
\hspace{0.12cm} 
       \begin{subfigure}[b]{0.48\linewidth}
             \includegraphics[width=\linewidth]{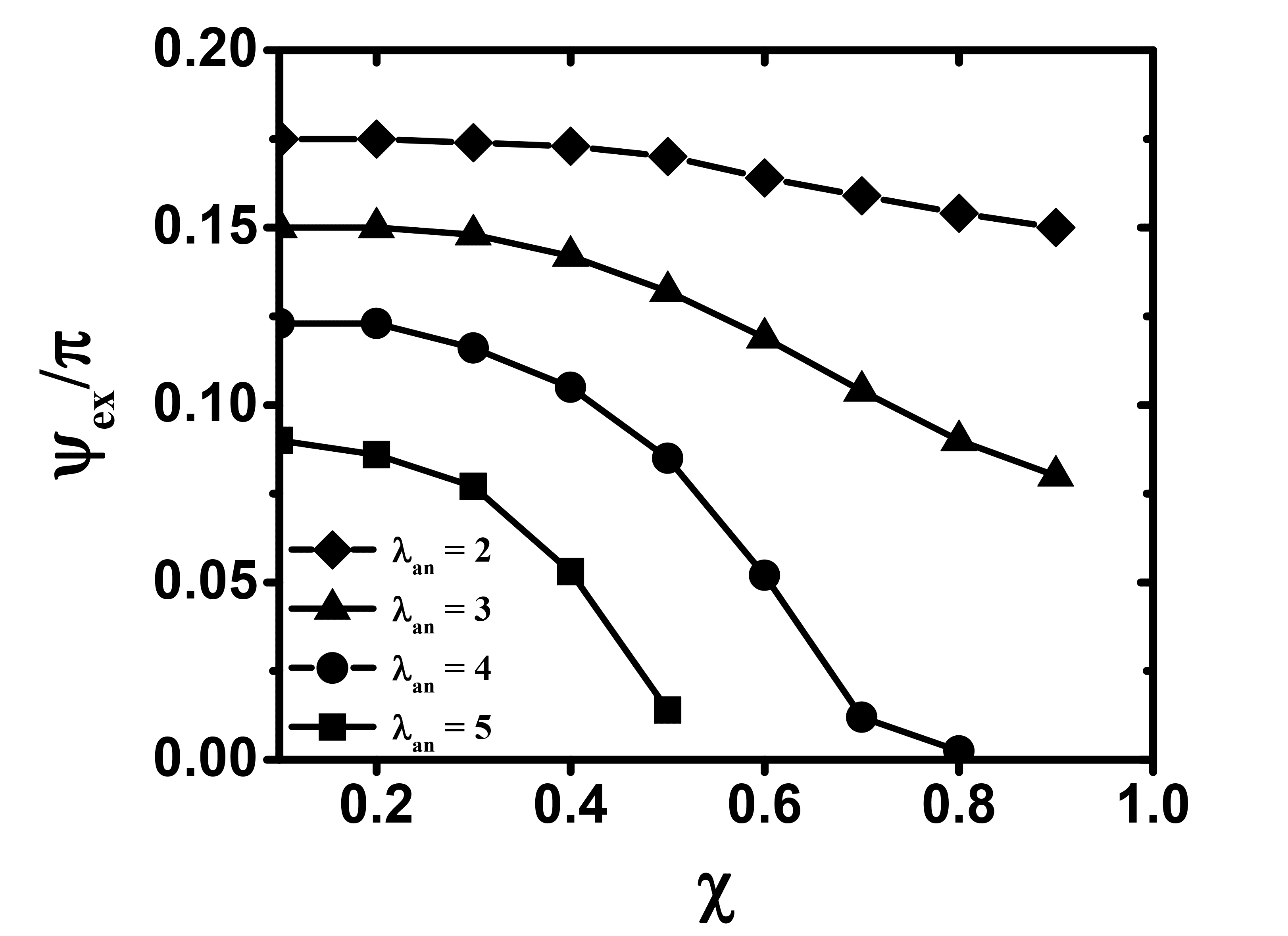}
             \caption{}
            \end{subfigure}
 \hspace{0.12cm}             
       \begin{subfigure}[b]{0.5\linewidth}
        \includegraphics[width=\linewidth]{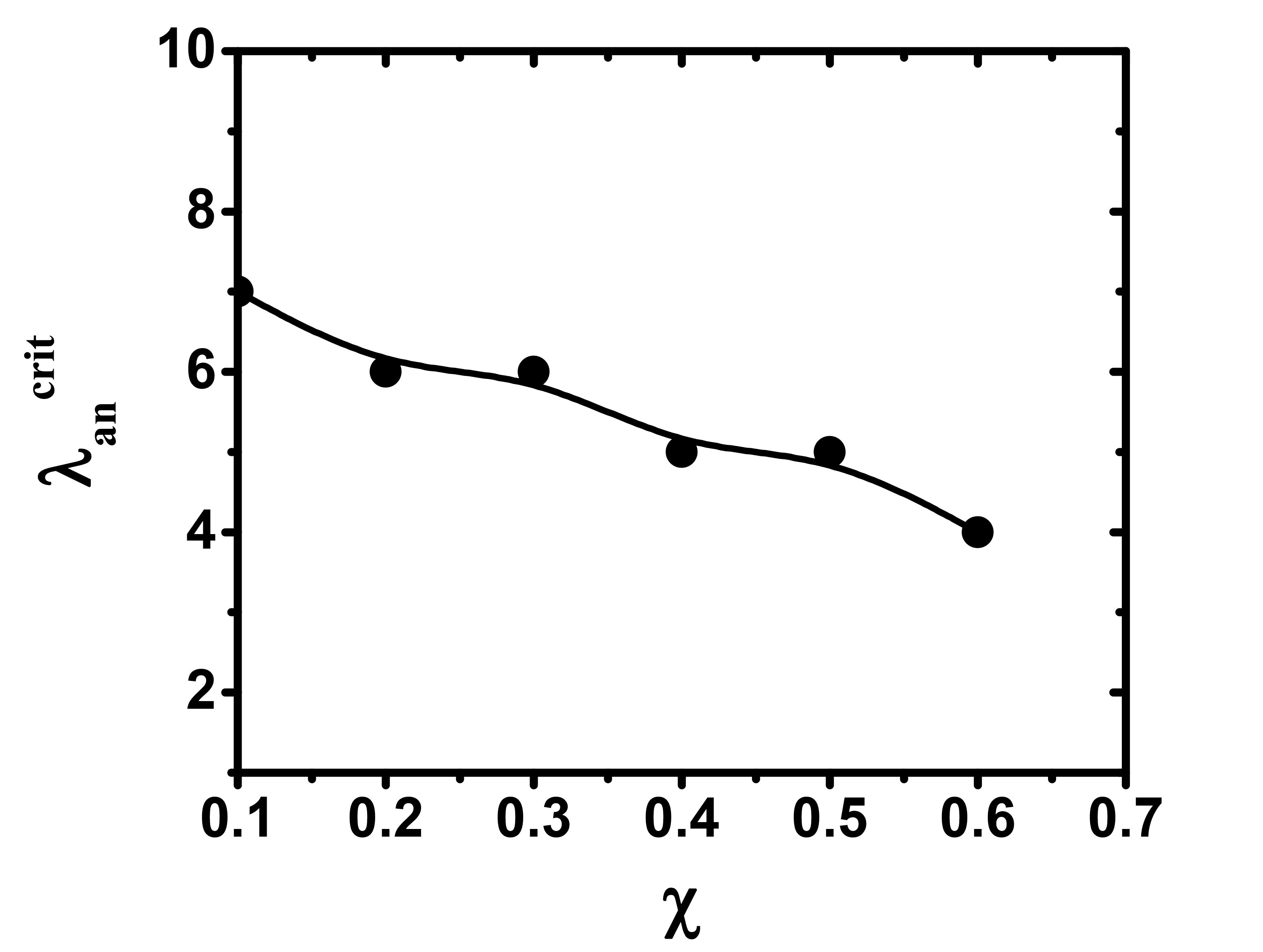}
        \caption{}
            \end{subfigure}                           
 \caption{Variation of inclination angle of the outer vesicle with (a) viscosity contrast across the outer vesicle ($\Lambda_{an}$) and (b) size of the inner vesicle ($\chi$); (c) Variation of the critical viscosity contrast ($\lambda_{an}^{crit}$) across the outer vesicle with the size of the inner vesicle, for vesicle with high inner viscosity using higher order theory ($\Delta_{in}=0.002, \Delta_{ex}=0.2, \lambda_{in}=1000, Ca=1$).}
    \label{viscinhuge}
\end{figure}
To validate our theoretical model with the results of a compound vesicle with a solid inclusion (\cite{vlah2011PRL}), the viscosity of the inner vesicle is increased significantly. In such a  system the high viscosity ratio across the inner vesicle  renders it into the TU regime for all capillary numbers. The outer vesicle then shows different dynamical modes.

Figure \ref{viscinhuge}a,b shows the variation of the inclination angle of the outer vesicle with viscosity contrast across it. The figure indicates that for any given $\lambda_{an}$, the inclination angle of the outer vesicle decreases with an increase in the size of the inner vesicle (fig \ref{viscinhuge}a,b) which is in agreement with literature predictions (\cite{vlah2011PRL}). However, a quantitative agreement is not seen with the results in \cite{vlah2011PRL}. While \cite{vlah2011PRL} results predict a critical size of the inner vesicle of $\chi=0.5$ for $\lambda_{an}=1$ for a TT-TU transition, on the other hand we observe a much gradual variation of the critical $\lambda_{an}^{crit}$ vs size of the inner vesicle ($\chi$) (fig. \ref{viscinhuge}c). This could be attributed to realising a rigid sphere as a limiting case of the viscosity of the inner vesicle going to infinity.  
  
\section{Conclusions}
In this work we present an elaborate investigation of the dynamics of a compound vesicle in shear flow. Our results are in a  fair qualitative agreement with the few experimental data presented by \cite{steinberg2014PRL}. The primary cause of quantitative discrepancy could be the high excess area of the vesicles in the experimental study. Our study though is restricted to small excess area which is used as a perturbation parameter. Our results indicate that the inclination angle of the outer vesicle has a weak dependence on the size of the inner vesicle ($\chi$), while the inner vesicle shows much higher sensitivity in the TT regime. The inclination angles of both the inner and the outer vesicle decrease with $\Lambda_{an}$, such that the critical $\Lambda_{an}^{crit}$ is greater for the inner vesicle for transition out of the TT regime. Elaborate phase diagrams in the $\Lambda_{an}-S$ coordinates indicate that the inner and outer vesicle show similar dynamics for most $\chi$'s in a large area of the phase space. A modification of the TU mode, which we term the IUS mode is observed for the inner vesicle. Significant differences between the dynamics of the inner and the outer vesicles though are observed at moderate $\chi$'s. While at lower $\chi$'s the tumbling dynamics is sensitive to the capillary number (or $S$), a nearly synchronized motion is observed at higher $\chi$'s. For a solid inclusion, a reduction in the inclination angle is observed with an increase in the size of the inner vesicle ($\chi$).  

An important drawback of the present work, and for that reason any asymptotic analysis in vesicle dynamics is the restriction to small excess area. Experiments though indicate use of much higher excess area. Therefore the results presented in this work, could get modified when extended to higher excess area, and extensive numerical simulations may be required. Nevertheless the results of the present work are in fair qualitative agreement with experiments. Experiments on the dynamics of  nucleate cells
and their comparison with the results obtained for compound vesicles will establish the relevance of such work in biology, and help make  progress in this extremely important problem of relevance.\\
     
\textbf{\large Acknowledgement}\\ 

Authors would like to acknowledge the Department of Science and Technology, India, for financial support.

\newpage
\bibliographystyle{model1-num-names}
\bibliography{arxivecompshear}

\newpage
\appendix
\section*{\LARGE Appendix}
\section{Expressions for $Z_i$ in the normal hydrodynamic stresses} \label{App1}
\begin{align}
\ Z_{in}|_{r=\chi}= &\frac{\lambda^{-2 + l}}{\sqrt{l}}(C^g_{lm1} (-1 + l) \sqrt{l} y_{lm1} \chi + C^g_{lm0} (\sqrt{l} y_{lm0} + 2 l^{3/2} y_{lm0} \chi^2 + l^{5/2} y_{lm0} (-1 + \chi^2) + l^2 \sqrt{1 + l} y_{lm2} (-1 + \chi^2)&\nonumber \\
& + l \sqrt{1 + l} y_{lm2} (1 + \chi^2)) - C^g_{lm2} (\sqrt{1 + l} y_{lm0} (3 + l (2 + l) (-1 + \chi^2)) + \sqrt{l} y_{lm2} (3 + \chi^2 + l (2 + l) (-1 + \chi^2))))&\\
\ Z_{an}|_{r=\chi}=&\frac{\chi^{-3 - l}}{\sqrt{l} \sqrt{1 + l}} (C^{da}_{lm2} \sqrt{l} (-4 \sqrt{1 + l} y_{lm2} - \sqrt{l} y_{lm0} (-4 + \chi^2) + 
l^{5/2} y_{lm0} (-1 + \chi^2) - l^2 \sqrt{1 + l} y_{lm2} (-1 + \chi^2))&\nonumber \\
&+ C^{da}_{lm0} \sqrt{l} (-2 l \sqrt{1 + l} y_{lm0} - l^2 \sqrt{1 + l} y_{lm0} - \sqrt{1 + l} y_{lm0} \chi^2 + l^2 \sqrt{1 + l} y_{lm0} \chi^2 - \sqrt{l} (1 + l) y_{lm2}(-2 +&\nonumber \\
    & l (-1 + \chi^2))) + \chi (-C^{da}_{lm1} \sqrt{l}\sqrt{1 + l} (2 + l) y_{lm1} + \chi^{2 l} 
          (C^{ga}_{lm2} (-(1 + l) y_{lm0} (3 + l (2 + l) (-1 + \chi^2)) - \sqrt{l} \sqrt{1 + l} y_{lm2} &\nonumber \\
            &(3 + \chi^2 + l (2 + l) (-1 + \chi^2))) + \sqrt{l} (C^{ga}_{lm1} (-1 + l) \sqrt{1 + l} y_{lm1} \chi+ C^{ga}_{lm0} (\sqrt{l} y_{lm2} (1 - l^2 + (1 + l)^2 \chi^2) &\nonumber \\
          &+ \sqrt{1 + l} y_{lm0} (1 - l^2 + l (2 + l) \chi^2))))))&\\
\ Z_{an}|_{r=1}=&C^{ga}_{lm0} y_{lm0} + 2 C^{ga}_{lm0} l y_{lm0} - 3 C^{ga}_{lm2} y_{lm0}/\sqrt{l/(1 + l)} + 3 C^{da}_{lm2} \sqrt{l/(1 + l)} y_{lm0} - (2 C^{da}_{lm1} + C^{ga}_{lm1}) y_{lm1}&\nonumber \\
        &- C^{da}_{lm1} l y_{lm1} + C^{ga}_{lm1} l y_{lm1}- 4 C^{da}_{lm2} y_{lm2} - 4 C^{ga}_{lm2} y_{lm2} + 2 C^{ga}_{lm0} \sqrt{l} \sqrt{1 + l} y_{lm2} - 
 C^{da}_{lm0} (y_{lm0} + 2 l y_{lm0}&\nonumber \\
    & - 2 \sqrt{l} \sqrt{1 + l} y_{lm2})&\\
\ Z_{ex}|_{r=1}=&C^\infty_{lm0} y_{lm0} + 2 C^\infty_{lm0} l y_{lm0} - 3 C^\infty_{lm2} y_{lm0}/\sqrt{l/(1 + l)} + 
 3 C^d_{lm2} \sqrt{l/(1 + l)} y_{lm0} - (2 C^d_{lm1} + C^\infty_{lm1}) y_{lm1} &\nonumber \\
    &- C^d_{lm1} l y_{lm1} + 
 C^\infty_{lm1} l y_{lm1} - 4 C^d_{lm2} y_{lm2} - 4 C^\infty_{lm2} y_{lm2} + 2 C^\infty_{lm0} \sqrt{l} \sqrt{1 + l} y_{lm2} - 
 C^d_{lm0} (y_{lm0} + 2 l y_{lm0}&\nonumber \\
    &- 2 \sqrt{l} \sqrt{1 + l} y_{lm2})&
\end{align}

\section{Coefficients in expression for pressure}\label{App2}
\begin{align}
&\ A^g_{in}=-\frac{(3+2l)\left(C_{lm2}^g (l+1)-C_{lm0}^g \sqrt{l(l+1)}\right)}{l}&\\
&\ A^g_{an}=-\frac{(3+2l)\left(C_{lm2}^{ga}(l+1)-C_{lm0}^{ga}\sqrt{l(l+1)}\right)}{l}&\\
&\ A^d_{an}=\frac{\sqrt{l}(-l+21)\left
(C_{lm2}^{da} \sqrt{l}+C_{lm0}^{da}\sqrt{l+1}\right)}{l+1}&\\
&\ A^g_{ex}=-\frac{(3+2l)\left(C^\infty_{lm2} (l+1)-C^\infty_{lm0} \sqrt{l(l+1)}\right)}{l}&\\
&\ A^d_{ex}=\frac{\sqrt{l}(-1+2l)(C_{lm0}^d+C_{lm1}^d+C_{lm2}^d\sqrt{l (l+1)})}{(l+1)^{3/2}}&
\end{align}

\section{Curvature and membrane stress calculations for inner and outer vesicle upto first order spherical harmonic approximation} \label{App3}
The surfaces of the outer and inner vesicles are defined by,
\begin{align}
&\ r_s^{ex}=\left(\alpha_{ex}+\sum_{lm}f_{lm}^{ex}Y_{lm}\right)  \label{rex}&\\
&\ r_s^{in}=\chi \left(\alpha_{in}+\sum_{lm}f_{lm}^{in}Y_{lm}\right) \label{rin},&
\end{align}
respectively.\\
From the volume conservation, we get $\int (r_s^{ex})^3d\Omega=4\pi$ and $\int (r_s^{in})^3d\Omega=4\pi\chi^3$ for the outer and inner vesicle respectively with
$\int Y_{lm} d\Omega=0, \int Y_{lm} Y_{lm}^* d\Omega=\delta_{lm}$.\\
The volume conservation yields
\begin{align}
&\ \int{r_s^3 d\Omega}= 4\pi \alpha^3+3\alpha\int \sum f_{lm} f_{lm}^* \label{area3}&
\end{align}
where
\begin{align}
&\alpha_{ex}=1-\frac{1}{4\pi}\sum f_{lm}^{ex} (f_{lm}^{ex})^*&
\end{align}
and
\begin{align}
&\alpha_{in}=1-\frac{1}{4\pi}\sum f_{lm}^{in} (f_{lm}^{in})^*&
\end{align}

\textit{Unit Normal Calculation}\\
The surface of a vesicle is described by,
\begin{align}
& r_s=\alpha+\sum f_{lm} Y_{lm}& \label{surves1}
\end{align}
where the surface function is defined as
\begin{align}
& F=r-r_s&
\end{align}
the gradient of surface function is given by
\begin{align}
&\nabla F=e_r-\nabla_\Omega r_s&
\end{align}
such that for outer vesicle ($r=1$)
\begin{align}
&\ n_{ex}=\frac{e_r-\sum f_{lm}^{ex}\sqrt{l(l+1)}y_{lm0}}{|\nabla F|}&
\end{align}
and for the inner vesicle ($r=\chi$)
\begin{align}
&\ n_{in}=\frac{e_r-\sum f_{lm}^{in}\sqrt{l(l+1)}y_{lm0}}{|\nabla F|}&
\end{align}
\textit{Area Constraint for Inner and Outer Vesicle}
\begin{align}
&\frac{A}{R_{ex}^2}=\int \frac{r_s^2 d\Omega}{\hat{e}_r.n}&
\end{align}
for outer vesicle
\begin{align}
&\frac{A_{ex}}{R_{ex}^2}=\int \frac{(r_s^{ex})^2 d\Omega}{\hat{e}_r.n_{ex}}=4\pi+\Delta_{ex} \label{consarea1}&
\end{align}
for inner vesicle
\begin{align}
&\frac{A_{in}}{R_{ex}^2}=\int \frac{(r_s^{in})^2 d\Omega}{\hat{e}_r.n_{in}}=4\pi \chi^2+\Delta_{in} \chi^2 \label{consarea2}&
\end{align}
yields
\begin{align}
&\Delta_{ex}=\sum \frac{(l+2)(l-1)}{2} f_{lm}^{ex} (f_{lm}^{ex})^*&
\end{align}
and
\begin{align}
&\Delta_{in}=\sum \frac{(l+2)(l-1)}{2} f_{lm}^{in} (f_{lm}^{in})^*&
\end{align}
\\
\textit{Curvature Calculation}\\
Outer vesicle curvature
\begin{align}
&\ 2H_{ex}=\nabla.n_{ex}=\frac{1}{r_s^{ex}}\left(2+l(l+1)f_{lm}^{ex} Y_{lm}\right)\left(1-\frac{l(l+1)}{2}f_{lm}^{ex} (f_{lm}^{ex})^* y_{lm0}.y_{lm0}^*\right)&
\end{align}
and
\begin{align}
&\nabla_{\Omega}^2 H_{ex}=-\frac{1}{2}\sum (l+2) (l-1)l(l+1)f_{lm}^{ex} Y_{lm}&
\end{align}
The inner vesicle curvature
\begin{align}
&\ H_{in}=\frac{1}{\chi}\left(1+\frac{(l+2) (l-1)}{2}\sum f_{lm}^{in} Y_{lm}\right)&
\end{align}
and
\begin{align}
&\nabla^2 H_{in}=-\frac{1}{2\chi}\sum (l-1)(l+2)l(l+1) f_{lm}^{in} Y_{lm}&
\end{align}
\\
\textit{Membrane Stress Calculation}\\
The total membrane stress (here, over-bar presents dimensional quantities)
\begin{align}
\ \bar{\tau}^m=2\bar{\sigma}^{in} \bar{H}^{in}_n-\left(2\kappa_b \nabla_s^2 \bar{H}^{in}\right)n-\nabla_s \bar{\sigma}^{in} \label{totmemstr}
\end{align}
using $\bar{\sigma}^{in}=\bar{\sigma}_0^{in}+\sum \sum \bar{\sigma}_{lm}^{in} Y_{lm}$
\begin{align}
&\nabla_s \bar{\sigma}^{in}=\frac{\sqrt{l(l+1)}}{r_s^{in}} \bar{\sigma}_{lm}^{in} y_{lm0}&
\end{align}

The normal stress term, from eq. \ref{totmemstr}
\begin{align}
&\bar{\tau}_n^{m,in}=\bar{\tau}_n^{bend,in}+\bar{\tau}_n^{tens,in}=-2\kappa_b \nabla_s^2 \bar{H}^{in} n+2\bar{\sigma}^{in}\bar{H}^{in}n&\\
&\bar{\tau}_n^{m,in}=\left(-\frac{\kappa_b}{\chi}\sum (l-1)(l+2)l(l+1) f_{lm}^{in} Y_{lm}\right)n
+2\left(\frac{\bar{\sigma}_{lm}^{in}Y_{lm}}{\chi}+\frac{\bar{\sigma}_0^{in}}{2\chi}\sum f_{lm}^{in} Y_{lm} (l+2) (l-1)\right)n&
\end{align}
Non-dimensionalization of membrane stress by $\dot{\gamma} \mu_{ex}$, i.e, $\bar{\tau}^m=\tau^m (\dot{\gamma} \mu_{ex})$, curvature by $R_{ex}$, i.e.,$ H=\bar{H}_{ex}/R_{ex} $, and tension by $\kappa_b/R_{ex}^2$, i.e., $\bar{\sigma}=\sigma (\kappa_b/R_{ex}^2)$, yields
\begin{align}
&\bar{\tau_n^{m,in}}=\frac{1}{\chi Ca}\left(2\sigma_{lm}^{in}+(\sigma_0^{in}+l(l+1))(l+2)(l-1) f_{lm}^{in} \right)y_{lm2}&
\end{align}
where $Ca=\frac{\dot{\gamma}\mu_{ex} R_{ex}^3}{\kappa_b}$\\

The tangential stress term, from eq. \ref{totmemstr}
\begin{align}
&\bar{\tau}_t^{m,in}=\bar{\tau}_{nu}^{tens}=-\nabla_s \bar{\sigma}^{in}=-\frac{1}{\chi Ca}\sqrt{l(l+1)} \sigma_{lm}^{in} y_{lm0}&
\end{align}

\section{Curvature and membrane stresses for the outer vesicle up to second order approximation in the spherical harmonics} \label{App4}
The mean curvature is given by
\begin{equation}
\ H=1+\frac{1}{2}(l(l+1)-2)F(\theta,\phi)-(l(l+1)-1)F(\theta,\phi)^2 \label{curvout_mean}
\end{equation}
The Gaussian curvature
\begin{align}
&\ K= 1+(l(l+1)-2)F(\theta,\phi)-3(l(l+1)-1)F(\theta,\phi)^2 -l(l+1)F(\theta,\phi)\left(-l(l+1)F(\theta,\phi)-\frac{d^2F(\theta,\phi)}{d\theta^2}\right)& \nonumber \\
&-\cot^2\theta \csc^2\theta \left(\frac{dF(\theta,\phi)}{d\phi}\right)^2+2\cot\theta \csc^2\theta
 \frac{dF(\theta,\phi)}{d\phi}\frac{d^2F(\theta,\phi)}{d\theta d\phi}-\left(-l(l+1)F(\theta,\phi)-\frac{d^2F(\theta,\phi)}{d\theta^2}\right)^2& \nonumber \\
  &-\csc^2\theta \left(\frac{d^2F(\theta,\phi)}{d\theta d\phi}\right)^2 & \label{curvou_gauss}
\end{align}
where $F(\theta,\phi)=\sum_{l=2}\sum_{m=-l}^{l} f_{lm}^{ex} Y_{lm}(\theta, \phi)$\\

Substitution of eq. \ref{curvout_mean} and \ref{curvou_gauss} into eq. \ref{mnstr} and \ref{mtstr} gives the expression for the normal as well as tangential stress on the outer vesicle in terms of uniform tension ($\sigma_0^{in}, \sigma_0^{ex}$) and non uniform tension ($\sigma_{2j}^{in},\sigma_{2j}^{ex}$) on each vesicle.

From eq. \ref{mnstr}, the normal membrane stress on the outer vesicle
\begin{align}
&\tau^{m,ex}_{2-2} = 
  \frac{1}{Ca} ((144 \sqrt{5 \pi} f_{20}^{ex} f_{2-2}^{ex} + 
  40 \sqrt{5 \pi} f_{20}^{ex} f_{2-2}^{ex} \sigma_0^{ex})/(14 \pi) + 4 f_{2-2}^{ex} (6 + \sigma_0^{ex}) + 2 \sigma_{2-2}^{ex}) y_{2-22}&
\end{align}
\begin{align}
&\tau^{m,ex}_{20}=\frac{1}{Ca} ((144 \sqrt{5 \pi} f_{22}^{ex} f_{2-2}^{ex}+40 \sqrt{5 \pi} f_{22}^{ex} f_{2-2}^{ex} \sigma_0^{ex}-4 \sqrt{5 \pi} (5 \sigma_0^{ex}+18) (f_{20}^{ex})^2)/(14\pi)+4 f_{20}^{ex} (6+\sigma_0^{ex})& \nonumber \\ &+2 \sigma_{20}^{ex}) y_{202}&
\end{align}
\begin{align}
&\tau^{m,ex}_{22} = 
  \frac{1}{Ca} ((144 \sqrt{5 \pi} f_{20}^{ex} f_{22}^{ex} + 
 40 \sqrt{5 \pi} f_{20}^{ex} f_{22}^{ex} \sigma_0^{ex})/(14 \pi) + 4 f_{22}^{ex} (6 + \sigma_0^{ex}) + 2 \sigma_{22}^{ex}) y_{222}&
\end{align}

From eq. \ref{mtstr}, the tangential membrane stress on the outer vesicle in terms of the non-uniform tension ($\sigma_{2j}^{ex}$) can be written as
\begin{align}
\tau^{m,ex}_{2m} = -\frac{\sigma_{2m}^{ex} \sqrt{(l (l + 1))} y_{2m0}} {Ca}
\end{align}

\section{Curvature and membrane stresses for the inner vesicle up to second order approximation in the spherical harmonics}\label{App5}

The mean curvature for the inner vesicle 
\begin{equation}
\ H=\frac{1}{\chi}+\frac{(l(l+1)-2)F(\theta,\phi)}{2\chi}-\frac{(l(l+1)-1)F(\theta,\phi)^2}{\chi}  \label{CurvsecMean}
\end{equation}
The Gaussian curvature for the inner vesicle
\begin{align}
&\ K= \frac{1}{\chi^2}+\frac{(l(l+1)-2)F(\theta,\phi)}{\chi^2}+\frac{1}{\chi^2}(-3(l(l+1)-1)F(\theta,\phi)^2-\cot^2\theta \csc^2\theta \left(\frac{dF(\theta,\phi)}{d\phi}\right)^2& \nonumber \\
 &+2\cot\theta \csc^2\theta
 \frac{dF(\theta,\phi)}{d\phi}\frac{d^2F(\theta,\phi)}{d\theta d\phi}-\csc^2\theta \left(\frac{d^2F(\theta,\phi)}{d\theta d\phi}\right)^2-\left(\cot\theta\frac{dF(\theta,\phi)}{d\theta}+\csc^2\theta\frac{d^2F(\theta,\phi)}{d\phi^2}\right)^2& \nonumber \\
  &
  -l(l+1)F(\theta,\phi)\left(\cot\theta\frac{dF(\theta,\phi)}{d\theta}+\csc^2\theta\frac{d^2F(\theta,\phi)}{d\phi^2}\right)) & \label{CurvsecGauss}
\end{align}
where $F(\theta,\phi)=\sum_{l=2}\sum_{m=-l}^{l} f_{lm}^{in} Y_{lm}(\theta, \phi)$  \\

From eq. \ref{mnstr}, the non-linear term in the normal stress, that is ($H^3-KH$) can be written in terms of spherical harmonics associated with $Y_{2-2}$, $Y_{20}$, $Y_{20}$ mode using eq. \ref{CurvsecMean} and eq. \ref{CurvsecGauss} as\\
\begin{align}
\ HnonY_{2-2}=&\frac{(3 f_{2-2}^{in} (4576 \sqrt{5} \pi^2 f_{20}^{in}))/(4004 \pi^{5/2} \chi^3)(3 f_{2-2}^{in} (4576 \sqrt{5} \pi^2 f_{20}^{in}))}{4004 \pi^{5/2} \chi^3} & \label{hnon1}\\  
\ HnonY_{20}=&\frac{3 (-4576 \sqrt{5} \pi^2 ((f_{20}^2)^{in} - 2 f_{2-2}^{in} f_{22}^{in}))}{8008 \pi^{5/2} \chi^3} & \label{hnon2}\\ 
\ HnonY_{22}=&\frac{(3 f_{22}^{in} (4576 \sqrt{5} \pi^2 f_{20}^{in}))/(4004 \pi^{5/2} \chi^3)(3 f_{22}^{in} (4576 \sqrt{5} \pi^2 f_{20}^{in}))}{4004 \pi^{5/2} \chi^3} &\label{hnon3} 
\end{align}

From eq. \ref{mnstr}, The Beltrami term in the normal stress, that is ($\nabla^2 H$) can be written in terms of spherical harmonics associated with $Y_{2-2}$, $Y_{20}$, $Y_{20}$ mode using eq. \ref{CurvsecMean} and eq. \ref{CurvsecGauss} as\\
\begin{align}
\ (\nabla_s^2H)_{Y2-2}=&-\frac{12 (7 + 5 \sqrt{5/\pi} f_{20}^{in}) f_{2-2}^{in}}{7\chi^3} & \label{belt1}\\  
\ (\nabla_s^2H)_{Y20}=&\frac{6 (14 \sqrt{5 \pi} f_{20}^{in} - 25 (f_{20}^2)^{in} + 50 f_{2-2}^{in} f_{22}^{in})}{7 \sqrt{5 \pi} \chi^3} &\label{belt2}\\ 
\ (\nabla_s^2H)_{Y22}=&-\frac{12 (7 + 5 \sqrt{5/\pi} f_{20}^{in}) f_{22}^{in}}{7\chi^3}& \label{belt3}
\end{align}

Substitution of eq. \ref{CurvsecMean} to \ref{belt3} into eq. \ref{mnstr} and \ref{mtstr} gives the expression for the normal as well as tangential stresses on the inner vesicle in terms of uniform tensions ($\sigma_0^{in}, \sigma_0^{ex}$) and non-uniform tensions ($\sigma_{2j}^{in},\sigma_{2j}^{ex}$) on each vesicle.

From eq. \ref{mnstr}, the normal membrane stresses on the inner vesicles for all the three modes ($m=-2,0,2$) become,
\begin{align}
 \tau^{m,in}_{2-2}=&\frac{1}{Ca}\left(\frac{\sqrt{5 \pi} (72 + 20 \sigma_0 \chi^2) f_{20}^{in} f_{2-2}^{in} }{7\pi \chi^3}+\frac{2\sigma_{2-2}}{\chi}+4 f_{2-2}^{in}\left(\frac{\sigma_0}{\chi}+\frac{6}{\chi^3}\right)\right) y_{2-20} &\\  
  \tau^{m,in}_{20}=&\frac{1}{Ca}\left(\frac{\sqrt{5 \pi}}{7 \pi \chi^3}(72 + 20 \sigma_0 \chi^2) f_{2-2}^{in} f_{22}^{in}+\frac{2\sigma_{20}}{\chi}+4 f_{20}^{in}\left(\frac{\sigma_0}{\chi}+\frac{6}{\chi^3}\right)-\sqrt{5\pi}\frac{(10\sigma_0 \chi^2+36)}{7\pi \chi^3}(f_{20}^2)^{in}\right) y_{200}&\\  
 \tau^{m,in}_{22}=&\frac{1}{Ca}\left(\frac{\sqrt{5 \pi} (72 + 20 \sigma_0 \chi^2) f_{20}^{in} f_{22}^{in} }{7\pi \chi^3}+\frac{2\sigma_{22}}{\chi}+4 f_{22}^{in}\left(\frac{\sigma_0}{\chi}+\frac{6}{\chi^3}\right)\right) y_{220}& 
\end{align}

From eq. \ref{mtstr}, the tangential membrane stress on the inner vesicle for all three modes ($m=-2,0,2$) in terms of non-uniform tension ($\sigma_{2j}^{in}$) on inner vesicle become,
\begin{align}
 \tau^{m,in}_{2m}=&-\frac{ \sigma_{2m}^{in} \sqrt{(l (l + 1))}}{\chi Ca} y_{2m0} &
\end{align}

\section{Coefficients obtained from the overall stress balance} \label{App6}
Tangential stress balance on the outer vesicle yields
\begin{align}
&\ \tau_{2-2,t}^{h,ex}-\lambda_{an} \tau_{2m,t}^{h,an}=\tau_{2m,t}^{m,ex} &\\
&\ \tau_{20, t}^{h,ex}-\lambda_{an} \tau_{20,t}^{h,an}=\tau_{20,t}^{m,ex} &\\
&\ \tau_{22,t}^{h,ex}-\lambda_{an} \tau_{2p, t}^{h,an}=\tau_{2p,t}^{m,ex} &
\end{align}
Solution of the above three equations gives the nonuniform tension in the outer vesicle for each mode
\begin{align}
&\sigma_{2-2}^{ex}=
 Ca (-40 C^\infty_{2-20} + 10 \sqrt{6} C^\infty_{2-22}  - \chi^3 (\sqrt{6} C^g_{2-22} (-25 + 61 \chi^2 - 40 \chi^4) + 
10 C^g_{2-20} (3 - 9 \chi^2 + 8 \chi^4)) (-1 + \lambda_{an}) & \nonumber \\
 &- 
\sqrt{6} C^{ga}_{2-22} (4 + \chi^3 (25 - 61 \chi^2  + 
40 \chi^4) (-1 + \lambda_{an}) + 6 \lambda_{an}) + 10 C^{ga}_{2-20} (\chi^3 (3 - 9 \chi^2 + 8 \chi^4)  (-1 + \lambda_{an}) & \nonumber \\
 &+ 2 (1 + \lambda_{an})))/(4 \sqrt{6})&
\end{align}
\begin{align}
&\sigma_{20}^{ex} = Ca (-40 C^\infty_{200}+10\sqrt{6}C^\infty_{202}-\chi^3(\sqrt{6} C^g_{202}(-25+61 \chi^2-40 \chi^4)+10 C^g_{200}(3-9\chi^2+8\chi^4)) (-1+\lambda_{an})& \nonumber \\
 &-\sqrt{6} C^{ga}_{202}(4 +\chi^3(25-61 \chi^2+40 \chi^4) (-1+\lambda_{an})+6 \lambda_{an})+10 C^{ga}_{200}(\chi^3 (3 -9\chi^2+8\chi^4) (-1+\lambda_{an})& \nonumber \\
          & +2(1+\lambda_{an})))/(4 \sqrt{6})& 
\end{align}
\begin{align}
&\sigma_{22}^{ex} = Ca (-40 C^\infty_{220} + 10 \sqrt{6} C^\infty_{222} - \chi^3 (\sqrt{6} C^g_{222} (-25 + 61 \chi^2 - 40 \chi^4) + 
10 C^g_{220} (3 - 9 \chi^2 + 8 \chi^4)) (-1 + \lambda_{an}) & \nonumber \\
   & - 
\sqrt{6} C^{ga}_{222} (4 + \chi^3 (25 - 61 \chi^2  + 
40 \chi^4) (-1 + \lambda_{an}) + 6 \lambda_{an}) + 10 C^{ga}_{220} (\chi^3 (3 - 9 \chi^2 + 8 \chi^4)(-1 + \lambda_{an}) & \nonumber \\
   &+ 2 (1 + \lambda_{an})))/(4 \sqrt{6})&
\end{align}
Tangential stress balance on inner vesicle
\begin{align}
&\ \lambda_{an} \tau_{2-2, t}^{h,an}-\lambda_{ie} \tau_{2-2, t}^{h,in}=\tau_{2-2,t}^{m,in} &\\
&\ \lambda_{an} \tau_{20, t}^{h,an}-\lambda_{ie} \tau_{20, t}^{h,in}=\tau_{20,t}^{m,in} &\\
&\ \lambda_{an} \tau_{22, t}^{h,an}-\lambda_{ie} \tau_{22, t}^{h,in}=\tau_{22,t}^{m,in} &
\end{align}
Solution of above three equations gives the nonuniform tension in the inner vesicle 
\begin{align}
&\sigma_{2-2}^{in} = (Ca \chi (5 C^{ga}_{2-20} (-5 + 9 \chi^2) (9 - 
7 \chi^2 - 14 \chi^5 + 12 \chi^7) \lambda_{an} - 2 C^g_{2-20} \chi (4 \chi (4 + 50 \chi^3 - 129 \chi^5 + 75 \chi^7) \lambda_{an}& \nonumber \\
& + \chi (4 - 25 \chi^3 + 51 \chi^5 - 30 \chi^7) \lambda_{ie})))/(\sqrt{6} (-5 + 9 \chi^2) (-4 + 25 \chi^3 - 51 \chi^5 + 30 \chi^7))&
\end{align}
\begin{align}
&\sigma_{20}^{in} = Ca \chi (5 C^{ga}_{200} (-5 + 9 \chi^2) (9 - 
7 \chi^2 - 14 \chi^5 + 12 \chi^7) \lambda_{an} - 2 C^g_{200} \chi (4 \chi (4 + 50 \chi^3 - 129 \chi^5 + 
75 \chi^7) \lambda_{an} & \nonumber \\
& + \chi (4 - 25 \chi^3 + 51 \chi^5 - 30 \chi^7) \lambda_{ie}))/(\sqrt{6} (-5 + 9 \chi^2) (-4 + 25 \chi^3 - 51 \chi^5 + 30 \chi^7))&
\end{align}
\begin{align}
&\sigma_{22}^{in} =(Ca \chi (5 C^{ga}_{220} (-5 + 9 \chi^2) (9 - 
7 \chi^2 - 14 \chi^5 + 12 \chi^7) \lambda_{an}- 2 C^g_{220} \chi (4 \chi (4 + 50 \chi^3 - 129 \chi^5 + 75 \chi^7) \lambda_{an}& \nonumber \\
&  + \chi (4 - 25 \chi^3 + 51 \chi^5 - 30 \chi^7) \lambda_{ie})))/(\sqrt{6} (-5 + 9 \chi^2) (-4 + 25 \chi^3 - 51 \chi^5 + 30 \chi^7))&
\end{align}
Normal stress balance ($j=-2,0,2$) on the inner and the outer vesicle
\begin{align}
& \lambda_{an} \tau_{2j, n}^{h,an}-\lambda_{ie} \tau_{2j, n}^{h,in}=\tau_{2j,n}^{m,in} &\\
& \tau_{2j, n}^{h,ex}-\lambda_{an} \tau_{2j, n}^{h,an}=\tau_{2j,n}^{m,ex} &
\end{align}
By solving these two simultaneous equations (one from inner and another from outer vesicle) for each mode, we get the expressions for $C^g_{2j0}, C^{ga}_{2j0}$ and use the incompressibility conditions (for three modes $j=-2,0,2$) given below to get the normal velocity components in each vesicle. Also all the nonuniform tension terms which are present in $C^g_{2j0}, C^{ga}_{2j0}$ are replaced by the expressions of nonuniform tension mentioned above in this subsection of the appendix.  \\
For $j=-2$ mode
\begin{align}
C^g_{2-22} = C^g_{2-20} (\sqrt{6} (-1 + 3 \chi^2))/(-5 + 9 \chi^2)
\end{align}
\begin{align}
& C^{ga}_{2-22} = \sqrt{6} (-16 C^g_{2-20} \chi^5 (1 + \chi) + 
 C^{ga}_{2-20} (-5 + 9 \chi^2) (2 + \chi (1 + \chi) (2 - 3 \chi^2 + 10 \chi^4)))/((-5 + 9 \chi^2) (4 + \chi (1 + \chi) (4& \nonumber \\
 & - 21 \chi^2 + 30 \chi^4)))&
\end{align}
For $j=0$ mode
\begin{align}
C^g_{202} = C^g_{200} (\sqrt{6} (-1 + 3 \lambda^2))/(-5 + 9 \lambda^2)
\end{align}
\begin{align}
& C^{ga}_{202} = (\sqrt{6} (-16 C^g_{200} \lambda^5 (1 + \lambda) + 
 C^{ga}_{200} (-5 + 9 \chi^2) (2 + \chi (1 + \chi) (2 - 3 \chi^2 + 10 \chi^4))))/((-5 + 9 \chi^2) (4 + \chi (1 + \chi) (4& \nonumber \\
 & - 21 \chi^2 + 30 \chi^4)))&
\end{align}
For $j=2$ mode
\begin{align}
C^g_{222} = C^g_{220} (\sqrt{6} (-1 + 3 \chi^2))/(-5 + 9 \chi^2)
\end{align}
\begin{align}
& C^{ga}_{222} = (\sqrt{6} (-16 C^g_{220} \chi^5 (1 + \chi) + 
 C^{ga}_{220} (-5 + 9 \chi^2) (2+ \chi (1 + \chi) (2 - 3 \chi^2 + 10 \chi^4))))/((-5 + 9 \chi^2) (4 + \chi (1 + \chi) (4& \nonumber \\
 & - 21 \chi^2 + 30 \chi^4)))&
\end{align}
The final expressions for the normal velocity components of the inner and the annular fluid are too long and so are not provided in the manuscript, but can be easily obtained from the relationships mentioned above.

\section{Final shape evolution equation}\label{App7}

After knowing the normal velocity components for the inner and the annular region ($C^g_{2j2}, C^{ga}_{2j2}$) from Appendix \ref{App6}, substituting them into eq. \ref{evoleq-1} and eq. \ref{evoleq-2} in section \ref{kinematiceq} of the manuscript.\\ 
For outer vesicle from eq. \ref{evoleq-2}\\
\begin{align}
&\dot{f}_{22}^{ex}=((-(35 Ca (\sqrt{6} C^\infty_{220} + 9 C^\infty_{222}) - 
        24 f_{22}^{ex} (7 (6 + \sigma_0^{ex}) +\nonumber \\ & 
           f_{20}^{ex} \sqrt{5/\pi} (18 + 
              5 \sigma_0^{ex}))) (16 \lambda_{an} \chi (-8 - 
          10 \chi^3 + 13 \chi^5 + 5 \chi^7) + \nonumber \\ &
       46 \lambda_{ie} \chi (-2 + 5 \chi^3 - 13 \chi^5 + 
          10 \chi^7)) + 
    384 f_{22}^{in} \chi (7 (6 + \sigma_0^{in} \chi^2) + \nonumber \\ &
       f_{20}^{in} \sqrt{5/\pi} (18 + 5 \sigma_0^{in} \chi^2)) (16 - 
       11 \lambda_{an} + \chi^2 (-8 + 3 \lambda_{an} + \nonumber \\ &
          4 (-1 + \lambda_{an}) \chi^3 (-1 + 
             3 \chi^2))))/(28 Ca \chi ((32 + 
         23 \lambda_{an}) (32 \lambda_{an} \nonumber \\ &+ 23 \lambda_{ie}) - 
      25 (-8 + \lambda_{an}) (16 \lambda_{an} - 
         23 \lambda_{ie}) \chi^3 - 
      84 (4 + \lambda_{an}) (8 \lambda_{an}\nonumber \\ & - 
         23 \lambda_{ie}) \chi^5 - 
      100 (2 + \lambda_{an}) (4 \lambda_{an} + 
         23 \lambda_{ie}) \chi^7 + 
      736 (-1 + \lambda_{an}) \nonumber \\ &(\lambda_{an} - \lambda_{ie}) \chi^{10})) +
  i f_{22}^{ex} \omega_{ex}) \label{exevol1}&
\end{align}
\begin{align}
& \dot{f}_{20}^{ex}=(-(35 Ca (\sqrt{6} C^\infty_{200} + 9 C^\infty_{202}) + 
       12 (-14 f_{20}^{ex} (6 + \sigma_0^{ex}) + \nonumber \\ &
          (f_{20}^{ex})^2 \sqrt{5/\pi} (18 + 5 \sigma_0^{ex}) - 
          2 f_{2-2}^{ex} f_{22}^{ex} \sqrt{5/\pi} (18 +5 \sigma_0^{ex}))) (16 \lambda_{an} \chi \nonumber \\ &(-8 - 
         10 \chi^3 + 13 \chi^5 + 5 \chi^7) + 
      46 \lambda_{ie} \chi (-2 + 5 \chi^3 - 13 \chi^5 + 
         10 \chi^7)) \nonumber \\ &+ 
   2688 f_{20}^{in} \chi (6 + \sigma_0^{in} \chi^2) (16 - 
      11 \lambda_{an} + \chi^2 (-8 + 3 \lambda_{an} + 
         4 (-1 + \lambda_{an})\nonumber \\ & \chi^3 (-1 + 3 \chi^2))) - 
   192 (f_{20}^{in})^2 \sqrt{5/\pi} \chi (18 + 5 \sigma_0^{in} \chi^2) (16 - 
      11 \lambda_{an} +\nonumber \\ & \chi^2 (-8 + 3 \lambda_{an} + 
         4 (-1 + \lambda_{an}) \chi^3 (-1 + 3 \chi^2))) + 
   384 f_{2-2}^{in} f_{22}^{in} \sqrt{5/\pi} \chi \nonumber \\ &(18 + 5 \sigma_0^{in} \chi^2) (16 - 
      11 \lambda_{an} + \chi^2 (-8 + 3 \lambda_{an} + 
         4 (-1 + \lambda_{an}) \chi^3 \nonumber \\ &(-1 + 
            3 \chi^2))))/(28 Ca \chi ((32 + 
        23 \lambda_{an}) (32 \lambda_{an} + 23 \lambda_{ie}) - 
     25 (-8 + \lambda_{an}) \nonumber \\ &(16 \lambda_{an} - 23 \lambda_{ie}) \chi^3 - 
     84 (4 + \lambda_{an}) (8 \lambda_{an} - 23 \lambda_{ie}) \chi^5 - 100 (2 + \lambda_{an})\nonumber \\ & (4 \lambda_{an} + 
        23 \lambda_{ie}) \chi^7 + 736 (-1 + \lambda_{an}) (\lambda_{an} - \lambda_{ie}) \chi^{10})) \label{exevol2}&
\end{align}
\begin{align}
& \dot{f}_{2-2}^{ex}=((-(35 Ca (\sqrt{6} C^\infty_{2-20} + 9 C^\infty_{2-22}) - 
        24 f_{2-2}^{ex} (7 (6 + \sigma_0^{ex}) +\nonumber \\ & 
           f_{20}^{ex} \sqrt{5/\pi} (18 + 
              5 \sigma_0^{ex}))) (16 \lambda_{an} \chi (-8 - 
          10 \chi^3 + 13 \chi^5 + 5 \chi^7) + \nonumber \\ &
       46 \lambda_{ie} \chi (-2 + 5 \chi^3 - 13 \chi^5 + 
          10 \chi^7)) + 
    384 f_{2-2}^{in} \chi (7 (6 + \sigma_0^{in} \chi^2) + \nonumber \\ &
       f_{20}^{in} \sqrt{5/\pi} (18 + 5 \sigma_0^{in} \chi^2)) (16 - 
       11 \lambda_{an} + \chi^2 (-8 + 3 \lambda_{an} + \nonumber \\ &
          4 (-1 + \lambda_{an}) \chi^3 (-1 + 
             3 \chi^2))))/(28 Ca \chi ((32 + 
         23 \lambda_{an}) (32 \lambda_{an} \nonumber \\ &+ 23 \lambda_{ie}) - 
      25 (-8 + \lambda_{an}) (16 \lambda_{an} - 
         23 \lambda_{ie}) \chi^3 - 
      84 (4 + \lambda_{an}) (8 \lambda_{an}\nonumber \\ & - 
         23 \lambda_{ie}) \chi^5 - 
      100 (2 + \lambda_{an}) (4 \lambda_{an} + 
         23 \lambda_{ie}) \chi^7 + 
      736 (-1 + \lambda_{an}) \nonumber \\ &(\lambda_{an} - \lambda_{ie}) \chi^{10})) +
  i f_{2-2}^{ex} \omega_{ex}) \label{exevol3}&
\end{align}
For inner vesicle from eq. \ref{evoleq-1}
\begin{align}
&\dot{f}_{22}^{in} =1/2 (-(((5 - 3 \chi^2) (-1 + 
          3 \chi^2) (10 \lambda_{an} (35 Ca (\sqrt{6}
                  C^\infty_{220} + 9 C^\infty_{222}) \nonumber \\ & - 
             24 f_{22}^{ex} (7 (6 + \sigma_0^{ex}) + 
                f_{20}^{ex} \sqrt{5/\pi} (18 + 5 \sigma_0^{ex}))) \chi^4 (-18 + 
             7 \chi^2 - 7 \chi^5  \nonumber \\ &+ 18 \chi^7) - 
          48 f_{22}^{in} \chi (-32 - 23 \lambda_{an} - 
             25 (-8 + \lambda_{an}) \chi^3 - 
             84 (4 + \lambda_{an}) \chi^5  \nonumber \\ &+ 
             100 (2 + \lambda_{an}) \chi^7 + 
             32 (-1 + \lambda_{an}) \chi^{10}) (7 (6 + \sigma_0^{in} \chi^2) + 
             f_{20}^{in} \sqrt{5/\pi}  \nonumber \\ &(18 + 
                5 \sigma_0^{in} \chi^2))))/(28 Ca \chi^6 ((32 + 
             23 \lambda_{an}) (32 \lambda_{an} + 23 \lambda_{ie}) - 
          25  \nonumber \\ &(-8 + \lambda_{an}) (16 \lambda_{an} - 
             23 \lambda_{ie}) \chi^3 - 
          84 (4 + \lambda_{an}) (8 \lambda_{an} - 
             23 \lambda_{ie}) \chi^5 -  \nonumber \\ &
          100 (2 + \lambda_{an}) (4 \lambda_{an} + 
             23 \lambda_{ie}) \chi^7 + 
          736 (-1 + \lambda_{an}) (\lambda_{an} - \lambda_{ie}) \
\chi^{10})))  \nonumber \\ &- ((-1 + \chi^2) (-5 + 
        9 \chi^2) (10 \lambda_{an} (35 Ca (\sqrt{6} C^\infty_{220} + 
              9 C^\infty_{222}) -  \nonumber \\ &24 f_{22}^{ex} (7 (6 + \sigma_0^{ex}) + 
              f_{20}^{ex} \sqrt{5/\pi} (18 + 5 \sigma_0^{ex}))) \chi^4 (-18 + 
           7 \chi^2 - 7 \chi^5  \nonumber \\ &+ 18 \chi^7) - 
        48 f_{22}^{in} \chi (-32 - 23 \lambda_{an} - 
           25 (-8 + \lambda_{an}) \chi^3 - 
           84 (4 + \lambda_{an}) \chi^5  \nonumber \\ &+ 
           100 (2 + \lambda_{an}) \chi^7 + 
           32 (-1 + \lambda_{an}) \chi^{10}) (7 (6 + \sigma_0^{in} \
\chi^2) + 
           f_{20}^{in} \sqrt{5/\pi}  \nonumber \\ &(18 + 
              5 \sigma_0^{in} \chi^2))))/(28 Ca \chi^6 ((32 + 
           23 \lambda_{an}) (32 \lambda_{an} + 23 \lambda_{ie}) - 
        25  \nonumber \\ &(-8 + \lambda_{an}) (16 \lambda_{an} - 
           23 \lambda_{ie}) \chi^3 - 
        84 (4 + \lambda_{an}) (8 \lambda_{an} - 
           23 \lambda_{ie}) \chi^5 -  \nonumber \\ &
        100 (2 + \lambda_{an}) (4 \lambda_{an} + 
           23 \lambda_{ie}) \chi^7 + 
        736 (-1 + \lambda_{an}) (\lambda_{an} - \lambda_{ie}) \
\chi^{10}))  \nonumber \\ &+ 2 i f_{22}^{in} \omega_{in})  \label{exevol4}&
\end{align}
\begin{align}
&\dot{f}_{20}^{in}=-((5 \lambda_{an} (35 Ca (\sqrt{6} C^\infty_{200} + 
           9 C^\infty_{202}) + 
        12 (-14 f_{20}^{ex} (6 + \sigma_0^{ex}) + \nonumber \\ & 
           (f_{20}^{ex})^2 \sqrt{5/\pi} (18 + 5 \sigma_0^{ex}) - 
           2 f_{2-2}^{ex} f_{22}^{ex} \sqrt{5/\pi} (18 + 5 \sigma_0^{ex}))) \chi^3 (-18  \nonumber \\ &+ 
        7 \chi^2 - 7 \chi^5 + 18 \chi^7) - 
     168 f_{20}^{in} (6 + \sigma_0^{in} \chi^2) (-32 - 23 \lambda_{an} - 
        25  \nonumber \\ &(-8 + \lambda_{an}) \chi^3 - 
        84 (4 + \lambda_{an}) \chi^5 + 
        100 (2 + \lambda_{an}) \chi^7 + 
        32 (-1 + \lambda_{an}) \chi^{10}) \nonumber \\ & + 
     12 (f_{20}^{in})^2 \sqrt{5/\pi} (18 + 5 \sigma_0^{in} \chi^2) (-32 - 23 \lambda_{an} - 
        25 (-8 + \lambda_{an}) \chi^3 -  \nonumber \\ &
        84 (4 + \lambda_{an}) \chi^5 + 
        100 (2 + \lambda_{an}) \chi^7 + 
        32 (-1 + \lambda_{an}) \chi^{10}) -  \nonumber \\ &
     24 f_{2-2}^{in} f_{22}^{in} \sqrt{5/\pi} (18 + 5 \sigma_0^{in} \chi^2) (-32 - 23 \lambda_{an} - 
        25 (-8 + \lambda_{an}) \chi^3 - \nonumber \\ & 
        84 (4 + \lambda_{an}) \chi^5 + 
        100 (2 + \lambda_{an}) \chi^7 + 
        32 (-1 + \lambda_{an}) \chi^{10}))/ \nonumber \\ &(7 Ca \chi^3 ((32 + 23 \lambda_{an}) (32 \lambda_{an} + 23 \lambda_{ie}) - 
       25 (-8 + \lambda_{an}) (16 \lambda_{an} -  \nonumber \\ &
          23 \lambda_{ie}) \chi^3 - 
       84 (4 + \lambda_{an}) (8 \lambda_{an} - 
          23 \lambda_{ie}) \chi^5 - 
       100 (2 + \lambda_{an}) (4 \lambda_{an} +  \nonumber \\ &
          23 \lambda_{ie}) \chi^7 + 
       736 (-1 + \lambda_{an}) (\lambda_{an} - \lambda_{ie}) \chi^{10}))) \label{exevol5}&
\end{align}
\begin{align}
&\dot{f}_{2-2}^{in}=1/2 (-(((5 - 3 \chi^2) (-1 + 
          3 \chi^2) (10 \lambda_{an} (35 Ca (\sqrt{6}
                  C^\infty_{2-20} + 9 C^\infty_{2-22}) \nonumber \\ & - 
             24 f_{2-2}^{ex} (7 (6 + \sigma_0^{ex}) + 
                f_{20}^{ex} \sqrt{5/\pi} (18 + 5 \sigma_0^{ex}))) \chi^4 (-18 + 
             7 \chi^2 - 7 \chi^5  \nonumber \\ &+ 18 \chi^7) - 
          48 f_{2-2}^{in} \chi (-32 - 23 \lambda_{an} - 
             25 (-8 + \lambda_{an}) \chi^3 - 
             84 (4 + \lambda_{an}) \chi^5  \nonumber \\ &+ 
             100 (2 + \lambda_{an}) \chi^7 + 
             32 (-1 + \lambda_{an}) \chi^{10}) (7 (6 + \sigma_0^{in} \chi^2) + 
             f_{20}^{in} \sqrt{5/\pi}  \nonumber \\ &(18 + 
                5 \sigma_0^{in} \chi^2))))/(28 Ca \chi^6 ((32 + 
             23 \lambda_{an}) (32 \lambda_{an} + 23 \lambda_{ie}) - 
          25  \nonumber \\ &(-8 + \lambda_{an}) (16 \lambda_{an} - 
             23 \lambda_{ie}) \chi^3 - 
          84 (4 + \lambda_{an}) (8 \lambda_{an} - 
             23 \lambda_{ie}) \chi^5 -  \nonumber \\ &
          100 (2 + \lambda_{an}) (4 \lambda_{an} + 
             23 \lambda_{ie}) \chi^7 + 
          736 (-1 + \lambda_{an}) (\lambda_{an} - \lambda_{ie}) \
\chi^{10})))  \nonumber \\ &- ((-1 + \chi^2) (-5 + 
        9 \chi^2) (10 \lambda_{an} (35 Ca (\sqrt{6} C^\infty_{2-20} + 
              9 C^\infty_{2-22}) -  \nonumber \\ &24 f_{2-2}^{ex} (7 (6 + \sigma_0^{ex}) + 
              f_{20}^{ex} \sqrt{5/\pi} (18 + 5 \sigma_0^{ex}))) \chi^4 (-18 + 
           7 \chi^2 - 7 \chi^5  \nonumber \\ &+ 18 \chi^7) - 
        48 f_{2-2}^{in} \chi (-32 - 23 \lambda_{an} - 
           25 (-8 + \lambda_{an}) \chi^3 - 
           84 (4 + \lambda_{an}) \chi^5  \nonumber \\ &+ 
           100 (2 + \lambda_{an}) \chi^7 + 
           32 (-1 + \lambda_{an}) \chi^{10}) (7 (6 + \sigma_0^{in} \
\chi^2) + 
           f_{20}^{in} \sqrt{5/\pi}  \nonumber \\ &(18 + 
              5 \sigma_0^{in} \chi^2))))/(28 Ca \chi^6 ((32 + 
           23 \lambda_{an}) (32 \lambda_{an} + 23 \lambda_{ie}) - 
        25  \nonumber \\ &(-8 + \lambda_{an}) (16 \lambda_{an} - 
           23 \lambda_{ie}) \chi^3 - 
        84 (4 + \lambda_{an}) (8 \lambda_{an} - 
           23 \lambda_{ie}) \chi^5 -  \nonumber \\ &
        100 (2 + \lambda_{an}) (4 \lambda_{an} + 
           23 \lambda_{ie}) \chi^7 + 
        736 (-1 + \lambda_{an}) (\lambda_{an} - \lambda_{ie}) \
\chi^{10}))  \nonumber \\ &+ 2 i f_{2-2}^{in} \omega_{in}) \label{exevol6}&
\end{align}
where $f_{20}^{ex}=\sqrt{A_{ex}/2} \sin\theta_{ex}, f_{20}^{in}=\sqrt{A_{in}/2} \sin\theta_{in}$ 

The uniform tensions on the two vesicles ($\sigma_0^{ex}, \sigma_0^{in}$) were estimated by using eq. \ref{exevol1} to \ref{exevol6} in area conservation constraint for two vesicles, i.e.,
\begin{align}
& \ f_{2-2}^{ex} \dot{f}_{22}^{ex} + f_{22}^{ex} \dot{f}_{2-2}^{ex} + f_{20}^{ex} \dot{f}_{20}^{ex}= 0 
 \label{areacon-ex}&\\ 
& \ f_{2-2}^{in} \dot{f}_{22}^{in} + f_{22}^{in} \dot{f}_{2-2}^{in} + f_{20}^{in} \dot{f}_{20}^{in} = 0 \label{areacon-in}&
\end{align}
Solving these two equations simultaneously (by substituting eq. \ref{exevol1} to \ref{exevol6}) gives the uniform tensions in the inner and outer vesicle ($\sigma_0^{in}, \sigma_0^{ex}$). \\
The final evolution is obtained by replacing all uniform tension expressions in eq. \ref{exevol1} to \ref{exevol6} from the expressions obtained above from eq. \ref{areacon-ex} and \ref{areacon-in}. 

\section{Appendix plots}\label{App8}
\begin{figure} [tp] %[tbp]
    \hspace{0.0cm}
    \begin{subfigure}[b]{0.28\linewidth}       \includegraphics[width=\linewidth]{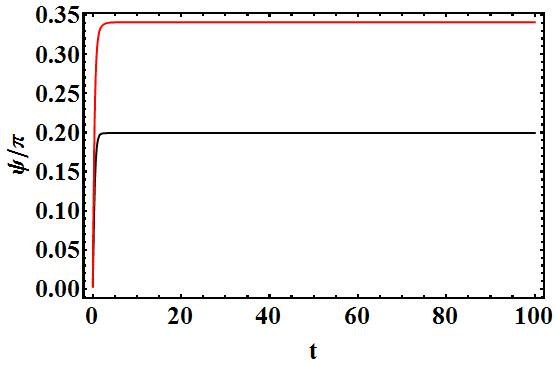}
     \caption{$\chi=0.3, \lambda_{an}=1, \Lambda_{an}=0.517$}
     \end{subfigure}
     \hspace{1.2cm}
      \begin{subfigure}[b]{0.28\linewidth}
      \includegraphics[width=\linewidth]{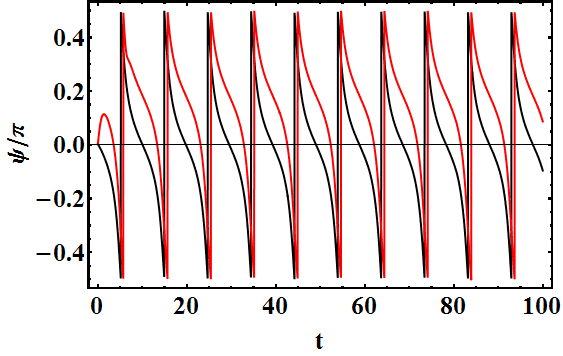}
      \caption{$\chi=0.3, \lambda_{an}=8, \Lambda_{an}=1.244$}
     \end{subfigure} 
 \hspace{1.2cm}
      \begin{subfigure}[b]{0.28\linewidth}
      \includegraphics[width=\linewidth]{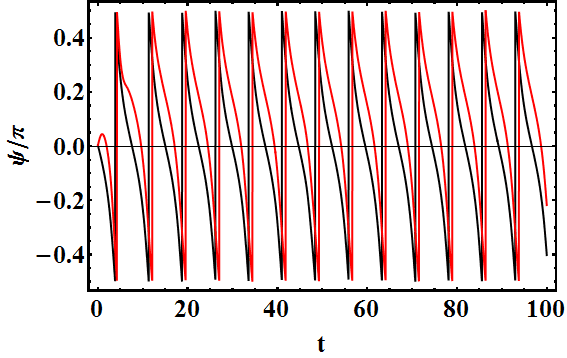}
      \caption{$\chi=0.3, \lambda_{an}=12, \Lambda_{an}=1.774$}
     \end{subfigure} 
    \hspace{0.12cm}
    \begin{subfigure}[b]{0.28\linewidth}       \includegraphics[width=\linewidth]{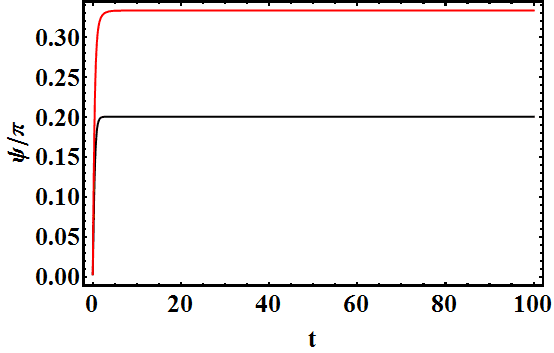}
     \caption{$\chi=0.5, \lambda_{an}=1, \Lambda_{an}=0.517$}
     \end{subfigure}
     \hspace{0.12cm}
      \begin{subfigure}[b]{0.28\linewidth}
      \includegraphics[width=\linewidth]{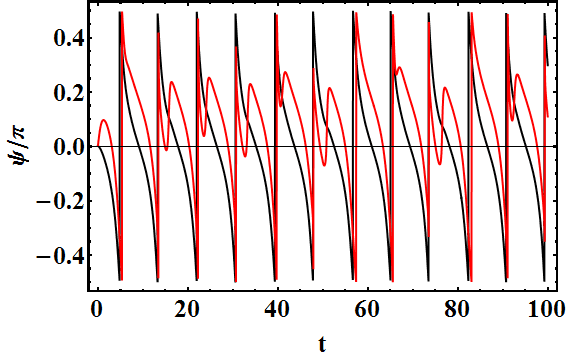}
      \caption{$\chi=0.5, \lambda_{an}=8, \Lambda_{an}=1.244$}
     \end{subfigure} 
 \hspace{0.12cm}
      \begin{subfigure}[b]{0.28\linewidth}
      \includegraphics[width=\linewidth]{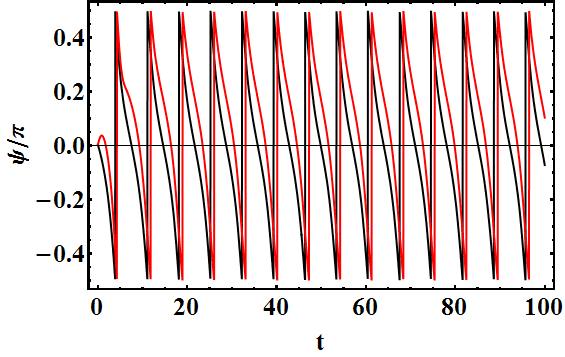}
      \caption{$\chi=0.5, \lambda_{an}=12, \Lambda_{an}=1.774$}
     \end{subfigure} 
     \hspace{0.12cm}
     \begin{subfigure}[b]{0.28\linewidth}       \includegraphics[width=\linewidth]{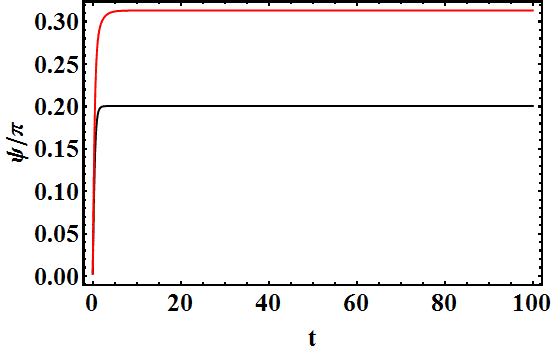}
      \caption{$\chi=0.7, \lambda_{an}=1, \Lambda_{an}=0.517$}
      \end{subfigure}
      \hspace{0.12cm}
       \begin{subfigure}[b]{0.28\linewidth}
       \includegraphics[width=\linewidth]{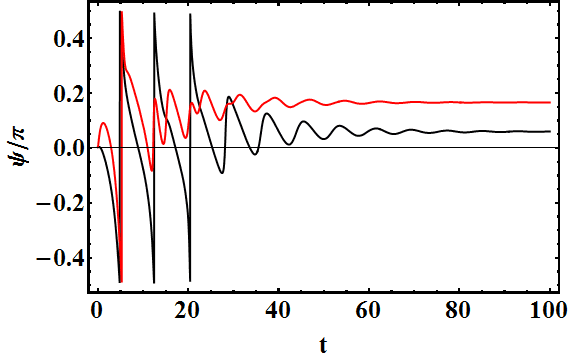}
       \caption{$\chi=0.7, \lambda_{an}=8, \Lambda_{an}=1.244$}
      \end{subfigure} 
  \hspace{0.12cm}
       \begin{subfigure}[b]{0.28\linewidth}
       \includegraphics[width=\linewidth]{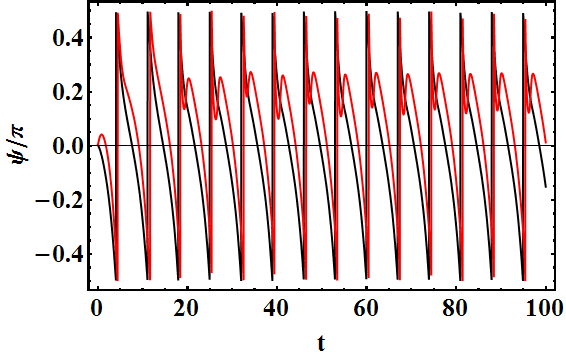}
       \caption{$\chi=0.7, \lambda_{an}=12, \Lambda_{an}=1.774$}
      \end{subfigure} 
    \hspace{0.12cm}
      \begin{subfigure}[b]{0.28\linewidth}       \includegraphics[width=\linewidth]{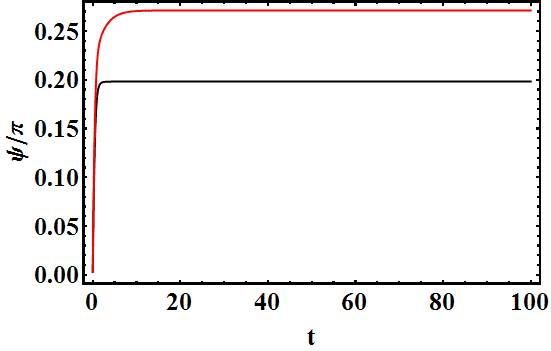}
       \caption{$\chi=0.9, \lambda_{an}=1, \Lambda_{an}=0.517$}
       \end{subfigure}
       \hspace{1.2cm}
        \begin{subfigure}[b]{0.28\linewidth}
        \includegraphics[width=\linewidth]{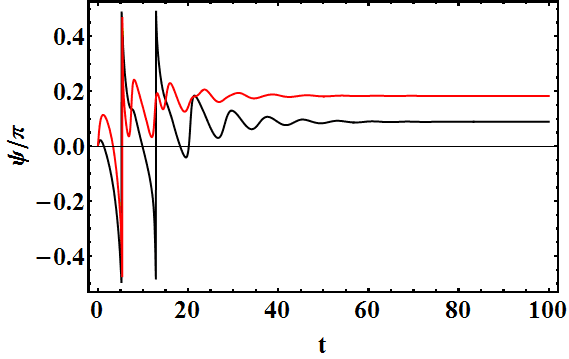}
        \caption{$\chi=0.9, \lambda_{an}=8, \Lambda_{an}=1.244$}
       \end{subfigure} 
   \hspace{1.2cm}
        \begin{subfigure}[b]{0.28\linewidth}
        \includegraphics[width=\linewidth]{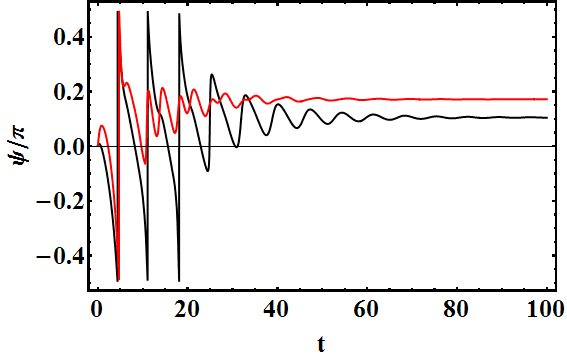}
        \caption{$\chi=0.9, \lambda_{an}=12, \Lambda_{an}=1.774$}
       \end{subfigure}                      
 \caption{Variation of the inclination angle of the outer (black) and the inner (red) vesicles with time for $\lambda_{in}=1, \Delta_{in}=\Delta_{ex}=0.2$ using the leading order theory.}
    \label{psivst_leading}
\end{figure}

\begin{figure} [ht] %[tbp]
\centering
      \begin{subfigure}[b]{0.6\linewidth}
      \includegraphics[width=\linewidth]{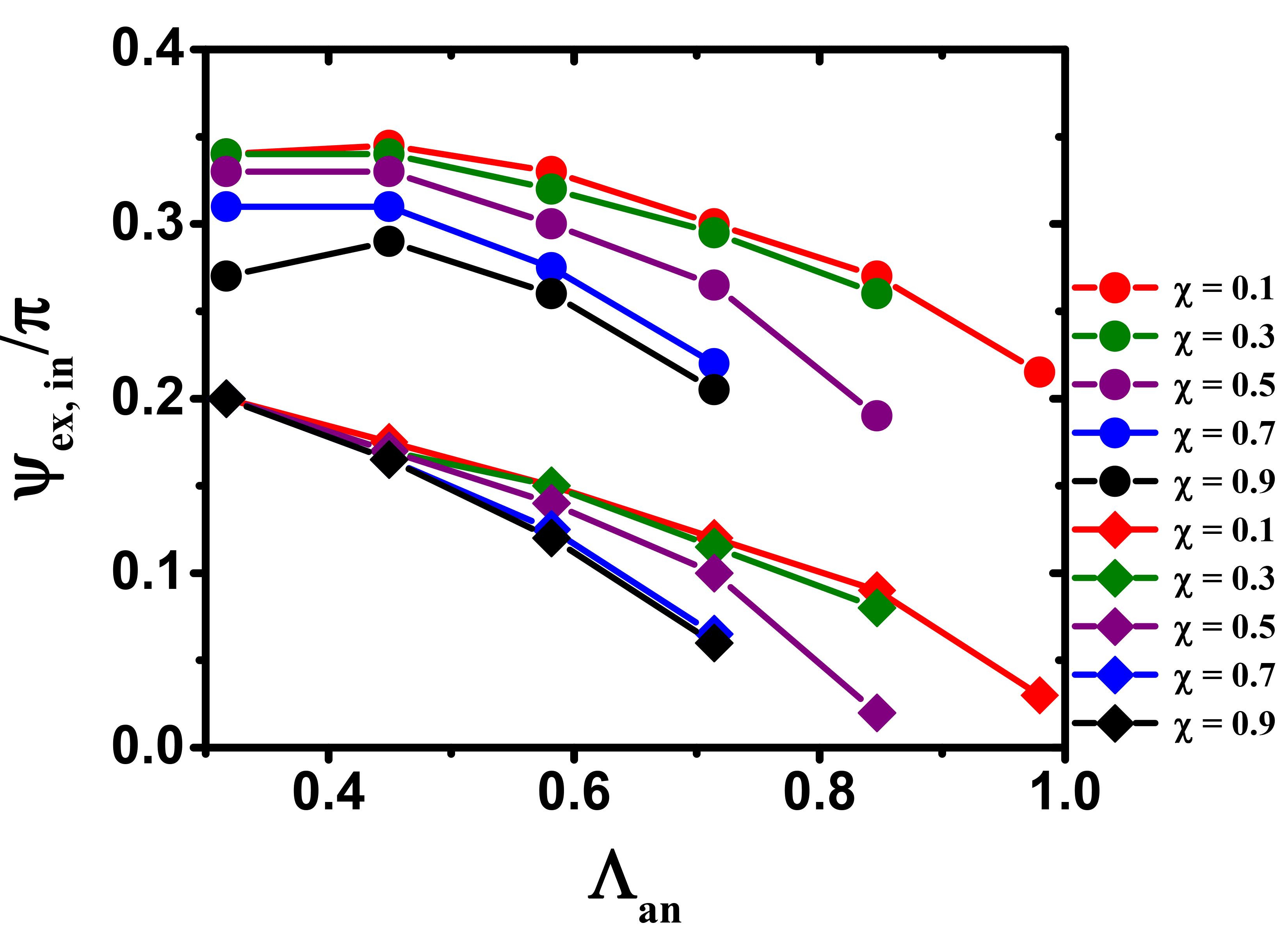}
     \end{subfigure}               
 \caption{Variation of the inclination angle of the inner (solid circles) and the outer (solid diamonds) vesicle with $\Lambda_{an}$ across the outer vesicle for different $\chi$ using leading order theory ($\Delta_{in}=\Delta_{ex}=0.2$). The $\Lambda_{an}$ is realized by varying $\lambda_{an}$}
    \label{inclinationangle_leadord}
\end{figure}

\begin{figure} [htbp] %[tbp]
   \centering
      \begin{subfigure}[b]{0.49\linewidth}       \includegraphics[width=\linewidth]{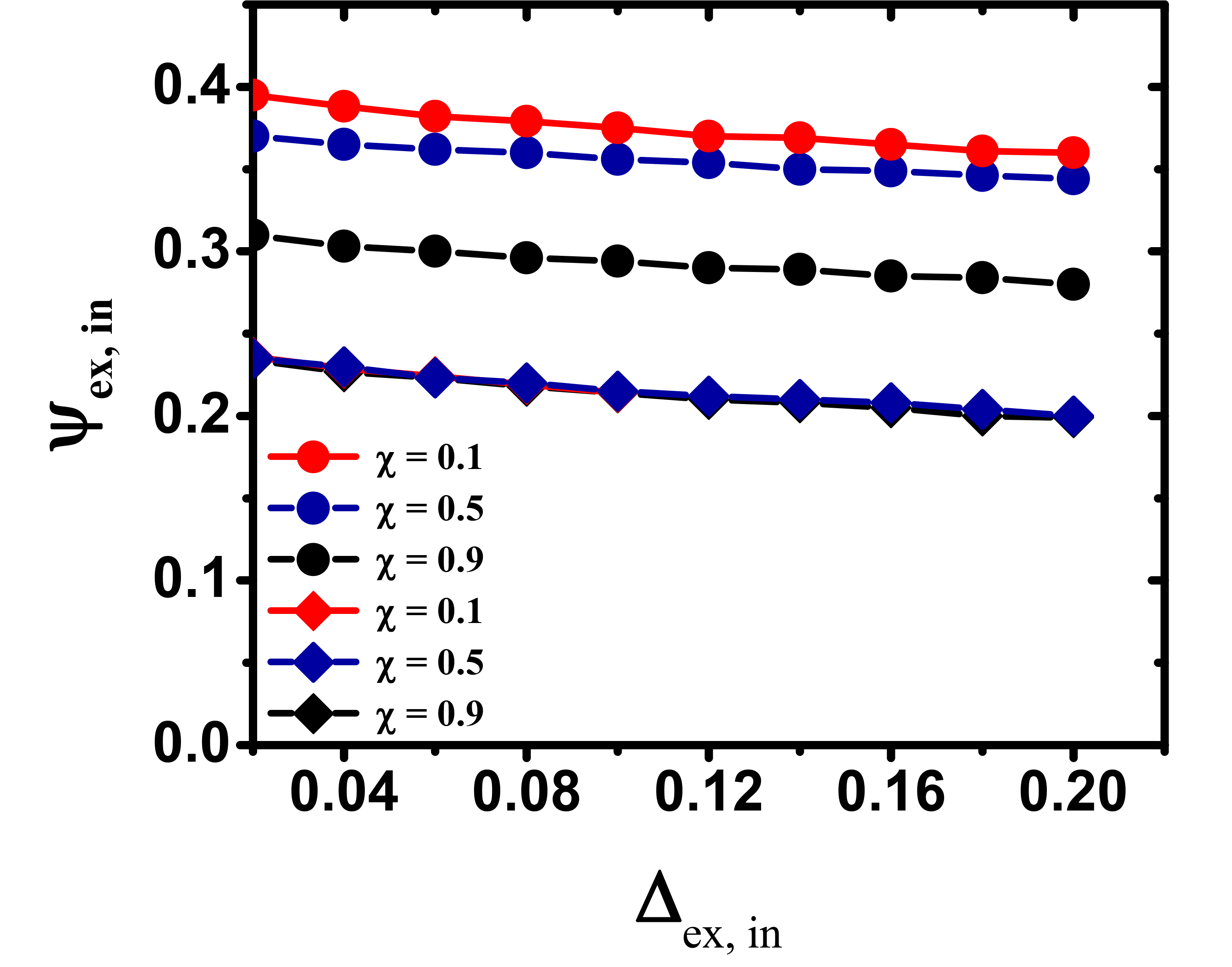}
       \caption{}
      \end{subfigure} 
      \hspace{0.cm}
   \begin{subfigure}[b]{0.49\linewidth}       \includegraphics[width=\linewidth]{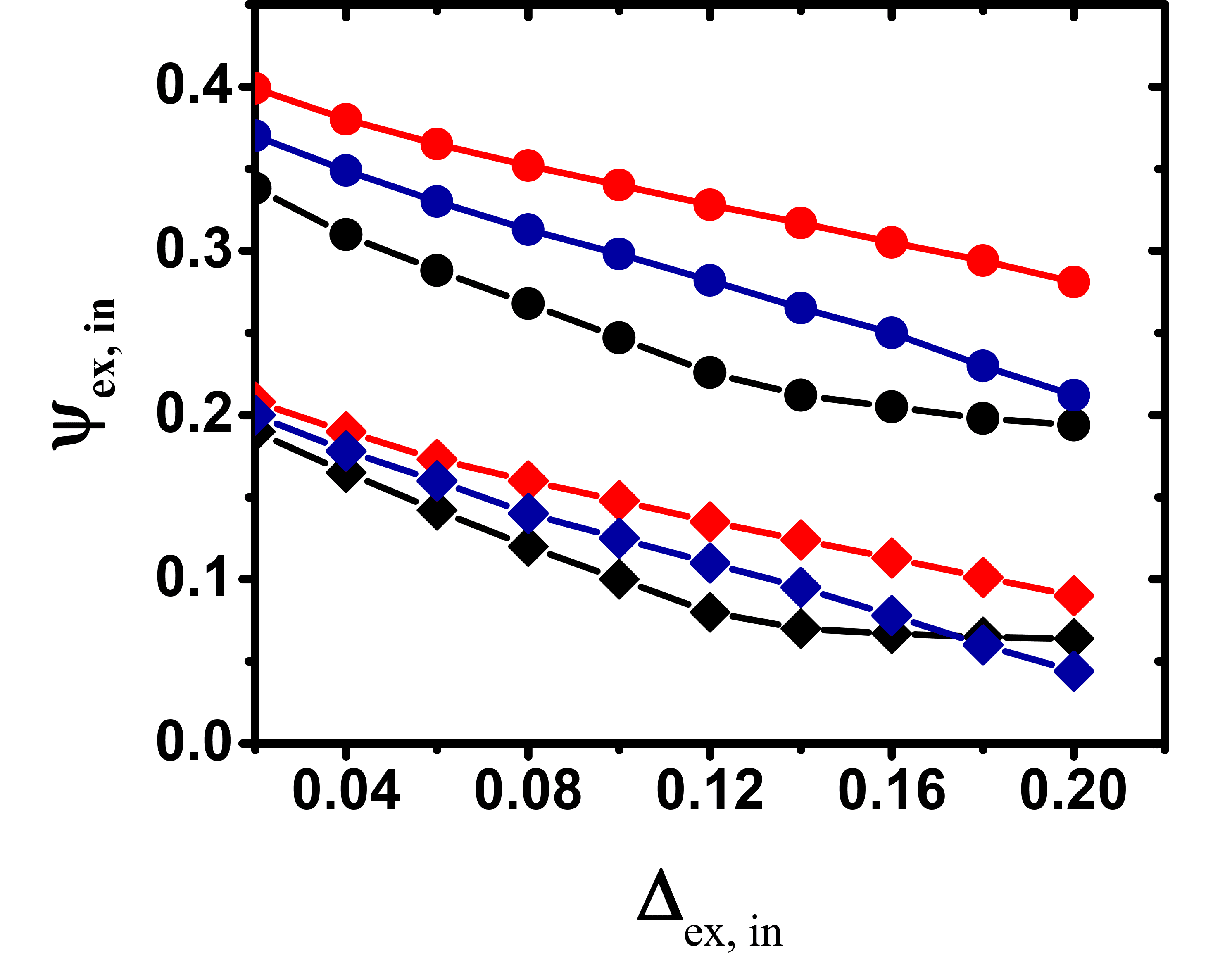}
   \caption{}
         \end{subfigure}                  
 \caption{Variation of inclination angle of the inner (solid circles) and the outer vesicle (solid diamonds) with excess area ($\Delta_{in}=\Delta_{ex}$) for (a) $\lambda_{an}=1$, (b) $\lambda_{an}=5$ ($Ca=1$).}
    \label{PSIvsA}
\end{figure}

\begin{figure} [bp] %[tbp]
   \centering
      \begin{subfigure}[b]{0.48\linewidth}       \includegraphics[width=\linewidth]{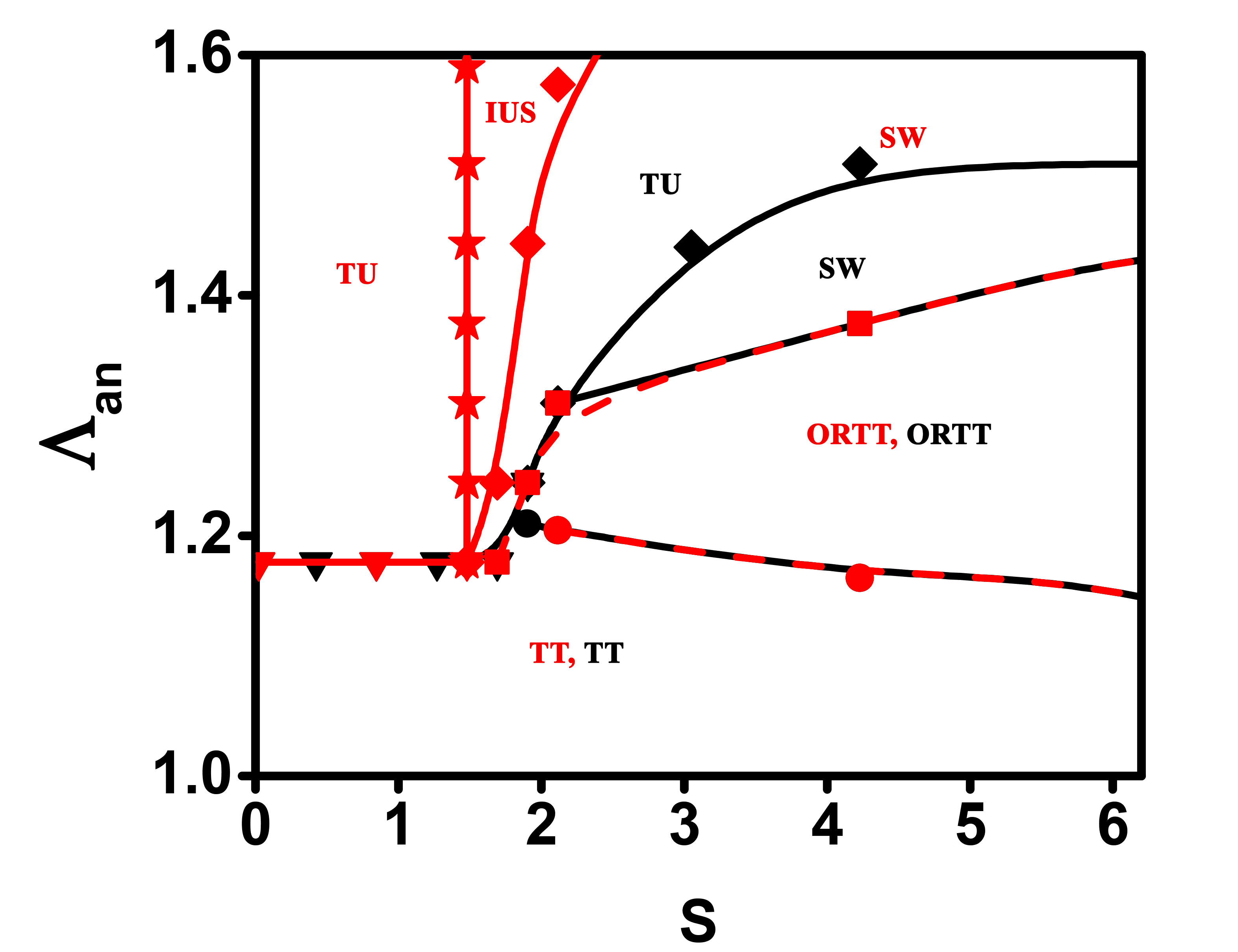}
       \caption{}
      \end{subfigure} 
      \hspace{0.2cm}
   \begin{subfigure}[b]{0.48\linewidth}       \includegraphics[width=\linewidth]{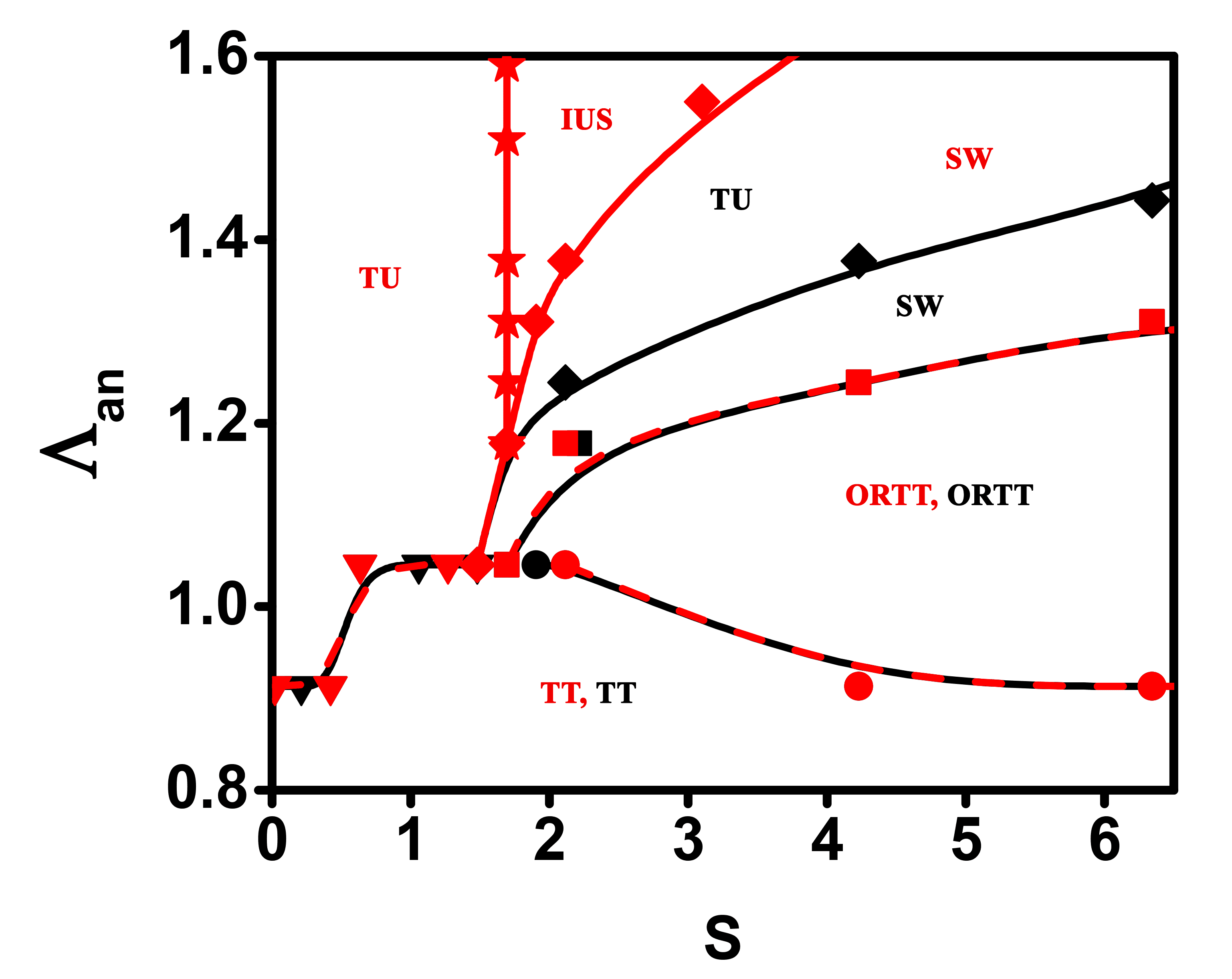}
   \caption{}
         \end{subfigure}                  
 \caption{Inset picture (low $Ca$ or $S$ region) of fig \ref{combinedPD}a, and fig \ref{combinedPD}b for (a) $\chi=0.1$, and (b) $\chi=0.5$, respectively.}
    \label{inset-phasediag}
\end{figure}

\begin{figure} [ht] %[tbp]
   \centering
      \begin{subfigure}[b]{0.6\linewidth}       \includegraphics[width=\linewidth]{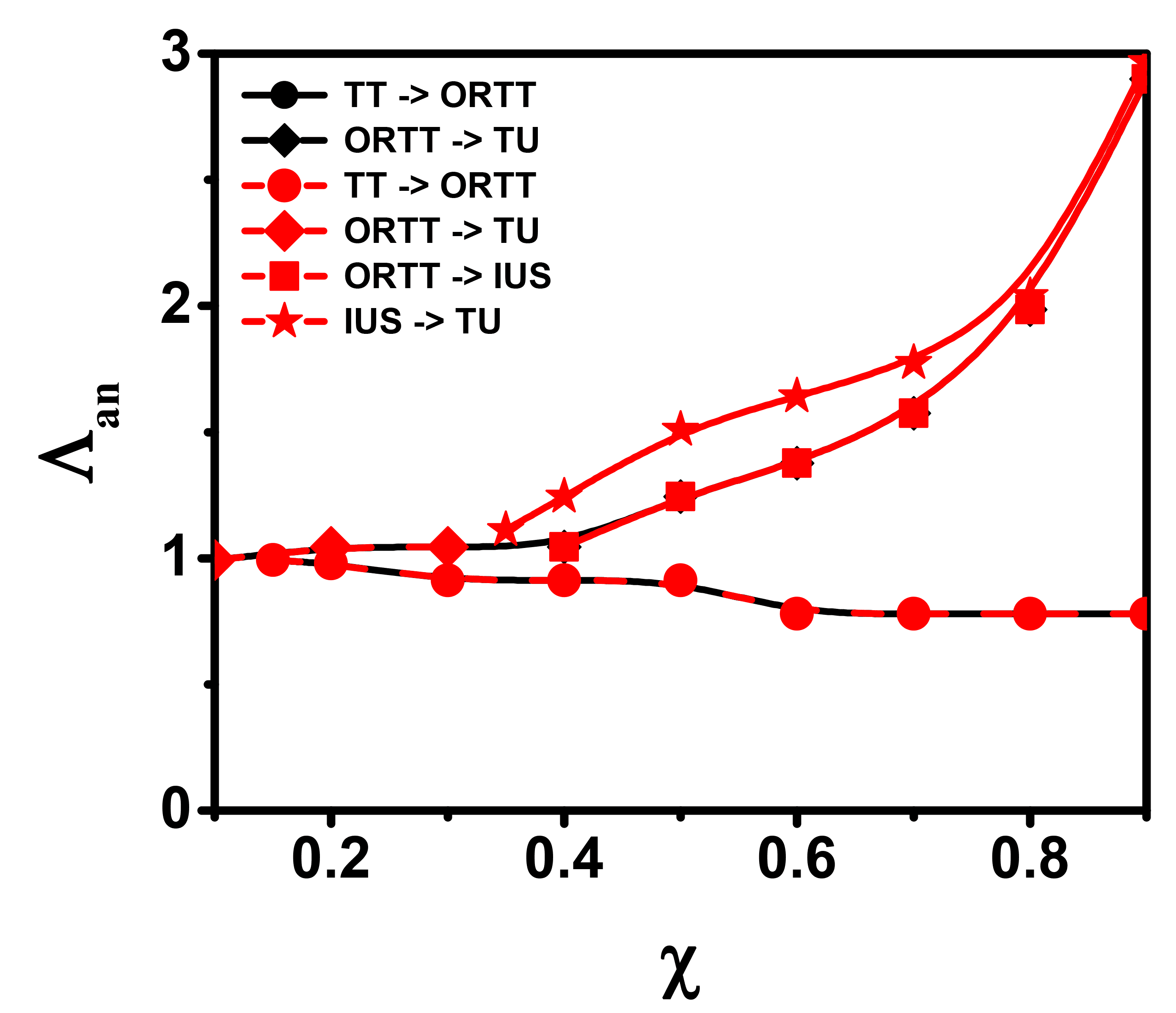}
      \end{subfigure}               
 \caption{$\Lambda_{an}$ vs $\chi$ phase diagram for the inner (red line) and the outer vesicle (black line) in linear shear flow ($\Delta_{in}=\Delta_{ex}=0.2, \lambda_{in}=1$) using leading order theory. All legends presents transition from bottom to top.}
    \label{viscoutvschi_leading}
\end{figure}

\end{document}